\documentclass[preprint,10pt,3p]{elsarticle}

\usepackage{graphicx}
\usepackage{amssymb}
\usepackage{paralist}
\usepackage{amsmath,mathtools}
\usepackage{float}
\usepackage{multirow}
\usepackage[table]{xcolor}
\usepackage{tikz}
\usepackage{csquotes}
\biboptions{sort&compress}
\usepackage{rotating}

\usepackage{url}
\usepackage{doi}
\usepackage{color}
\usepackage{graphicx}
\usepackage{subcaption}
\usepackage{bbm}
\usepackage{nomencl}
\usepackage{algorithmicx,algorithm,algpseudocode}
\makenomenclature

\allowdisplaybreaks

\newcommand{\diag}{\mathop{\rm diag}}

\hypersetup{
	colorlinks   = true, 
	urlcolor     = blue, 
	linkcolor    = blue, 
	citecolor    = blue 
}

\journal{Transportation Research Part C: Emerging Technologies}
\begin{document}
\begin{frontmatter}


\title{Noise-Aware and Equitable Urban Air Traffic Management:\\An Optimization Approach}





\author{Zhenyu Gao\footnote{Corresponding Author. Postdoctoral Fellow, Department of Aerospace Engineering and Engineering Mechanics}, Yue Yu\footnote{Postdoctoral Fellow, Oden Institute for Computational Engineering and Sciences}, Qinshuang Wei\footnote{Postdoctoral Fellow, Oden Institute for Computational Engineering and Sciences}, Ufuk Topcu\footnote{Temple Foundation Endowed Professorship No. 1}, and John-Paul Clarke\footnote{Professor and Ernest Cockrell, Jr. Memorial Chair in Engineering}}

\address{The University of Texas at Austin, Austin, TX 78712, United States}

\begin{abstract}

Urban air mobility (UAM), a transformative concept for the transport of passengers and cargo, faces several integration challenges in complex urban environments. Community acceptance of aircraft noise is among the most noticeable of these challenges when launching or scaling up a UAM system. Properly managing community noise is fundamental to establishing a UAM system that is environmentally and socially sustainable. In this work, we develop a holistic and equitable approach to manage UAM air traffic and its community noise impact in urban environments. The proposed approach is a hybrid approach that considers a mix of different noise mitigation strategies, including limiting the number of operations, cruising at higher altitudes, and ambient noise masking. We tackle the problem through the lens of network system control and formulate a multi-objective optimization model for managing traffic flow in a multi-layer UAM network while concurrently pursuing demand fulfillment, noise control, and energy saving. Further, we use a social welfare function in the optimization model as the basis for the efficiency-fairness trade-off in both demand fulfillment and noise control. We apply the proposed approach to a comprehensive case study in the city of Austin and perform design trade-offs through both visual and quantitative analyses. 

\end{abstract}


\end{frontmatter}



\section{Introduction}\label{sec:intro}


Urban air mobility (UAM) is a game-changing concept set to transform urban transportation by extending the current system into a third domain -- the sky. The advancement of UAM hinges on several key factors: novel aircraft configurations, advanced battery technologies, strides in autonomy and robotics, augmented air traffic management systems, and innovative business models~\cite{gao2023noise}. As envisioned, UAM will primarily operate electric vertical take-off and landing (eVTOL) aircraft, hybrid-electric ultra-short takeoff and landing (eSTOL) aircraft, and unmanned aerial vehicles (UAVs) at lower altitudes within or traversing metropolitan areas. These operations will support a range of urban services, including passenger mobility, cargo delivery, infrastructure monitoring, public safety, and emergency response~\citep{cohen2021urban}. Undoubtedly, UAM will revolutionize the transport of passengers and cargo and promote public welfare, particularly in currently underserved local and regional environments.

However, incorporating UAM into existing urban environments introduces a range of intricate challenges. Drawing from the lessons of past urban helicopter services and recent analyses of UAM's operational constraints, it becomes clear that community acceptance, system safety, and infrastructure are critical obstacles during the implementation or scale-up of UAM systems~\cite{vascik2018analysis,vascik2017constraint,pons2022understanding}. To promote community/societal acceptance of UAM, solutions to mitigate its community noise impact are of utmost importance~\cite{vascik2017evaluation,pons2022understanding,gao2023noise}. Given that UAM operates in closer proximity to urban residents, the range of noises generated by UAM aircraft poses significant health risks, including psychological discomfort, sleep disruption, and an increased risk of cardiovascular diseases~\cite{gao2022multi}. Therefore, modeling and mitigation of UAM noise are now at the forefront of research in the aerospace community~\cite{rizzi2022prediction,rizzi2021comparison,rimjha2021urban,ng2022noise,yunus2023efficient,bian2021assessment}. 

Overall, there are many operational strategies for mitigating UAM's community noise impact, such as: (1) limiting the number of operations, (2) cruising at higher altitudes, and (3) ambient noise masking which concentrates flights over less noise-sensitive areas. Practical approaches to implementing these noise mitigation strategies include air traffic flow management and flight trajectory planning. In addition, UAM will operate as a network system, wherein aerial vehicles navigating through flight corridors that link nodes comprising vertiports and waypoints. On the control of UAM traffic flow in a network system, existing studies~\cite{kai2022vertiport,yu2023vertiport,wei2023risk,willey2021uamnetwork,wang2022complexity,yu2023alpha} have explored various objectives, such as demand fulfillment, congestion relief, risk management, traffic complexity reduction, and fairness. 
Of the three identified strategies for UAM noise mitigation, most of the current research has centered exclusively on one specific strategy and has not yet rigorously incorporated the intriguing concept of ambient noise masking into a solution.
A more holistic solution would be to employ a hybrid approach and utilize a mix of different noise mitigation strategies. This requires a more integrated approach that manages UAM operations in the context of control of network systems, which remains an unaddressed research gap in the literature. 

In this study, we aim to develop an optimization approach that considers a mix of all three noise mitigation strategies, with an emphasis on ambient noise masking, to reduce UAM's community noise impact in urban environment. We consider a scenario where an urban area consists of many communities, and the ambient noise level for each community is accessible through real-world data. We then manage the UAM traffic flow within a multi-layer network over the city to control UAM's noise impact, while striving to accommodate more service demand and consume less energy. Our goal is to ensure that the impact of UAM noise is both (1) limited -- not exceeding a maximum threshold above each community's ambient noise level, and (2) equitable -- maintaining fairness in the increased community noise levels. We summarize the four main contributions of this work as follows:
\begin{enumerate}
    \item \textit{Formulating an optimization problem} which leverages a hybrid approach to manage UAM's community noise impact. This multi-objective optimization formulation assigns air traffic flow in a UAM network while concurrently pursuing demand fulfillment, noise control, and energy saving. It can also perform efficiency-fairness trade-off in both demand fulfillment and noise control using a social welfare function. 
    \item \textit{Proposing a numerical method} that solves the optimization problem which has nonsmooth and nonconvex functions in both its objectives and constraints. This method replaces the nonsmooth functions with linear differentiable ones by introducing auxiliary variables and iteratively approximates the nonconvex function via the convex-concave procedure.
    \item \textit{Applying the proposed methodology} to a real-world application case for the city of Austin and demonstrating its efficacy. This case study includes real-world geospatial and ambient noise data for 292 census tracts (communities) in Austin and a multi-layer UAM network design that contains three cruising altitudes, 19 vertiports, 270 links, and 62 origin-destination (O-D) pairs.
    \item \textit{Conducting design trade-offs} in multiple aspects via visual and quantitative analytics. These include (i) intuitive data visualizations, (ii) efficiency-fairness trade-offs for demand fulfillment and noise increase, (iii) three-dimensional trade-offs between the average demand fulfillment, noise increase, and energy consumption, and (iv) detailed comparisons between various candidate design schemes.
\end{enumerate}

We organize the remainder of the paper as follows. In Section~\ref{sec:lit} we review the literature in three relevant streams and identify research gaps. In Section~\ref{sec:noise} we introduce details of noise measures, aircraft noise modeling, and community noise concerns. In Section~\ref{sec:network} we describe the multi-layer Austin UAM network for the case study. In Section~\ref{sec:pf} we formulate a network model and an optimization model, and a procedure for solving the optimization model. In Section~\ref{sec:results} we present results of the case study via design visualizations and quantitative comparisons. In Section~\ref{sec:remarks} we discusses the limitations and extensions of the study before Section~\ref{sec:conclusions} concludes the paper.

\section{Background and Literature Review}\label{sec:lit}

\subsection{UAM Operational Challenges}

By elevating urban transportation into the sky, UAM will unlock numerous benefits for the evolution of urban life and development. Yet its integration into urban environments also presents a variety of complex challenges. In the past, urban helicopter services struggled with challenges in community acceptance, safety concerns, and financial viability~\cite{vascik2018analysis}. Recently, researchers from MIT~\cite{vascik2017constraint,vascik2017evaluation} deduced that community acceptance of aircraft noise, availability of takeoff and landing area, and scalability of air traffic control are the three most stringent constraints during the growth of a UAM system. Similarly, an analysis by European researchers~\cite{pons2022understanding} identified five areas of implementation and operational challenges for UAM: technology, societal considerations, infrastructure, services, and policies. Every thorough examination of UAM's operational challenges places community/societal acceptance high on the list of stringent obstacles. 

Among the aspects crucial for societal acceptance, community noise stands out as a particularly prominent factor. Study~\cite{vascik2017evaluation} pinpointed aircraft noise as the only UAM operational constraint that has the highest severity level across all development stages; study~\cite{pons2022understanding} listed noise reduction as a challenge necessitating the most extended timeframe for resolution. Given the essential need for UAM systems to be environmentally and socially sustainable, addressing community noise has emerged as one of the top priorities.

\subsection{UAM Noise Modeling and Mitigation}

Efforts to model and mitigate UAM aircraft noise are now at the forefront of research in the aerospace community. Precise modeling of aircraft noise is essential for identifying effective noise mitigation solutions~\cite{gao2022probabilistic}. UAM aircraft generate different types of noise, with the primary source being rotor Blade-Vortex Interaction (BVI) noise generated from unsteady pressure fluctuations on blades~\cite{yu2000rotor}. On the noise modeling approach, the current practice predominantly depends on acoustic simulation due to the scarcity of measurement data for novel eVTOL aircraft configurations. Researchers from NASA~\cite{rizzi2022prediction,rizzi2021comparison} leveraged advanced acoustic models to predict the noise patterns of UAM aircraft and generated noise data for aviation noise analysis tools~\cite{rizzi2023modeling}. Such results have informed several studies on UAM's community noise impact in urban environment~\cite{rimjha2021urban,ng2022noise,rizzi2022second}. Other studies conducted detailed modeling of UAM's noise footprints in complex urban environments, such as those near vertiports~\cite{yunus2023efficient} and 3D flight trajectories~\cite{bian2021assessment}.

There are three main types of operational strategy for UAM noise mitigation: (1) limiting the number of operations, (2) maintaining a greater distance from objects, and (3) ambient noise masking which concentrates flights over less noise-sensitive areas such as areas with large ambient noise levels. While the first strategy focuses on limiting the number of operations/services, the latter two strategies focus on adjusting flight trajectory. Noise-aware flight trajectory planning has been an effective approach for noise reduction regardless of aircraft technologies. Within commercial aviation, noise abatement trajectories such as the continuous descent approach (CDA)~\cite{clarke2004continous} and the noise abatement departure procedures (NADP)~\cite{lim2020noise} have been extensively studied. Similar ideas have also been applied to design flight trajectories for helicopter~\cite{greenwood2017helicopter} and UAM~\cite{bian2021assessment} operations. Most recently, a new paradigm~\cite{gao2023noise} was introduced for noise mitigation through urban airspace management, i.e., defining noise-sensitive no-fly zones for urban noise control.

\subsection{UAM Network and Traffic Management}

The future UAM system, when operational in urban and suburban environments, is expected to function as a network system. Network design and air traffic management are two crucial problems to ensure effective and sustainable UAM operations. Both problems encompass multiple objectives in the design process and have garnered interest from researchers across a range of fields such as aerospace engineering, transportation, operations research, and urban planning. Existing studies in the literature have explored UAM network design and traffic flow management for various objectives, such as serving passenger demand~\cite{kai2022vertiport}, minimizing ground traffic congestion~\cite{yu2023vertiport}, managing traffic overflows due to physical and operational disruptions~\cite{wei2023risk}, incorporating the multi-leg operational strategy~\cite{willey2021uamnetwork}, reducing air traffic complexity~\cite{wang2022complexity}, and serving the demands of urban communities in a fair manner~\cite{yu2023alpha}.

In both air traffic management and the larger scope of operations management, the importance of equity and fairness is on the rise. Through incorporating fairness consideration, planners strive to equitably distribute both merits (e.g., resources) and downsides (e.g., delays) of a system among all participants. In the field of air transportation, the concept of fairness has been integrated into optimization problems related to service distribution~\cite{yu2023alpha}, delay management~\cite{montlaur2020flight,chin2021efficient}, airport slot allocation~\cite{ribeiro2018optimization, zografos2019bi}, conflict resolution~\cite{barnhart2012equitable}, and separation management~\cite{guo2022air}. In other important fields such as healthcare, fairness has also been considered in problems including kidney transplantation~\cite{bertsimas2013fairness} and healthcare facility location~\cite{shehadeh2023equity}. In future system design problems involving multiple participants/parties, the efficiency-fairness trade-off~\cite{bertsimas2012efficiency} is likely to emerge as a new norm.

\subsection{Research Gaps}

After an in-depth review of the literature in the three most pertinent research streams, we have identified the following research gaps:
\begin{enumerate}
    \item Within the three UAM noise mitigation strategies, the majority of existing studies have focused solely on the effects of operation volume and the distance between aircraft and ground structures or buildings. The potential of ambient noise masking in managing UAM noise remains an unexamined area so far.
    \item The existing approaches to noise-aware UAM operations design have solely focused on a single noise mitigation strategy. To the best of our knowledge, there has been no prior exploration into the use of a mixed approach of different noise mitigation strategies.
    \item No previous research has investigated UAM noise management problem in the context of control and optimization of network systems. Our effort marks the first step toward combining UAM noise management with control of network system.
\end{enumerate}

To address these research gaps, in this work we consider a noise-aware UAM traffic flow management problem in a multi-layer service network. Specifically, we simultaneously consider a mix of all three noise mitigation strategies, emphasizing on a novel strategy ambient noise masking, to control community noise in the urban environment. In addition, we also consider multiple trade-offs involving demand fulfillment, noise control, energy consumption, and equity in various aspects.

\section{UAM and Community Noise}\label{sec:noise} 

The noise consideration in this work refers to the unwanted sound produced by UAM aircraft or its components in flight. For the typical configurations of UAM aircraft, the three primary sources of noise are engine noise (from both propeller and mechanical components), aerodynamic noise, and aircraft systems noise~\cite{gao2022statistics}. UAM noise can cause community annoyance, disrupt sleep, and could elevate the risk of cardiovascular disease in urban residents~\cite{basner2017aviation}. Unlike conventional aviation noise---which primarily affects people living in the vicinity of airports---UAM noise could broadly impact most communities throughout a city. In this section, we introduce the computation of UAM aircraft noise, metrics for both single and cumulative noise events, and aspects of community noise in urban environments.

\subsection{Single Event and Cumulative Noise Measures}

Before computing single event noise and the cumulative effect of multiple noise events, we first introduce multiple fundamental concepts in noise and discuss the differences among them. Decibel (dB) is the unit used to quantify sound intensity. Decibels are measured on a logarithmic scale, reflecting how our ears perceive sound pressures. A-weighted decibel (dBA) is the unit that considers the way the human ear perceives different pitches or frequencies. It emphasizes the portions of the frequency spectrum most sensitive to our hearing and is used by the Federal Aviation Administration (FAA) to measure aviation noise. Maximum sound level (Lmax) is the maximum sound level in the duration of a particular noise event.

\begin{figure}[htbp]
	\centering
        \includegraphics[width=0.55\textwidth]{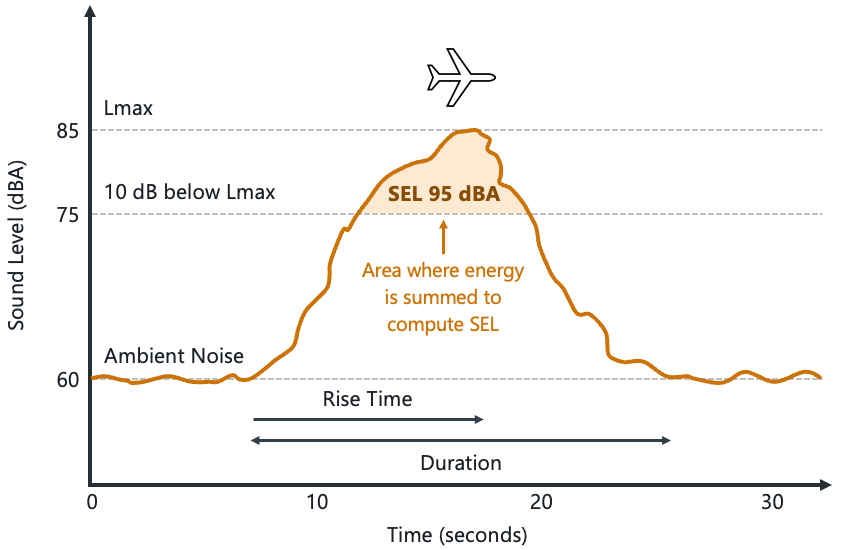}
	\caption{Single event aircraft noise: sound exposure level (SEL), maximum noise level (Lmax), and duration}
	\label{fig:sel}
\end{figure}

On the noise evaluation of a single event, sound exposure level (SEL) is the most commonly used noise measure for an aircraft flyover. It captures both the intensity and duration of a sound event in one numerical quantity and compresses the event into a reference duration of one second:
\begin{equation}
    \text{SEL} = 10 \log_{10} \left(\frac{1}{T_0} \sum_{i=1}^n 10^{L_i/10}\right)
\end{equation}
where $T_0 = 1 s$ is the reference duration, $n$ is the number of seconds during the measurement period, and $L_i$ is the sound level (in dBA) for the i-th one-second time period. Figure~\ref{fig:sel} illustrates the computation of SEL for a single event, where the measurement period covers the portion of the duration when the sound level is within 10 dB below Lmax. 

Equivalent sound level (Leq) measures the average acoustic energy during a specific time frame to account for the cumulative effect of multiple noise events, such as the aggregate sound level at a location with aircraft flyovers throughout the day. Given the individual SEL values of various noise events, Leq is obtained by:
\begin{equation}
    \text{Leq} = 10 \log_{10} \left(\sum_{i=1}^n 10^{\text{SEL}_i/10}\right) - 10 \log_{10} \left(\frac{T}{T_0}\right) 
\end{equation}
where $\text{SEL}_i$ is the SEL value for the i-th event, $n$ is the number of events, $T$ is the time interval, and $T_0 = 1 s$ is the SEL reference duration. In the scope of this study, we are interested in the cumulative noise levels over either 1 hour or 24 hours. When $T$ equals to 3,600 and 86,400 seconds respectively, we have:
\begin{equation}\label{eqn:leq1h}
    \text{Leq}_{1h}  = 10 \log_{10} \left(\sum_{i=1}^n 10^{\text{SEL}_i/10}\right) - 35.56
\end{equation}
\begin{equation}
    \text{Leq}_{24h}  = 10 \log_{10} \left(\sum_{i=1}^n 10^{\text{SEL}_i/10}\right) - 49.37
\end{equation}

Day-night average sound level (DNL) and community noise equivalent level (CNEL) are both similar to Leq yet differ in how a noise event is treated during evening and nighttime. To account for higher noise sensitivity levels, DNL adds a 10 dBA ``penalty'' to each nighttime event (10 pm to 7 am); CNEL adds an additional 4.77 dBA penalty to each evening event (7 pm to 10 pm). CNEL is obtained by:
\begin{equation}
    \text{CNEL}  = 10 \log_{10} \left(\sum_{i=1}^n 10^{\text{SEL}_i/10}+\sum_{i=n+1}^m 10^{(\text{SEL}_i+4.77)/10}+\sum_{i=m+1}^r 10^{(\text{SEL}_i+10)/10}\right) - 49.37
\end{equation}
where the three terms in the bracket represent day, evening, and night events, respectively. 

\subsection{UAM Aircraft Noise Modeling}\label{sec:aircraftnoisemodeling}

Aircraft noise modeling has been a crucial problem in aerospace engineering and air transportation. The approach for noise computation depends on the modeling range and terrain complexity. When modeling the noise exposure of an aircraft operation in complex urban terrain, such as downtown areas, high-fidelity methods such as acoustic ray tracing~\cite{gao2023noise} are necessary to capture complex acoustics phenomena such as atmospheric absorption, reflection, diffraction, and scattering over the ground surfaces and buildings. On the other hand, the modeling of aircraft noise on larger scales, such as community noise exposure in urban and suburban areas, often employs more efficient and generalized aircraft noise models, such as noise spheres and noise-power-distance (NPD) data. For our study, NPD data is the favorable choice for modeling the community noise of UAM in a city.

NPD data describes the relationship between the noise level of an aircraft and its slant distance to the receiver. Currently, NPD data are available for close to 300 fixed-wing aircraft and 26 helicopter types~\cite{volpe2022uam}. For helicopters, the engine power variable in NPD data is replaced by operational mode (i.e., departure, approach, level-fly) due to different type characteristics. For each aircraft type, the NPD data come from either noise certification tests or controlled tests executed under stringent international standard procedures~\cite{gao2022minimax}. Separate sets of NPD data are available to account for different operational modes and in-flight directivity (left, right, and centre).

Since there is no existing NPD data for emerging UAM aircraft designs like eVTOL aircraft, recent studies and models on UAM noise have mainly adopted two approaches to estimate NPD data for potential UAM aircraft configurations. The first approach utilizes existing NPD data for helicopters as basis and applies noise reduction or scaled-down to account for the expectation or prediction that eVTOL aircraft are quieter than existing helicopters. For example, a recent work~\cite{rimjha2021urban} started with the NPD data of the Robinson R44 helicopter and applied two reduction levels on the baseline noise profile: 10 dBA and 15 dBA; another work~\cite{woodcock2022development} selected the Bell 206 helicopter as a reference vehicle for specifying a dBA offset. The second approach conducts acoustic simulation which can capture the 3D directivity of noise near a UAM aircraft configuration~\cite{rizzi2021comparison}. The simulation result is then processed and converted into noise sphere or NPD data for modeling UAM's community noise exposure.

\begin{figure}[h!]
     \centering
     \begin{subfigure}[b]{0.4\textwidth}
         \centering
         \includegraphics[width=\textwidth]{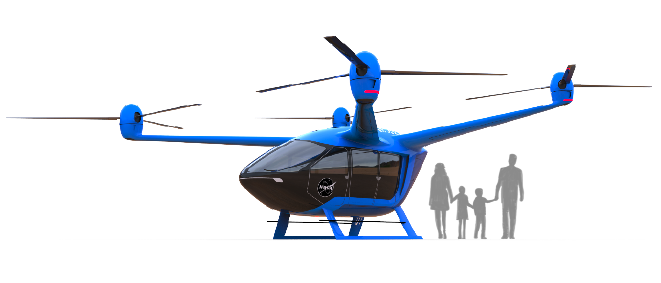}
         \caption{The NASA RVLT quadrotor reference vehicle}
         \label{fig:nasavehicle}
     \end{subfigure}
     \hspace{0.5cm}
     \begin{subfigure}[b]{0.55\textwidth}
         \centering
         \includegraphics[width=\textwidth]{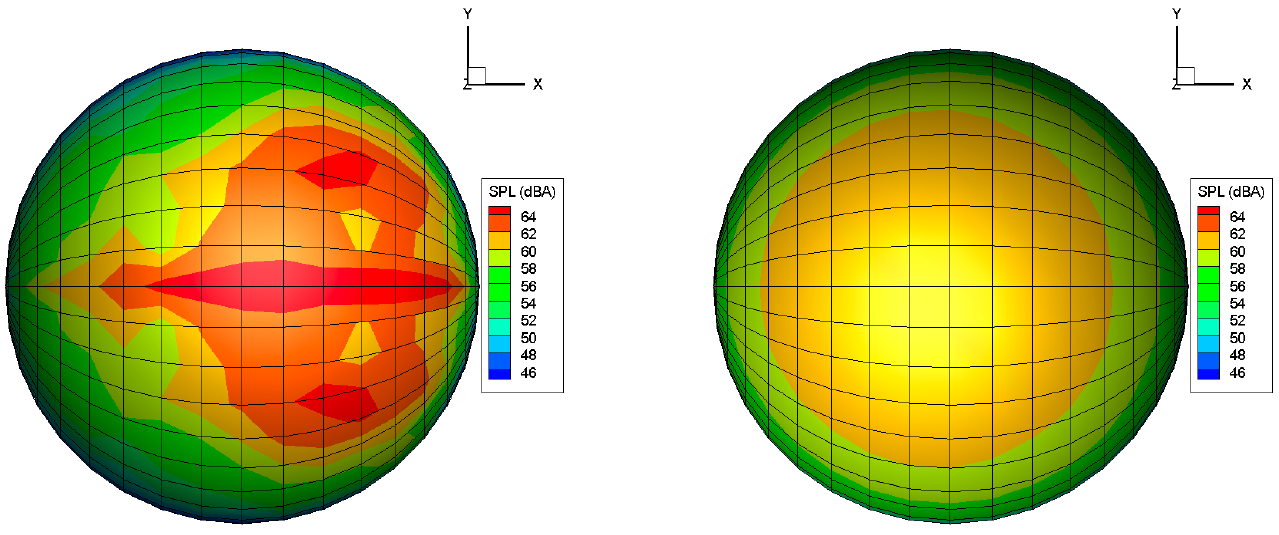}
         \caption{The example noise hemispheres generated by simulation}
         \label{fig:noisesphere}
     \end{subfigure}
     \begin{subfigure}[b]{0.95\textwidth}
         \centering
         \includegraphics[width=\textwidth]{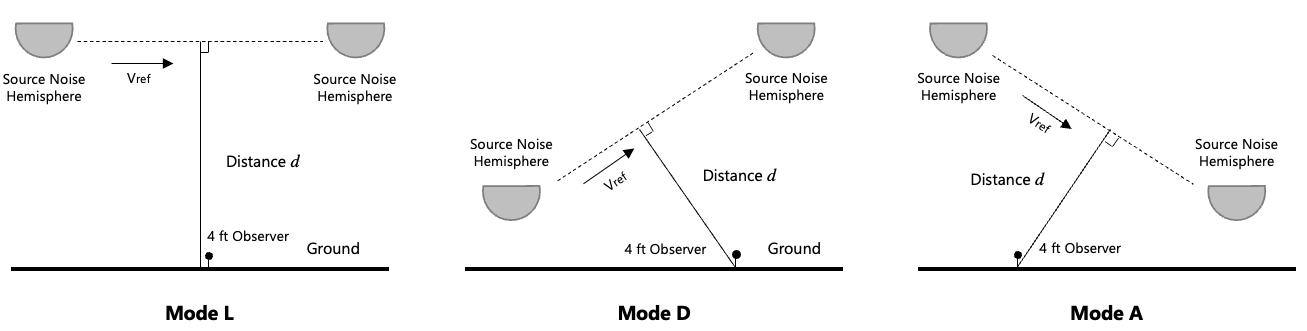}
         \caption{Noise simulation scenarios to obtain SEL NPD data for the three modes}
         \label{fig:noisesimulations}
     \end{subfigure}
        \caption{The vehicle configuration considered in this study and noise simulation methods (sources: \cite{rizzi2022prediction,rizzi2023modeling})}
        \label{fig:vehicle}
\end{figure}

\begin{figure}[htbp]
	\centering
        \includegraphics[width=0.425\textwidth]{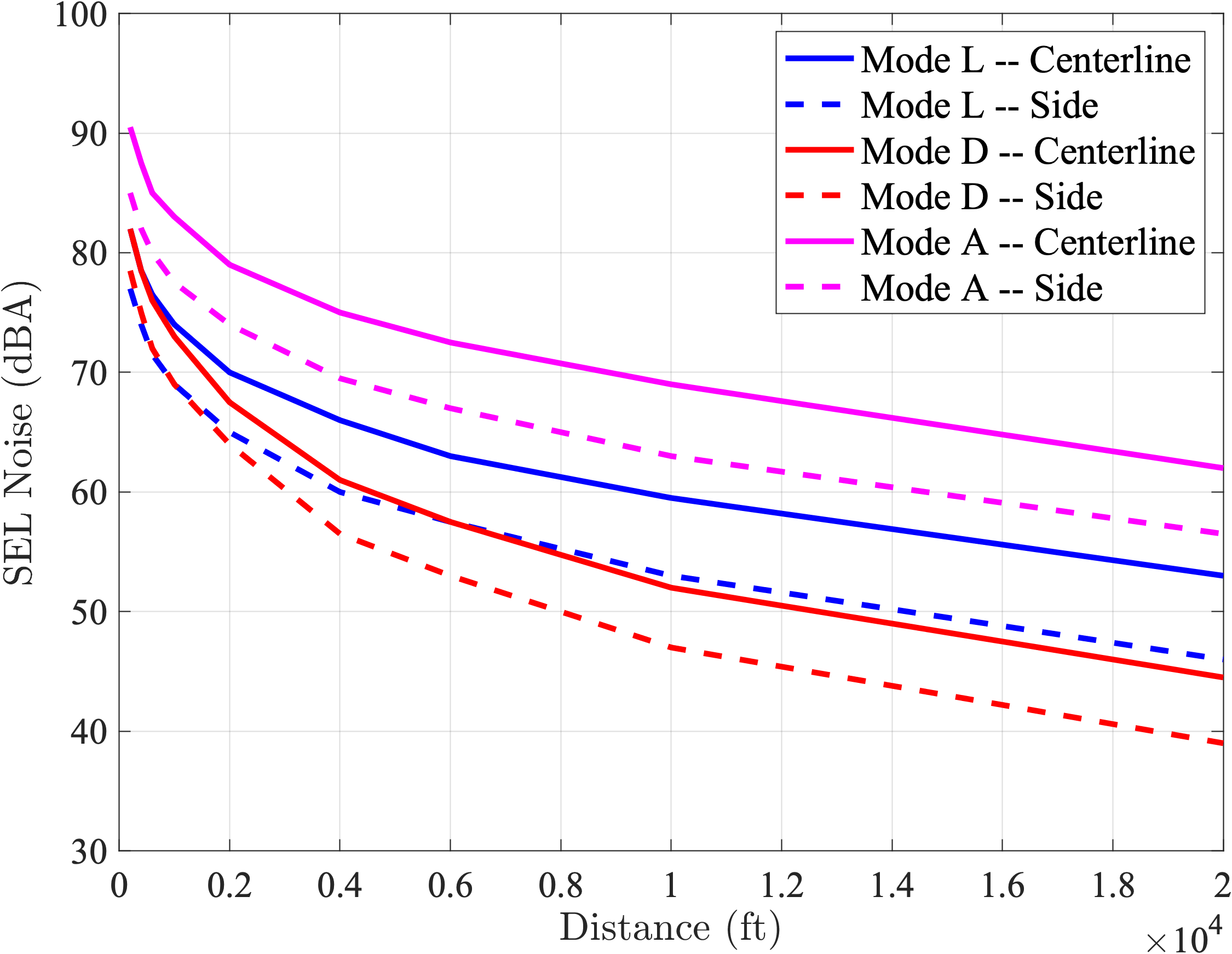}
        \hspace{1cm}
        \includegraphics[width=0.42\textwidth]{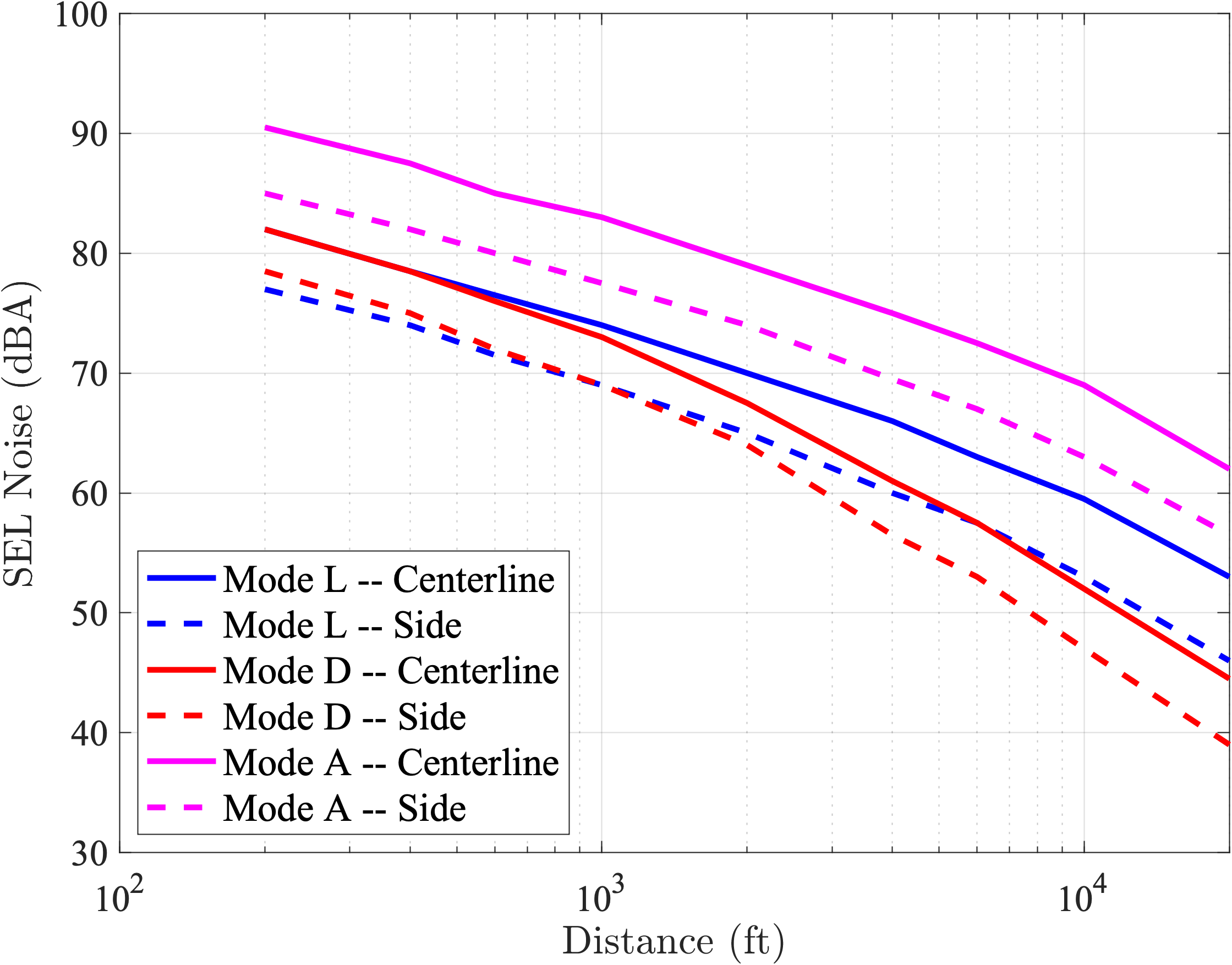}
	\caption{Simulated NPD data for the NASA RVLT quadrotor vehicle, under normal (left) and log (right) distance scales}
	\label{fig:npds}
\end{figure}

Researchers from NASA Langley Research Center~\cite{rizzi2022prediction,rizzi2023modeling} have used acoustic modeling tools and flight simulation to generate NPD data for novel UAM aircraft configurations. In this study, we adopt the noise models for the NASA Revolutionary Vertical Lift Technology (RVLT) quadrotor reference vehicle~\cite{silva2018vtol}, shown in Figure~\ref{fig:nasavehicle}. This all-electric vehicle has four three-bladed rotors, can carry up to six passengers, and can operate at a maximum airspeed of 109 knots true airspeed (KTAS). The noise data of this vehicle configuration is generated using ANOPP2's Aeroacoustic ROtor Noise (AARON) tool. Figure~\ref{fig:noisesphere} shows examples of the noise hemispheres for the RVLT quadrotor vehicle, which are then used in flight simulations to generate NPD data for different operational modes. Figure~\ref{fig:noisesimulations} displays the simulation scenarios for generating SEL NPD data for the three operational modes: level flyover (L), and (dynamic) departure (D) and approach (A). For each operational mode, measurements are collected along the centerline below the track and at 45 degrees azimuth angles to capture lateral directivity. Figure~\ref{fig:npds} displays the resulting NPD data for the NASA RVLT quadrotor vehicle. 


Because NPD data is only available at certain distance levels from 200 ft to 20,000 ft, practitioners often use logarithmic interpolation to estimate noise levels at other distance levels. In this work, we fit regression models to the NPD curves in Figure~\ref{fig:npds}. For each operational mode and measurement position combination, our regression model has the following form:
\begin{equation}\label{eqn:npds}
    L_{\text{AE}}(d) = a_0 + a_1 \log_{10}{d} + a_2 (\log_{10}{d})^2
\end{equation}
where $L_{\text{AE}}$ is noise in A-weighted SEL and $d \in [200, 20000]$ is distance in ft. Figure~\ref{fig:fittednpds} visualizes the goodness-of-fit for the six NPD curves, where each regression model is plotted with the original NPD data. Table~\ref{tbl:fittednpds} contains the regression coefficients for the six NPD curves in Figure~\ref{fig:fittednpds}.\\

\begin{minipage}{\textwidth}
  \hspace{0.1cm}
  \begin{minipage}[b]{0.42\textwidth}
    \centering
    \includegraphics[width=\textwidth]{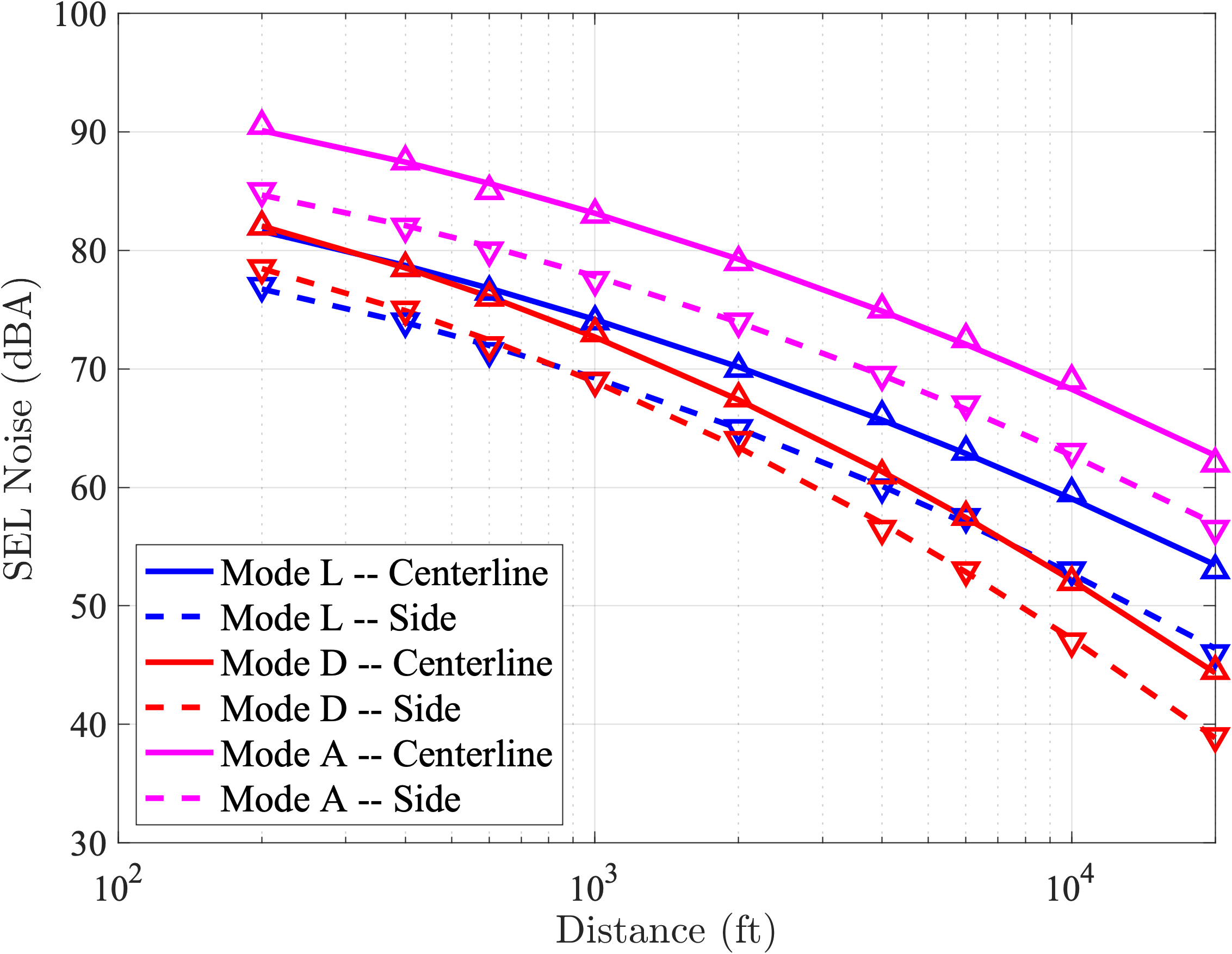}
    \captionof{figure}{Goodness-of-fit for NPD curves}
    \label{fig:fittednpds}
  \end{minipage}
  \hspace{1cm}
  \begin{minipage}[b]{0.45\textwidth}
    \centering
    \small
    \begin{tabular}{llll}
    \hline \hline
    Condition       & $a_0$    & $a_1$    & $a_2$    \\ \hline
    Mode L - Centerline & 88.09 & 3.21  & -2.62 \\ 
    Mode L - Side   & 78.01 & 7.26  & -3.39 \\ 
    Mode D - Centerline & 84.05 & 8.76  & -4.18 \\ 
    Mode D - Side   & 77.34 & 11.34 & -4.72 \\ 
    Mode A - Centerline & 93.35 & 5.17  & -2.86 \\ 
    Mode A - Side   & 85.55 & 6.83  & -3.14 \\ \hline \hline
    \end{tabular}
    \vspace{1cm}
    \captionof{table}{Regression coefficients for NPD curves}
    \label{tbl:fittednpds}
    \end{minipage}
\end{minipage}\\\\

To better model the lateral sound patterns, we add two additional lateral adjustments to the NPD data: (i) lateral attenuation adjustment, and (ii) lateral directivity adjustment~\cite{woodcock2022development}. The lateral attenuation adjustment takes into account ground reflection and refraction. First, we approximate the general ground attenuation component using the following equation:
\begin{equation}
    E_g (l) = 
    \begin{cases}
        11.83 \cdot \left(1-e^{-0.0009 \cdot l}\right), & \text{if}~ 0 \leq l \leq 3,000~\text{ft}\\
        10.86, & \text{if}~ l > 3,000~\text{ft}
    \end{cases}
\end{equation}
where $l$ is the lateral sideline distance (distance between the lateral receiver and the centerline). Then, the refraction component can be computed using the following equation:
\begin{equation}
    \Lambda (\beta) = 
    \begin{cases}
        10.86, & \text{if}~ \beta \leq 0\\
        1.137 - 0.0229 \cdot \beta + 9.72 \cdot e^{-0.142 \cdot \beta}, & \text{if}~ 0^{\circ} < \beta \leq 50^{\circ}\\
        0, & \text{if}~ 50^{\circ} < \beta \leq 90^{\circ}
    \end{cases}
\end{equation}
where $\beta$ is the elevation angle of the aircraft with respect to the horizontal plane of the receiver. With these two components, the overall lateral attenuation adjustment can be computed as:
\begin{equation}
    L_{\text{LA-ADJ}}(l,\beta) = \frac{E_g (l) \cdot \Lambda (\beta)}{10.86}
\end{equation}

The lateral directivity adjustment takes into account the change in directivity through linear interpolation between the two NPD curves where $\beta$ equals to 90$^{\circ}$ and 45$^{\circ}$, respectively. It can be computed as:
\begin{equation}
    L_{\text{LD-ADJ}}(d,\beta) = \left(L_{\text{AE}, 90^{\circ}}(d)-L_{\text{AE}, 45^{\circ}}(d)\right) \cdot \left( \frac{90-|\beta|}{90-45} \right)
\end{equation}

Finally, with NPD curves and the two additional lateral adjustments, the SEL noise that a flight operation poses to a receiver on the ground is calculated using the following equation:
\begin{equation}
    L(d,l,\beta) = L_{\text{AE}, 90^{\circ}}(d) - L_{\text{LD-ADJ}}(d,\beta) - L_{\text{LA-ADJ}}(l,\beta)
\end{equation}

\subsection{Community Noise}

Noise pollution in urban communities is a complex problem influenced by a variety of sources and their variabilities. A rigorous treatment of this problem, including the measurement, prediction, and control of community noise requires a systematic approach. When operating UAM services in urban environments, a crucial factor in noise control is recognizing that different communities within the city exhibit varying levels of noise sensitivity. Urban communities exhibit different noise acceptance levels due to two primary factors: (1) land use, and (2) ambient noise. The FAA's 14 CFR Part 150\footnote{FAA, Airport Noise Compatibility Planning (14 CFR Part 150), \url{https://www.faa.gov/airports/environmental/airport\_noise}} classifies land use into five categories: residential, public use, commercial use, manufacturing and production, and recreational. Within each category, there are 3 to 6 detailed sub-categories that may have various sensitivities to noise intrusions. The ambient noise in an urban space refers to the background sound produced by numerous sources in urban areas such as transportation, construction, human activity, aircraft, and natural phenomena. 

In this work, we consider ambient noise as the primary factor for determining the noise acceptance level of a community in the city for two reasons. First, even within a small area in a city, land use could be highly heterogeneous. For example, a small downtown neighborhood could well be a mixture of residential, public use, and commercial use sites; a suburban neighborhood could contain industrial, commercial use, and public use lands that are intertwined. Second, one of the most effective noise mitigation strategies is ``ambient noise masking'', i.e., concentrating flights over less noise-sensitive areas such as areas with large traffic volume and industrial parks~\cite{vascik2017evaluation}.

\begin{table}[htbp]
\centering
\caption{The five categories of urban residential areas, their daytime noise levels in dBA, and the amounts of correction added to measured CNEL (source: \cite{epa1971community}).}
\begin{tabular}{c|cc|c}
\hline \hline
\multirow{2}{*}{Description} & \multicolumn{2}{c|}{Daytime Median Noise Level in dBA} & \multirow{2}{*}{\begin{tabular}[c]{@{}c@{}}Correction Added to\\ Measured CNEL in dB\end{tabular}} \\ \cline{2-3}
                             & \multicolumn{1}{c|}{Typical Range}       & Average       &                                                                                                              \\ \hline
Quiet Suburban   & \multicolumn{1}{c|}{41 to 45}            & 43            & +10                                                                                                           \\ 
Normal Suburban  & \multicolumn{1}{c|}{46 to 50}            & 48            & +5                                                                                                            \\ 
Urban            & \multicolumn{1}{c|}{51 to 55}            & 53            & 0                                                                                                            \\ 
Noisy Urban      & \multicolumn{1}{c|}{56 to 60}            & 58            & -5                                                                                                           \\ 
Very Noisy Urban & \multicolumn{1}{c|}{61 to 65}            & 63            & -10                                                                                                          \\ \hline \hline
\end{tabular}
\label{tbl:comm}
\end{table}

We next look at the classification of urban communities based on ambient noise and community reactions to noise levels above ambient noise. On the former, Table~\ref{tbl:comm} contains how the U.S. Environmental Protection Agency (EPA)~\cite{epa1971community} classified urban residential areas into five categories, according to their ambient noise levels. The five classes of urban communities in Table~\ref{tbl:comm} are widely adopted; the daytime median noise level in dBA is also representative of a community's ambient noise level. Considering the increased urban noise levels since these standards were first made, we use the upper bounds of the typical ranges in Table~\ref{tbl:comm}, namely, 45 dBA, 50 dBA, 55 dBA, 60 dBA, and 65 dBA to represent the ambient noise levels of the five community classes. Table~\ref{tbl:delta} displays the result of a related study that models the expected community reaction as a function of noise exposure above the ambient noise level. We estimate that when UAM's cumulative noise is 10 dB above the ambient noise, widespread complaints would happen; and the reaction becomes threats of legal action when the difference increases to 20 dB. We will use information from Table~\ref{tbl:comm} and Table~\ref{tbl:delta} as evidence to build evaluation functions for community noise annoyance.

\begin{table}[htbp]
\centering
\caption{Summary of expected community reaction and approximate annoyance as a function of CNEL and ambient noise levels (source: \cite{epa1971community}).}
\begin{tabular}{c|cc|c|c}
\hline \hline
\multirow{2}{*}{\begin{tabular}[c]{@{}c@{}}Expected\\ Community\\ Reaction\end{tabular}} & \multicolumn{2}{c|}{\begin{tabular}[c]{@{}c@{}} Difference Between\\ CNEL and Average Daytime\\ Median Noise Level in dB\end{tabular}} & \multirow{2}{*}{\begin{tabular}[c]{@{}c@{}}Approximate \%\\ Very Much\\ Annoyed\end{tabular}} & \multirow{2}{*}{\begin{tabular}[c]{@{}c@{}}Approximate \%\\ Little or Not\\ Annoyed\end{tabular}} \\ \cline{2-3}
                                                                                         & \multicolumn{1}{c|}{Mean}                                                            & Range of Data                                                           &                                                                                                      &                                                                                                          \\ \hline
No reaction                                                                              & \multicolumn{1}{c|}{2}                                                               & 0 to 8                                                                  & 10                                                                                                   & 45                                                                                                       \\ 
Sporadic complaints                                                                      & \multicolumn{1}{c|}{6}                                                               & 3 to 8                                                                  & 5                                                                                                    & 37                                                                                                       \\ 
Widespread complaints                                                                    & \multicolumn{1}{c|}{11}                                                              & 7 to 19                                                                 & 0                                                                                                    & 26                                                                                                       \\ 
Threats of legal action                                                                  & \multicolumn{1}{c|}{21}                                                              & 18 to 24                                                                & -5                                                                                                   & 14                                                                                                       \\ 
Vigorous action                                                                          & \multicolumn{1}{c|}{28}                                                              & 23 to 34                                                                & -10                                                                                                  & 7                                                                                                        \\ \hline \hline
\end{tabular}
\label{tbl:delta}
\end{table}

\begin{figure}[h!]
     \centering
     \begin{subfigure}[b]{0.325\textwidth}
         \centering
         \includegraphics[width=\textwidth]{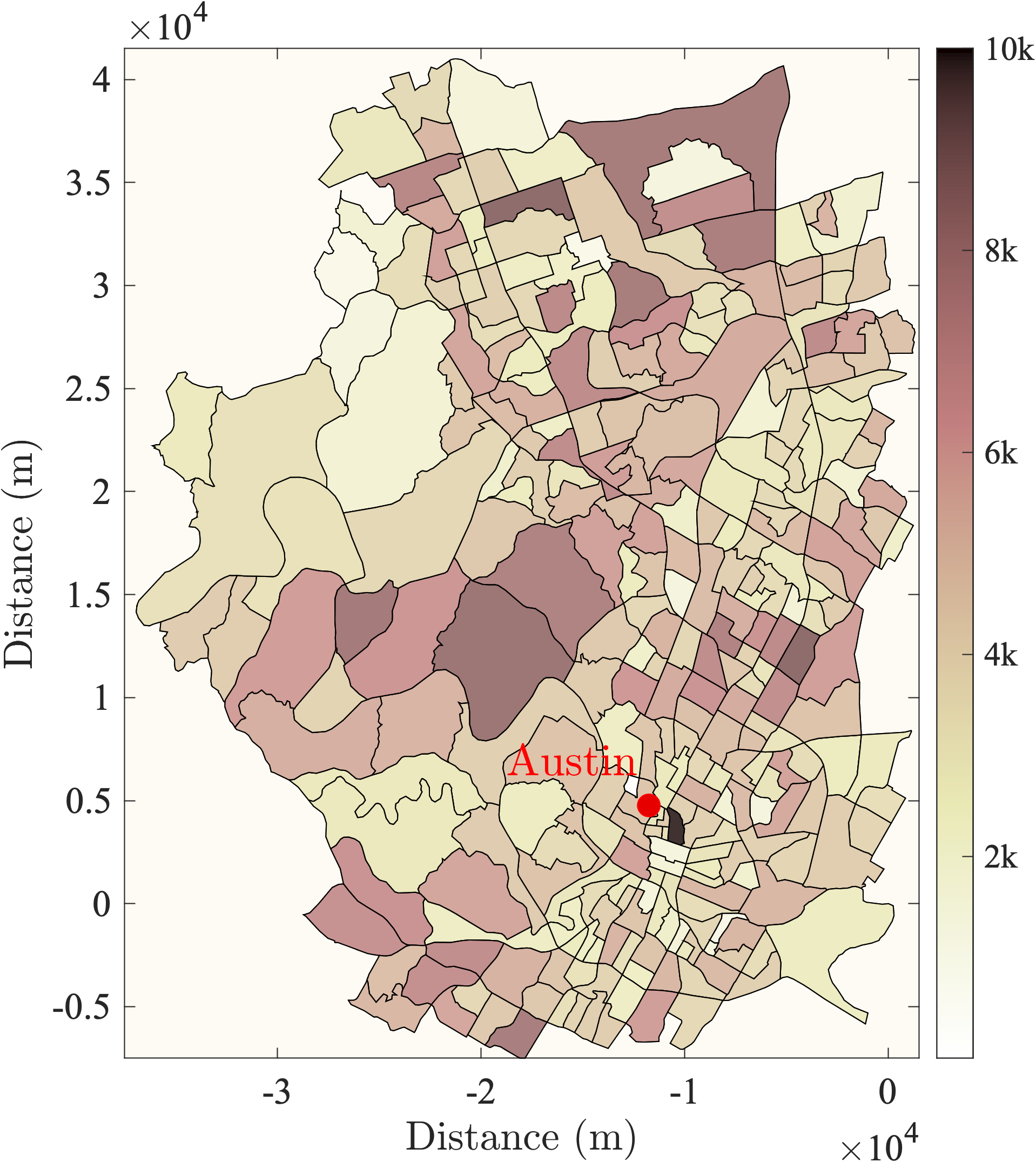}
         \caption{The Austin communities and their total populations}
         \label{fig:auscomm1}
     \end{subfigure}
     \begin{subfigure}[b]{0.325\textwidth}
         \centering
         \includegraphics[width=\textwidth]{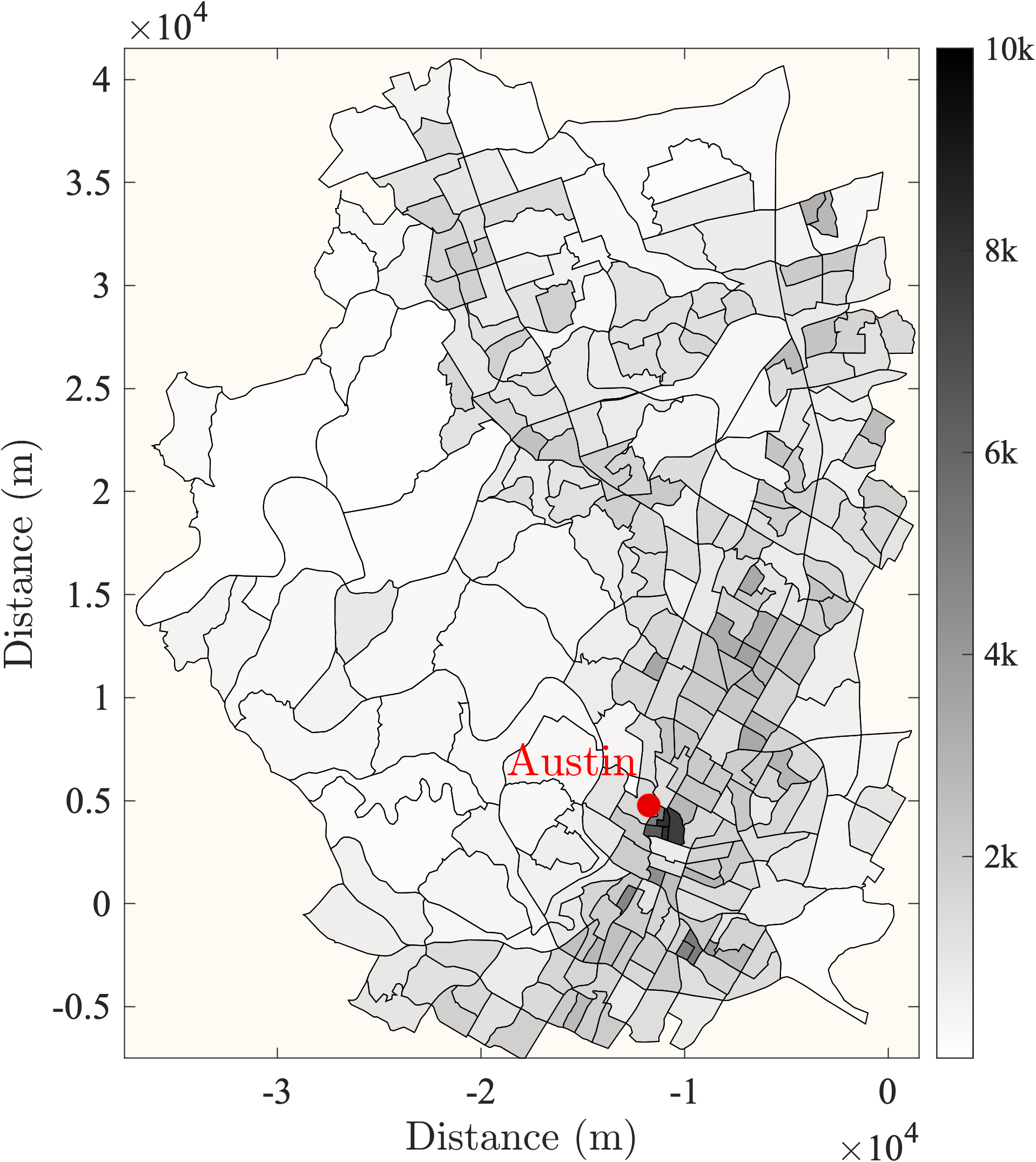}
         \caption{The Austin communities and their populations densities}
         \label{fig:auscomm2}
     \end{subfigure}
     \begin{subfigure}[b]{0.325\textwidth}
         \centering
         \includegraphics[width=\textwidth]{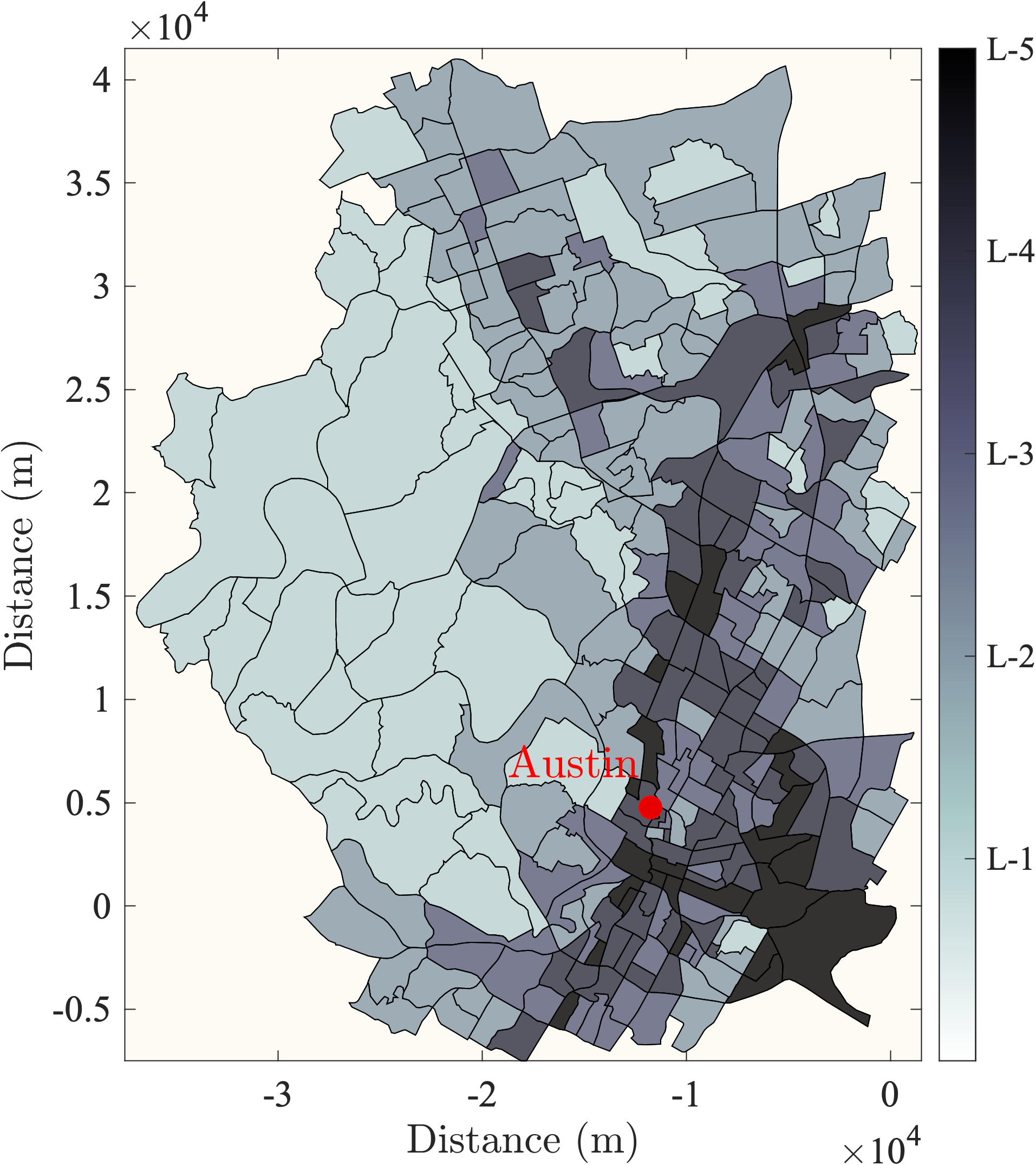}
         \caption{The Austin communities classified by ambient noise level}
         \label{fig:auscomm3}
     \end{subfigure}
        \caption{Partition and aspects of the 292 communities in the city of Austin}
        \label{fig:auscomm}
\end{figure}

In this study we choose the city of Austin as the use case. To partition an urban area into communities, the two most common methods are zip code and census tract. Zip code areas are determined based on factors including mail volume, geographical features, and delivery routes. The city of Austin can be partitioned into 46 zip code areas. Zip code is not a natural organization of communities in every city and should be evaluated on a case-by-case basis. Although zip code can provide a fine separation of communities in the city of Austin, we choose census tract in this study for two reasons. First, census tract provides partition at a more detailed level. Figure~\ref{fig:auscomm1} displays the partition of 292 census tracts in the city of Austin, where the position of downtown Austin is marked by the red dot. Second, it is more convenient to find ambient noise and population data for a census tract, as shown in Figure~\ref{fig:auscomm2} and Figure~\ref{fig:auscomm3}. To obtain the ambient noise levels of the 292 Austin communities, we refer to the open source data National Transportation Noise Exposure Map~\cite{seto2023national}, which contains the most recent noise levels modeled by the Bureau of Transportation Statistics (BTS) for road, rail and aviation traffic. We then utilize the noise exposure data to classify the communities into five classes, following the guidelines in Table~\ref{tbl:comm}. In the community noise classification result in Figure~\ref{fig:auscomm3}, a darker color indicates a higher ambient noise level. Population is another critical factor in the evaluation of aviation noise exposure. Figure~\ref{fig:auscomm1} displays the 5-year (2016-2020) total population estimates from the Census Bureau's American Community Survey (ACS).

\section{The Multi-layer UAM Network}\label{sec:network}

\begin{figure}[h!]
     \centering
     \begin{subfigure}[b]{0.36\textwidth}
         \centering
         \includegraphics[width=\textwidth]{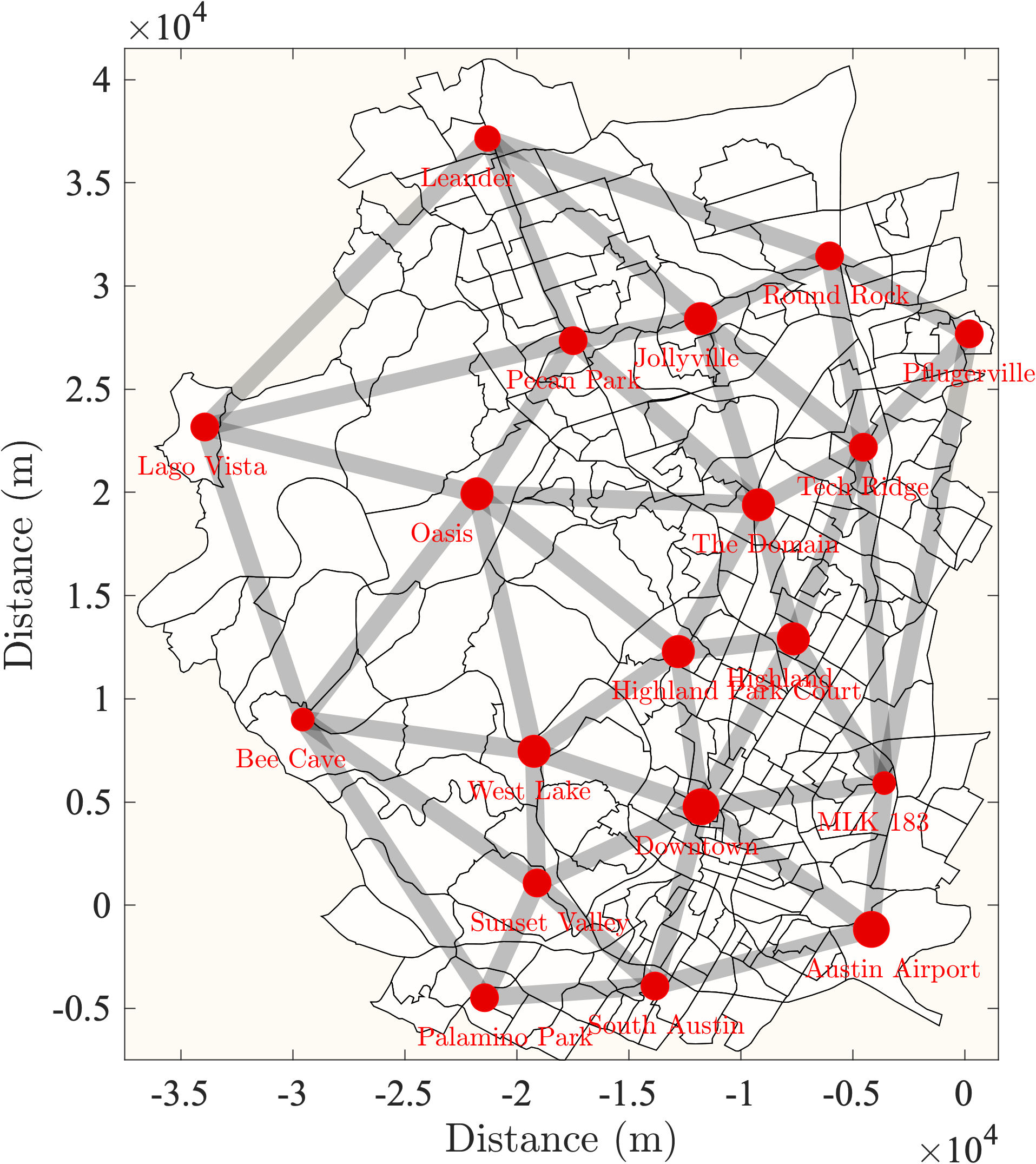}
         \caption{The Austin UAM service network}
         \label{fig:austinnetwork}
     \end{subfigure}
     \hspace{2.5cm}
     \begin{subfigure}[b]{0.35\textwidth}
         \centering
         \includegraphics[width=\textwidth]{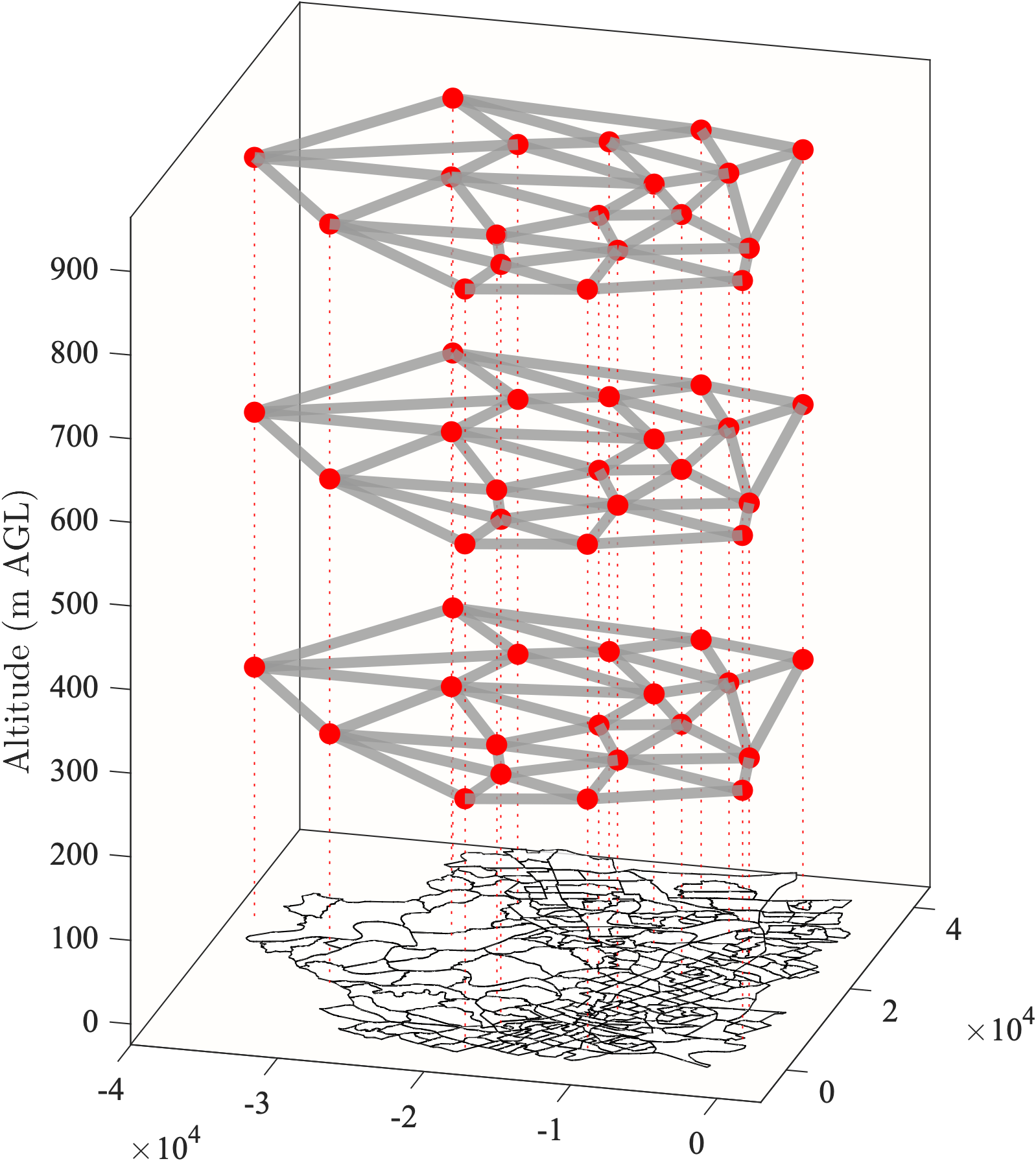}
         \caption{The three-layer flight corridors}
         \label{fig:3dnetwork}
     \end{subfigure}
        \caption{Schematic of the three-layer UAM network design for the city of Austin}
        \label{fig:3daustinnetwork}
\end{figure}

We choose the city of Austin as the case study to demonstrate our overall approach. Figure~\ref{fig:austinnetwork} displays a UAM network in Austin which contains 19 destinations (vertiports), 45 undirectional links (90 directional links) in each layer, 62 origin-destination (O-D) pairs, and 292 communities. This UAM network is an outcome of a systematic design that takes into account the following factors: (1) the current Austin public transportation network, (2) the future Austin Public Transit Plan\footnote{A New Transit Plan for Austin, \url{https://www.projectconnect.com/}}, (3) the current city traffic patterns, (4) the locations of communities and city sub-centers, (5) the distribution of city residents, and (6) features of urban topography. Each node in the network represents an actual location within a city center/sub-center that is geographically and societally reconfigurable into a vertiport. We select the set of O-D pairs through a preliminary demand analysis that involves the public transportation plans and the functions of communities surrounding each node. The set of links represents flight corridors between vertiports and has a simplified topological structure that has no intersection between the links and can efficiently serve all O-D pairs, i.e., there are multiple routes for each O-D pair, with the distance of each route being marginally longer than the direct straight-line distance between the O-D pair. 

The concept of flight corridors, the 3D sections of airspace designated for aerial vehicle operations~\cite{muna2021air}, is classic and long-established in air transportation. For both commercial aviation and UAM, these virtual highways in the sky can provide navigation, reduce the complexity and increase predictability of airspace, structure air traffic to enhance operational safety and efficiency, and minimize the impact on other operations. As urban air traffic management literature~\cite{wang2021air,muna2021air,wang2022complexity} and multiple UAM airspace design concepts~\cite{bauranov2021designing} have envisioned, flight corridors are integral parts of the UAM system infrastructure. 

Multi-layer network or system~\cite{zhao2019vulnerability,muna2021air,rodriguez2022multilayered,wang2022complexity} is another prevailing concept in air transport system modeling and planning. In this work, a multi-layer network has a significant implication -- it provides flight corridors at multiple altitudes such that UAM operations can mitigate community noise impact through adjusting cruising altitude.

Figure~\ref{fig:3dnetwork} displays the three-layer UAM network in Austin, where the three cruising altitudes are 1,000 ft, 2,000 ft, and 3,000 ft above ground level (AGL), respectively. We use 500 ft mean sea level (MSL) as the reference altitude for Austin and the resulting cruising altitudes in MSL are 1,500 ft, 2,500 ft, and 3,500 ft, respectively. The design of such a multi-layer network and flight corridors in urban airspace must consider several factors. The three most influential factors are urban topography (e.g., natural terrains, buildings), controlled airspace around airport (e.g., Class B, C, D airspace), and prohibited and restricted areas (e.g., restricted military area). In the Austin scenario, the lowest level of Figure~\ref{fig:3dnetwork} can maintain a 250 ft clearance distance from any objects on the ground. The Austin-Bergstrom International Airport locates at the southeast corner of Figure~\ref{fig:austinnetwork}, whose Class C airspace intersects part of the UAM network. We assume that UAM aircraft can operate within designated sections of the Austin airport's Class C airspace through `geofencing', a method that establishes virtual boundaries to separate UAM operations from those of commercial and general aviation. In Section~\ref{sec:pf}, we explore the management of UAM traffic flow within the network in Figure~\ref{fig:3dnetwork}.

\section{Problem Formulation}\label{sec:pf}

\subsection{Network Model}

In this subsection we introduce the basic definitions of an air mobility network, the additional considerations in this model, and the constraints in the optimization problem.

\subsubsection{Nodes, Links, Routes, and Communities}

We use a directed graph $\mathcal{G}$ to model the air mobility network. The network $\mathcal{G}$ consists of a set of nodes $\mathcal{N} = \{1, 2, ..., n_n\}$, where each node corresponds to either a vertiport (take-off/landing site on the ground) or a way point (intersection of flight corridors in the air), and a set of links $\mathcal{L} = \{1, 2, ..., n_l\}$, where each link corresponds to a flight corridor. Within the set of nodes, we let $\mathcal{V} = \{1, 2, ..., n_v\}$ denote a set of vertiports in the network. Note that in a $k$-level air mobility network, we have $n_n = k\cdot n_v$. A link is an ordered pair of distinct nodes, where the first and second nodes of the link correspond to its ``tail'' and ``head'', respectively. We let $\mathcal{R} = \{1, 2, ..., n_r\}$ denote a set of routes in the network, where each route is a sequence of links connected in a ``head-to-tail'' fashion. We let $\mathcal{O} = \{1, 2, ..., n_o\}$ denote a set of O-D pairs which have demand for air mobility services in the city. Lastly, we let $\mathcal{C} = \{1, 2, ..., n_c\}$ denote a set of communities in the city.

\subsubsection{Incidence Matrices}

We use incidence matrices to encode the topology of network $\mathcal{G}$ and the relationships between elements in the network. First, we use the node-link incidence matrix $E \in \mathbb{R}^{n_n \times n_l}$ to describe the incidence relationship between nodes and links. The entry $[E]_{ij}$ in matrix $E$ associates node $i$ and link $j$ as follows:
\begin{equation}
    [E]_{ij} = 
    \begin{cases}
        1, & \text{if node $i$ is the head of link $j$,}\\
        -1, & \text{if node $i$ is the tail of link $j$,}\\
        0, & \text{otherwise.}
    \end{cases}
\end{equation}

We then use the link-route incidence matrix $F \in \mathbb{R}^{n_l \times n_r}$ to describe the incidence relationship between links and routes. The entry $[F]_{ij}$ in matrix $F$ associates link $i$ and route $j$ as follows:
\begin{equation}
    [F]_{ij} = 
    \begin{cases}
        1, & \text{if route $j$ contains link $i$,}\\
        0, & \text{otherwise.}
    \end{cases}
\end{equation}

We use the O-D pair-route incidence matrix $H \in \mathbb{R}^{n_o \times n_r}$ to describe the incidence relationship between O-D pairs and routes. The entry $[H]_{ij}$ in matrix $H$ associates O-D pair $i$ and route $j$ as follows:
\begin{equation}
    [H]_{ij} = 
    \begin{cases}
        1, & \text{if route $j$ serves the demand of O-D pair $i$,}\\
        0, & \text{otherwise.}
    \end{cases}
\end{equation}

Lastly, to formulate the vertiport capacity constraints later, we use the destination-route incidence matrix $J \in \mathbb{R}^{n_v \times n_r}$ to describe the incidence relationship between destinations and routes. Specifically, the entry $[J]_{ij}$ in matrix $J$ associates destination $i$ and route $j$ as follows:
\begin{equation}
    [J]_{ij} = 
    \begin{cases}
        1, & \text{if node $i$ is the destination of route $j$,}\\
        0, & \text{otherwise.}
    \end{cases}
\end{equation}

\subsubsection{Network Flow Constraints}
We introduce two sets of network flow constraints to describe the relation between the traffic flow on each link, each route, and traverse each node.
The first set of network flow constraints ensures the flow balance conditions at the nodes. They state the conservation of flow, i.e., the flow (total number of flights) out of a node equals to the flow into a node at any time. This flow conservation condition is strictly true at each way point. Although each vertiport has a capacity (the physical space that can contain a certain number of aircraft), we still apply the flow conservation condition because the magnitude of capacity is relatively small compared to the number of flights in the long run. We use a link-flow vector $\bold{y} \in \mathbb{R}^{n_l}$ to denote the number of flights on different links per unit time, where the entry $[\bold{y}]_{i}$ denotes the number of flights that fly through link $i$. We then formulate the flow balance constraints as follows:
\begin{equation}\label{eqn:fc1}
    E \bold{y} = \boldsymbol{0}_{n_n}, \quad \bold{y} \geq \boldsymbol{0}_{n_l}
\end{equation}
where we let $\boldsymbol{0}_{n}$ denote the $n$-dimensional zero vector for any \(n\in\mathbb{N}\).

Next, we introduce the overlapping link constraints. In air mobility network operations, multiple routes can share a common link. Therefore, the number of flights assigned to each route is constrained by the overlapping links. We further define a route-flow vector $\bold{z} \in \mathbb{R}^{n_r}$ where each entry $[\bold{z}]_{i}$ denotes the number of flights that fly route $i$. With the link-route incidence matrix $F$, we write the overlapping link constraints as follows:
\begin{equation}\label{eqn:fc2}
    F \bold{z} = \bold{y}, \quad \bold{z} \geq \boldsymbol{0}_{n_r}
\end{equation}

\subsubsection{Capacity Constraints}

To ensure safe and feasible operations in the air mobility network, we must also consider the capacity constraints at the links, vertiports, and way points. For example, the throughput of a vertiport per unit of time is limited by its feasible space and air traffic control conditions nearby. Therefore, there exists an upper bound of the total inflow (number of arrival flights) that each vertiport can take. We use $\bold{c}_v \in \mathbb{R}_{\geq 0}^{n_v}$ to denote the vertiport capacity vector, where the entry $[\bold{c}_v]_i$ denotes the maximum inflow entering vertiport $i$ per unit time. We formulate the vertiport capacity constraints as follows:
\begin{equation}\label{eqn:cc1}
    J \bold{z} \leq (1-\epsilon) \bold{c}_v\, .
\end{equation}
where $\epsilon \in [0, 1]$ is a tolerance parameter to control the safety margin. Similarly, we use $\bold{c}_l \in \mathbb{R}_{\geq 0}^{n_l}$ to denote the link capacity vector, where the entry $[\bold{c}_l]_i$ denotes the maximum number of flights link $i$ can take per unit time. We then formulate the link capacity constraints as follows:
\begin{equation}\label{eqn:cc2}
    \bold{y} \leq (1-\epsilon) \bold{c}_l\, .
\end{equation}

Lastly, we consider the way point capacity constraints. Here, we further define a matrix $K \in \mathbb{R}^{n_n \times n_l}$ such that each entry in $K$ is $[K]_{ij} = \max\{[E]_{ij}, 0\}$. We use $\bold{c}_w \in \mathbb{R}_{\geq 0}^{n_n}$ to denote the way point capacity vector, where the entry $[\bold{c}_w]_i$ denotes the maximum inflow entering way point $i$ per unit time. We then formulate the way point capacity constraints as follows:
\begin{equation}\label{eqn:cc3}
    K \bold{y} \leq (1-\epsilon) \bold{c}_w\, .
\end{equation}

\subsection{Noise Impact Matrix}

The main objective of this work is to control the UAM system's noise impact through regulating the traffic flow in the links of the air mobility network $\mathcal{G}$. Therefore, a core component in this process is a matrix that describes the relationships between link traffic flow and its noise impacts on the communities. We use matrix $N \in \mathbb{R}^{n_l \times n_c}$ to describe this relationship, where the entry $[N]_{ij}$ denotes the magnitude of single event noise (in SEL) that a flyover through link $i$ poses to community $j$. We use the noise modeling approach described in Section~\ref{sec:aircraftnoisemodeling} to compute all entries in matrix $N$. This process also takes into account the ambient noise level of a community, such that $[N]_{ij} = 0$ if the magnitude of single event noise from link $i$ to community $j$ is less than or equal to the ambient noise level of community $j$. Hence, matrix $N$ is expected to be a sparse matrix whose majority of the entries are zero.

With the noise impact matrix $N$ and link flow vector $\bold{y}$, we can now compute the cumulative noise that a certain air traffic flow condition poses to each community in the city. We then use $\bold{n} \in \mathbb{R}^{n_c}$ to denote the cumulative community noise result, where the entry $[\bold{n}]_j$ denotes the cumulative noise at community $j$. According to Equation~\eqref{eqn:leq1h}, the $j$-th entry is
\begin{equation}
    [\bold{n}]_j = 10 \log_{10} \left(\sum_{i=1}^{n_l} y_i \cdot 10^{N_{ij}/10}\right) - 10 \log_{10} \left(\frac{T}{T_0}\right)\, .
\end{equation}

We then further define a processed matrix $M \in \mathbb{R}^{n_l \times n_c}$ such that each of its element $[M]_{ij} = 10^{[N]_{ij}/10}$. The cumulative community noise vector $\bold{n}$ is then computed as:
\begin{equation}\label{eqn: noise function}
    \bold{n} = 10 \log_{10} \left(M^\top \bold{y}\right) - 10 \log_{10} (T/T_0) \cdot \boldsymbol{1}_{n_c}\, .
\end{equation}

\subsection{Energy Consumption}

Cruising at a higher altitude serves as an effective method to reduce the UAM noise impact on communities, yet this approach comes at a price. A prominent price is the increased energy consumption. Climbing to and descending from higher altitudes consumes more energy, potentially reducing the overall efficiency of the UAM system. Since energy is among the crucial aspects of sustainability, we also consider energy consumption in the trade-off study. Section~\ref{sec:ec} includes a detailed analysis of the energy consumption of eVTOL aircraft in a standard flight profile with five segments: vertical takeoff, climb, cruise, descent, and vertical landing. The main takeaway is that, in the range of UAM missions, flying at a higher altitude often results in increased energy consumption. In our case study for the city of Austin, depending on the route distance, cruising at 2,000 ft rather than 1,000 ft AGL consumes 7.28\% to 18.34\% more energy; cruising at 3,000 ft rather than 1,000 ft AGL consumes 14.70\% to 36.97\% more energy.

To incorporate energy consumption in the optimization model, we use $\bold{p} \in \mathbb{R}^{n_r}$ to denote the additional energy consumption due to cruising at altitudes higher than 1,000 ft AGL, where the entry $[\bold{p}]_i$ denotes the additional energy consumption of route $i$. $[\bold{p}]_i = 0$ if route $i$ is at 1,000 ft AGL. In our study, we use ``average additional energy consumption'' $p_a$ below as the indicator of the energy efficiency of the system:
\begin{equation}\label{eqn:ec}
    p_a \coloneqq \frac{\bold{p}^\top \bold{z}}{\boldsymbol{1}_{n_r}^\top \bold{z}} \,.
\end{equation}
where the denominator $\boldsymbol{1}_{n_r}^\top \bold{z}$ is the total route flow in the network. In the optimization problem, one can incorporate $p_a$ either in the objective function or in the constraints.

\subsection{Fairness and Social Welfare Function}

Incorporating fairness criteria into optimization models has become a topic of rising interest. In the design and operation of UAM systems, fairness embodies social equity, a key factor in fostering societal acceptance of UAM. Compared to traditional optimization objectives such as cost and efficiency, fairness presents a more diverse array of understanding and formulation possibilities. 

Overall, formulating and solving an optimization model with fairness consideration involves three steps: (1) select an appropriate fairness/inequality criterion according to the context; (2) incorporate the fairness criterion in a social welfare function (SWF), which serves as either the objective function or a constraint in the optimization formulation; (3) convert the optimization formulation into a computationally tractable mathematical programming model. 

Throughout this section, we let $\bold{x}$ denote the system design outcome and $\bold{u}$ denote the associated ulitity outcome. We assume that $\bold{x}$ yields a distribution of utilities $\bold{u} = \{u_1,...,u_n\}$ to parties $1,...,n$, which are determined by a vector-valued utility function $\bold{u} = U(\bold{x}) = \left(U_1(\bold{x}),...,U_n(\bold{x})\right)$~\cite{chen2023guide}. A SWF $\mathcal{F} (\bold{u})$ aggregates the utility vector $\bold{u}$ into a scalar value representing the desirability of utilities $\bold{u}$. In this study, we favor SWFs that facilitate a trade-off between efficiency and fairness. Such SWFs are designed to achieve a balance between \textit{utilitarian} (maximizing the total or average welfare across all parties -- system efficiency) and \textit{maximin} (maximizing the welfare of the least advantaged parties -- system fairness) criteria. Here we discuss two SWFs that can achieve this trade-off: fairness-threshold, and alpha-fairness.

This fairness-threshold SWF is initially utilitarian, but switches to maximin criterion for certain groups when unfairness surpasses a predefined level. In the 2-party setting, the fairness-threshold SWF is
\begin{equation}\label{eqn: threshold swf}
    \mathcal{F}_\Delta (u_1, u_2) = 
    \begin{cases}
        \min \{u_1, u_2\} + \Delta, & \text{if}~ |u_1 - u_2| \geq \Delta\\
        \frac{1}{2}(u_1 + u_2), & \text{otherwise}
    \end{cases}
\end{equation}
where parameter $\Delta$ is a pre-defined threshold that influences the fairness features of the solution. A recent work~\cite{elcci2022structural} generalized the fairness-threshold SWF to $n$ parties:
\begin{equation}\label{eqn: threshold swf 2}
    \mathcal{F}_\Delta (\bold{u}) = \Delta + \frac{1}{n}\sum_{i=1}^n \min \{u_i - \Delta, u_{\min}\}
\end{equation}
where \(u_{\min}\coloneqq \min_j u_j\). In \eqref{eqn: threshold swf 2}, the utilitarian criterion is applied to parties whose utilities are within $\Delta$ of the lowest utility, and that other parties whose utilities are above this level are not viewed as deserving greater utility under the current distribution. To this end,
function \(\mathcal{F}_\Delta (\bold{u})\) evaluates the sum of \(n_c\) terms, one for each party. For the \(i\)-th party, if \(u_i-u_{\min}\leq \Delta\), this term takes the value \(u_i\), otherwise it takes the value \(u_{\min}\).

Existing literature has extensively investigated the theoretical properties of another popular option, the alpha-fairness SWF~\cite{lan2010fairness,bertsimas2012efficiency,bertsimas2013fairness}, which has the form
\begin{equation}
    \mathcal{F}_\alpha(\bold{u}) = 
    \begin{cases}
        \displaystyle \sum_{i = 1}^n \frac{1}{1-\alpha} u_i^{1-\alpha}, & \text{if $\alpha>0$, $\alpha \neq 1$}\\
        \displaystyle \sum_{i = 1}^n \log(u_i), & \text{if $\alpha=1$}
    \end{cases}
\end{equation}
where $\alpha$ is a control parameter that can change the nature of the function from utilitarian criterion ($\alpha = 0$) to maximin criterion ($\alpha \rightarrow \infty$). While both fairness-threshold and alpha-fairness SWFs introduce a parameter $p$ for efficiency-fairness trade-off, we mainly apply the fairness-threshold SWF in Section~\ref{sec: opt model} because it offers improved interpretability of the objective function, since the parameter $\Delta$ carries significant implications for policy-making.


Inequality measure is another crucial element in formulating and assessing fairness in a decision-making process. In scenarios where the objective is to establish an egalitarian distribution for utilities $\bold{u}$, system designers often achieve greater fairness by reducing inequality. Many statistics are available for quantifying inequality. In this study, we employ the Gini coefficient $\text{Gini}(\bold{u}) = (1/2 n^2 \bar{u}) \sum_{i=1}^n \sum_{j=1}^n |u_i - u_j|$ as the primary inequality measure due to its broad acceptance in economics, sociology, and engineering.

\subsection{The Optimization Model}
\label{sec: opt model}

With the network model, operational constraints, noise impact computation, energy consumption computation, and SWF, we formally formulate an optimization model to manage UAM traffic flow within the three-layer network in Figure~\ref{fig:3dnetwork}. We further assume that the overall demand for UAM services exceeds supply due to the limited UAM resources and/or network throughput. 

\begin{table}[h!]
\centering
\caption{The complete list of parameters and decision variables in the optimization model}
\begin{tabular}{lll}
\hline \hline
\multirow{13}{*}{\begin{tabular}[c]{@{}l@{}}Network\\ Parameters\end{tabular}}
& $M \in \mathbb{R}^{n_l \times n_c}$ & The processed noise impact matrix \\
                                                             & $E \in \mathbb{R}^{n_n \times n_l}$ & The node-link incidence matrix \\
                                                             & $F \in \mathbb{R}^{n_l \times n_r}$ & The link-route incidence matrix \\
                                                             & $H \in \mathbb{R}^{n_o \times n_r}$ & The O-D pair-route incidence matrix \\
                                                             & $J \in \mathbb{R}^{n_v \times n_r}$ & The destination-route incidence matrix \\
                                                             & $K \in \mathbb{R}^{n_n \times n_l}$ & The inflow indication matrix \\
                                                             & $\bold{a} \in \mathbb{R}_{\geq 0}^{n_c}$ & A vector with community ambient noise levels \\
                                                             & $\bold{e} \in \mathbb{R}_{\geq 0}^{n_o}$ & A vector with the estimated demands for all O-D pairs \\
                                                             & $\bold{c}_v \in \mathbb{R}_{\geq 0}^{n_v}$ & The vertiport capacity vector \\
                                                             & $\bold{c}_l \in \mathbb{R}_{\geq 0}^{n_v}$ & The link capacity vector \\
                                                             & $\bold{c}_w \in \mathbb{R}_{\geq 0}^{n_n}$ & The way point capacity vector \\ 
                                                             & $\bold{p} \in \mathbb{R}_{\geq 0}^{n_r}$ & The additional energy consumption vector \\ \hline
\multirow{4}{*}{\begin{tabular}[c]{@{}l@{}}Design\\ Parameters\end{tabular}}
                                                             & $\Delta_{n, max} \in \mathbb{R}_{\geq 0}$ & The maximum allowable noise above the ambient level \\
                                                             & $T \in \mathbb{R}_{\geq 0}$ & The time interval (in seconds)\\
                                                             & $m_u \in \mathbb{R}_{\geq 0}$ & The upper bound of average increased noise \\
                                                             & $p_u \in \mathbb{R}_{\geq 0}$ & The upper bound of additional energy consumption \\ \hline
\multirow{5}{*}{\begin{tabular}[c]{@{}l@{}}Decision\\ Variables\end{tabular}} & $\bold{y} \in \mathbb{R}^{n_l}$ & The link flow vector \\ 
                                                             & $\bold{z} \in \mathbb{R}^{n_r}$ & The route flow vector  \\
                                                             & $\bold{d} \in \mathbb{R}^{n_o}$ & The demand fulfillment vector  \\
                                                             & $\bold{n} \in \mathbb{R}^{n_c}$ & The absolute community noise impact vector \\
                                                             & $\bold{n}' \in \mathbb{R}^{n_c}$ & The increased community noise impact vector \\ \hline \hline
\end{tabular}
\label{tbl:opt}
\end{table}

We introduce our proposed optimization model, which includes three core components: decision variables, objective, and constraints. The overall objective is to serve more demand while limiting community noise impact and energy consumption. In addition, we seek fairness in two aspects: demand fulfillment and community noise impact. On the former, we aim to achieve a fair distribution of the percentage of demand satisfied across all O-D pairs. On the latter, we aim to achieve a fair distribution of community noise annoyance/reaction, which is mainly indicated by the noise increase beyond a community's ambient noise level (see Table~\ref{tbl:delta}), across all communities. Table~\ref{tbl:opt} is a complete list of all parameters and decision variables in the optimization model.

In the multi-objective optimization formulation~\eqref{opt: multi}, the objective function is a convex combination of two SWFs for demand fulfillment $\bold{d}$ and community noise increase $\bold{n}'$ respectively. 
\begin{subequations}\label{opt: multi}
\begin{align}
\underset{\bold{y}, \bold{z}, \bold{d}, \bold{n}, \bold{n}'}{\mbox{maximize}} \quad & \omega \mathcal{F}_{\Delta_1}(\bold{d}) + (1-\omega) \mathcal{F}_{\Delta_2}\left(\boldsymbol{1}_{n_c}-\frac{1}{\Delta_{n, max}}\bold{n}'\right) \label{eqn: obj fun 1}\\
\textrm{subject to} \quad &  E \bold{y} = \boldsymbol{0}_{n_n}, ~ F \bold{z} = \bold{y}, \label{eqn: flow constr 1}\\
       & J \bold{z} \leq (1-\epsilon) \bold{c}_v, ~ \bold{y} \leq (1-\epsilon) \bold{c}_l, ~ K \bold{y} \leq (1-\epsilon) \bold{c}_w, \label{eqn: cap constr 1}\\
       & \bold{d} = \text{diag}(\bold{e})^{-1} H \bold{z}, \label{eqn: demand constr 1}\\
       & \bold{n} = 10 \log_{10} \left(M^\top \bold{y}\right) - 10 \log_{10} (T/T_0) \cdot \boldsymbol{1}_{n_c}, \label{eqn: noise constr 1}\\
       & \bold{n}' = \max\{\bold{n} - \bold{a}, \boldsymbol{0}_{n_c}\}, \label{eqn: noise inc constr 1}\\
       & \bold{p}^\top \bold{z}/\boldsymbol{1}_{n_r}^\top \bold{z} \leq p_u, \label{eqn: energy constr 1}\\
       & \bold{d} \leq \boldsymbol{1}_{n_o}, ~ \bold{n}' \leq \Delta_{n, max} \cdot \boldsymbol{1}_{n_c}, \label{eqn: upper constr 1}\\
       & \bold{y} \geq \boldsymbol{0}_{n_l}, ~ \bold{z} \geq \boldsymbol{0}_{n_r}. \label{eqn: non neg constr 1}
\end{align}
\end{subequations}
where in the objective function~\eqref{eqn: obj fun 1}, a parameter $\omega \in [0, 1]$ controls the relative weights between the two objectives for design trade-offs. Both objectives use the fairness-threshold SWF from~\eqref{eqn: threshold swf 2}. 

Optimization model~\eqref{opt: multi} encompasses the following sets of constraints. The constraints in \eqref{eqn: flow constr 1}, first introduced in \eqref{eqn:fc1} and \eqref{eqn:fc2}, ensure the conservation of traffic flow at nodes and links. The constraints in \eqref{eqn: cap constr 1}, first introduced in \eqref{eqn:cc1}, \eqref{eqn:cc2}, and \eqref{eqn:cc3}, ensure the satisfaction of capacity constraints at vertiports, links, and waypoints. The constraints in \eqref{eqn: demand constr 1} define the vector of demand fulfillment ratio for all O-D pairs, denoted by $\bold{d}$, where $\bold{e}$ is a vector with the estimated demands for all O-D pairs. Each entry $[\bold{d}]_i$ in the range of $[0,1]$. The constraints in \eqref{eqn: noise constr 1}, first introduced in \eqref{eqn: noise function}, capture the relation between the cumulative community noise and link traffic flow. The constraints in \eqref{eqn: noise inc constr 1} describe the relation between the cumulative community noise \(\bold{n}\), the community noise increase, denoted by vector \(\bold{n}'\), and the community ambient noise levels denoted by vector $\bold{a}$ where the entries $[\bold{a}]_i$ and $[\bold{n}']_i$ denote the ambient noise level (in dB) and noise increase for community $i$, respectively. In particular, the entry $[\bold{n}']_i$ denotes the magnitude to which the cumulative UAM noise posed to the community $i$, $[\bold{n}]_i$, is higher than the community's ambient noise $[\bold{a}]_i$. Should the cumulative UAM noise be less than the community's ambient noise level, it becomes imperceptible to the residents, resulting in $[\bold{n}']_i = 0$. The constraints in \eqref{eqn: energy constr 1}, first introduced in \eqref{eqn:ec}, upper-bound the additional energy consumption due to cruising at higher altitudes. The constraints in \eqref{eqn: upper constr 1} upper bound constraints the demand fulfillment ratio and the noise increase. Entries in $\bold{d}$ should not exceed 1 as achieved supply should not exceed demand. In addition, for each community, we set an upper limit on the increased noise level as the limitation of community noise impact. In this study, we choose $\Delta_{n, max} = 25$ dB as the maximum allowable UAM noise above the ambient level. Finally, The constraints in \eqref{eqn: non neg constr 1} are nonnegativity constraints imposed on decision variables $\bold{y}$ and $\bold{z}$, also from~\eqref{eqn:fc1} and \eqref{eqn:fc2}.

\subsection{Solving the Optimization Problem}

Optimization problem \eqref{opt: multi} is challenging to solve due to the nonsmooth and nonconvex functions that appear in its objective and constraints. In particular, the pointwise maximum function appeared in \eqref{eqn: obj fun 1} and the noise increase function in \eqref{eqn: noise inc constr 1} are both piecewise linear and nonsmooth. On the other hand, the logarithm function appeared in \eqref{eqn: noise constr 1} is nonlinear and nonconvex, making optimization~\eqref{opt: multi} a nonconvex nonlinear program not solvable by any commercial linear program solvers.   

We propose a numerical method that solves optimization~\eqref{opt: multi} by iteratively solving linear programs. This method replaces the nonsmooth functions in \eqref{opt: multi} with linear differentiable ones by introducing auxiliary variables. Furthermore, it iteratively approximates the logarithm function in \eqref{opt: multi} via the \emph{convex-concave procedure}, an algorithm designed for a special class of nonlinear programs with structured nonconvexity. This algorithm guarantees convergence to local optimality and has empirically demonstrated rapid convergence in a wide range of applications. For further details on the convex-concave procedure, we refer the interested readers to \cite{lipp2016variations} and references therein.

We start with transforming the nonsmooth functions in \eqref{opt: multi} into linear ones using additional auxiliary variables. For the noise increase function, we introduce an auxiliary variable \(\bold{w}\in\mathbb{R}^{n_c}\), and make the following transformation 
\begin{equation}
    \bold{n}'=\bold{w}.
\end{equation}

In addition, we add the following constraints to \eqref{opt: multi}
\begin{equation}
    \bold{w}\geq \bold{n}-\bold{a}, ~\bold{w}\geq \mathbf{0}_{n_c}.
\end{equation}

For function \(\mathcal{F}_{\Delta_1}(\bold{d})\), we introduce two auxiliary variable \(\bold{u}\in\mathbb{R}^{n_o}\) and \(d_{\min}\in\mathbb{R}\), and make the following transformation
\begin{equation}
    \mathcal{F}_{\Delta_1}(\bold{d})=\Delta_1 +\frac{1}{n_o}\mathbf{1}_{n_o}^\top \bold{u}.
\end{equation}

In addition, we add the following constraints to \eqref{opt: multi}
\begin{equation}
    \bold{u}\leq \bold{d}-\Delta_1 \mathbf{1}_{n_o}, ~ \bold{u}\leq d_{\min}\mathbf{1}_{n_o}, ~ d_{\min}\mathbf{1}_{n_o}\leq \bold{d}.
\end{equation}

Similarly, for function \(\mathcal{F}_{\Delta_2}(\mathbf{1}_{n_c}-\frac{1}{\Delta_{n, \max}}\bold{n'})\), we introduce two auxiliary variable \(\bold{v}\in\mathbb{R}^{n_c}\) and \(s_{\min}\in\mathbb{R}\), and make the following transformation
\begin{equation}
\mathcal{F}_{\Delta_2}\left(\mathbf{1}_{n_c}-\frac{1}{\Delta_{n, \max}}\bold{n}'\right)=\Delta_2 +\frac{1}{n_c}\mathbf{1}_{n_c}^\top \bold{v}.
\end{equation}

In addition, we add the following constraints to \eqref{opt: multi}
\begin{equation}
    \bold{v}\leq \mathbf{1}_{n_c}-\frac{1}{\Delta_{n, \max}}\bold{n}'-\Delta_2 \mathbf{1}_{n_c}, ~ \bold{v}\leq s_{\min}\mathbf{1}_{n_c}, ~ s_{\min}\mathbf{1}_{n_c}\leq \mathbf{1}_{n_c}-\frac{1}{\Delta_{n, \max}}\bold{n}'.
\end{equation}

The idea of the above transformations is to replace pointwise maximum and minimum functions with upper and lower bounds with additional auxiliary variables, respectively. We note that these transformations are lossless since the objective in \eqref{opt: multi} aim to maximize \(\mathcal{F}_\Delta(\bold{d})\) and \(\mathcal{F}_{\Delta_2}(\mathbf{1}_{n_c}-\frac{1}{n, \max}\bold{n}')\). 

Next, we approximate the nonlinear nonconvex logarithmic function in \eqref{opt: multi} via linearization. Given a linearization point \(\hat{\bold{y}}\), we can make the following approximation
\begin{equation}
    \begin{aligned}
        \bold{n} & = 10 \log_{10} \left(M^\top \bold{y}\right) - 10 \log_{10} (T/T_0) \cdot \boldsymbol{1}_{n_c}\\
        &\approx A(\hat{\bold{y}}) (\bold{y}-\hat{\bold{y}})+b(\hat{\bold{y}})
    \end{aligned}
\end{equation}
where 
\begin{subequations}\label{eqn: linear app}
    \begin{align}
        A(\hat{\bold{y}}) & \coloneqq \frac{10}{\ln(10)} \diag(M\hat{\bold{y}})^{-1}M,\\
        b(\hat{\bold{y}}) & \coloneqq 10 \log_{10} \left(M^\top \hat{\bold{y}}\right) - 10 \log_{10} (T/T_0) \cdot \boldsymbol{1}_{n_c}.
    \end{align}
\end{subequations}

With the above steps, we approximate optimization~\eqref{opt: multi} using the following linear program:
\begin{subequations}\label{opt: app lp}
\begin{align}
\underset{\substack{\bold{y}, \bold{z}, \bold{d}, \bold{w}\\ \bold{u},  d_{\min}, \bold{v},  s_{\min}}}{\mbox{maximize}} \quad & \frac{\omega}{n_o} \mathbf{1}_{n_o}^\top  \bold{u}+\frac{1-\omega}{n_c} \mathbf{1}_{n_c}^\top \bold{v}\\
\textrm{subject to} \quad &  E \bold{y} = \boldsymbol{0}_{n_n}, ~ F \bold{z} = \bold{y}, ~ \bold{d} = \text{diag}(\bold{e})^{-1} H \bold{z},\\
       & J \bold{z} \leq (1-\epsilon) \bold{c}_v, ~ \bold{y} \leq (1-\epsilon) \bold{c}_l, ~ K \bold{y} \leq (1-\epsilon) \bold{c}_w,\\
       & \bold{w} \geq  A(\hat{\bold{y}})(\bold{y}-\hat{\bold{y}})+b(\hat{\bold{y}}), ~ \bold{w}\geq \mathbf{0}_{n_c},\\
       & \frac{1}{n_c}\mathbf{1}_{n_c}^\top\bold{w} \leq m_u, ~ \bold{w} \leq \Delta_{n, max} \cdot \boldsymbol{1}_{n_c}, ~ \bold{p}^\top \bold{z}/\boldsymbol{1}_{n_r}^\top \bold{z} \leq p_u,~ \bold{d} \leq \boldsymbol{1}_{n_o}, \\
       & \bold{u}\leq \bold{d}-\Delta_1 \mathbf{1}_{n_o}, ~ \bold{u}\leq d_{\min}\cdot\mathbf{1}_{n_o}, ~ d_{\min}\cdot\mathbf{1}_{n_o}\leq \bold{d},\\
       & \bold{v}\leq \mathbf{1}_{n_c}-\frac{1}{\Delta_{n, \max}} \bold{w}-\Delta_2 \cdot \mathbf{1}_{n_c}, ~ \bold{v}\leq s_{\min}\cdot\mathbf{1}_{n_c}, ~ s_{\min}\cdot\mathbf{1}_{n_c}\leq \bold{w},\\
       & \bold{y} \geq \boldsymbol{0}_{n_l}, ~ \bold{z} \geq \boldsymbol{0}_{n_r}.
\end{align}
\end{subequations}

Notice that the quality of approximating \eqref{opt: multi} with \eqref{opt: app lp}  depends heavily on the choice of \(\hat{\bold{y}}\) in \eqref{eqn: linear app}. 

To solve optimization~\eqref{opt: multi} using the approximating linear program in \eqref{opt: app lp}, we propose the convex-concave procedure, an iterative method to solve difference-of-convex programs via linearization \cite{lipp2016variations}. We summarize this procedure in Algorithm~\eqref{alg: ccp}. The idea is to solve the linear program in \eqref{opt: app lp}, then use the solution to construct a new linearization point and the corresponding linear program, and finally repeat this procedure until convergence.

\begin{algorithm}
\caption{Convex-concave procedure}
\label{alg: ccp}
\begin{algorithmic}
\Require The parameters listed in Table~\ref{tbl:opt}, an initial linearization point \(\hat{\bold{y}}\), stopping tolerance \(\epsilon\).
\State \(f_0=-\infty\), \(f_1=\infty\), \(k=1\).
\State Compute \(A(\hat{\bold{y}})\) and \(b(\hat{\bold{y}})\) using \eqref{eqn: linear app}.
\While{\(|f^{k+1}-f^k|>\epsilon\)}
\State Solve the linear program in \eqref{opt: app lp}, obtain optimal solution \((\bold{y}^k, \bold{z}^k, \bold{d}^k, \bold{w}^k, \bold{u}^k, , d_{\min}^k, \bold{v}^k, , s_{\min}^k)\) with optimal value \(f^k\). 
\State Compute \(A(\hat{\bold{y}})\) and \(b(\hat{\bold{y}})\) using \eqref{eqn: linear app} with \(\hat{\bold{y}}=\bold{y}^k\).
\State \(k=k+1\)
\EndWhile
\Ensure \((\bold{y}^k, \bold{z}^k, \bold{d}^k)\)
\end{algorithmic}
\end{algorithm}

\section{Results}\label{sec:results}

In this section we present and discuss detailed results of the case study of the city of Austin. We consider an urban air traffic management problem for a $T = 3,600$ seconds (1 hour) time interval. We conduct trade-off studies between the three aspects -- demand fulfillment, noise control, and energy consumption, with efficiency and fairness trade-off considerations in the first two aspects. We first display some representative results via data visualizations on flow and noise distributions, followed by quantitative comparisons using multiple metrics.

\begin{figure}[h!]
    \centering
    \includegraphics[width=0.425\textwidth]{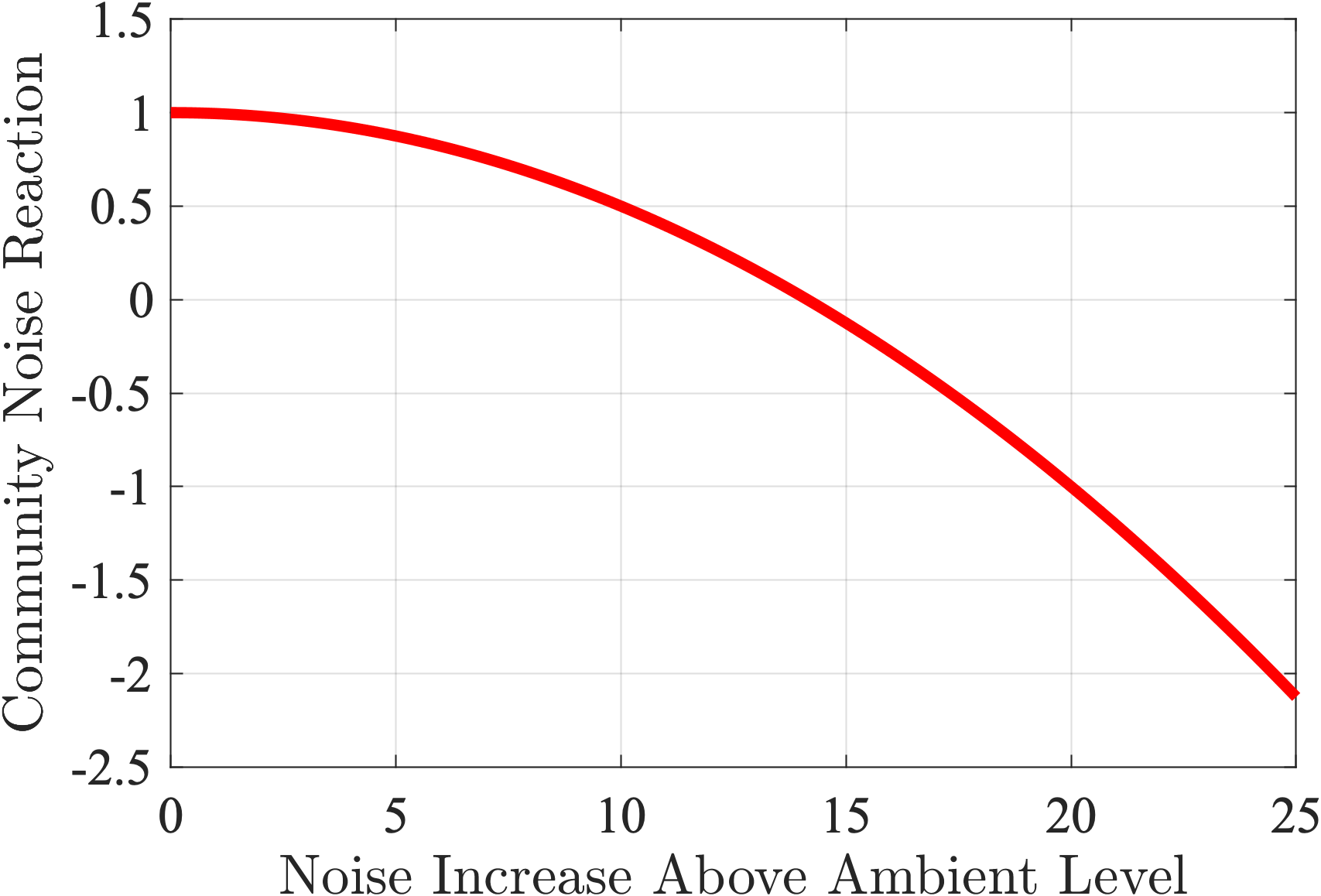}
    \caption{The community noise reaction score function}
    \label{fig:commreact}
\end{figure}

\subsection{Result Visualizations}

\begin{figure}[h!]
     \centering
     \begin{subfigure}[b]{\textwidth}
         \centering
         \includegraphics[width=0.226\textwidth]{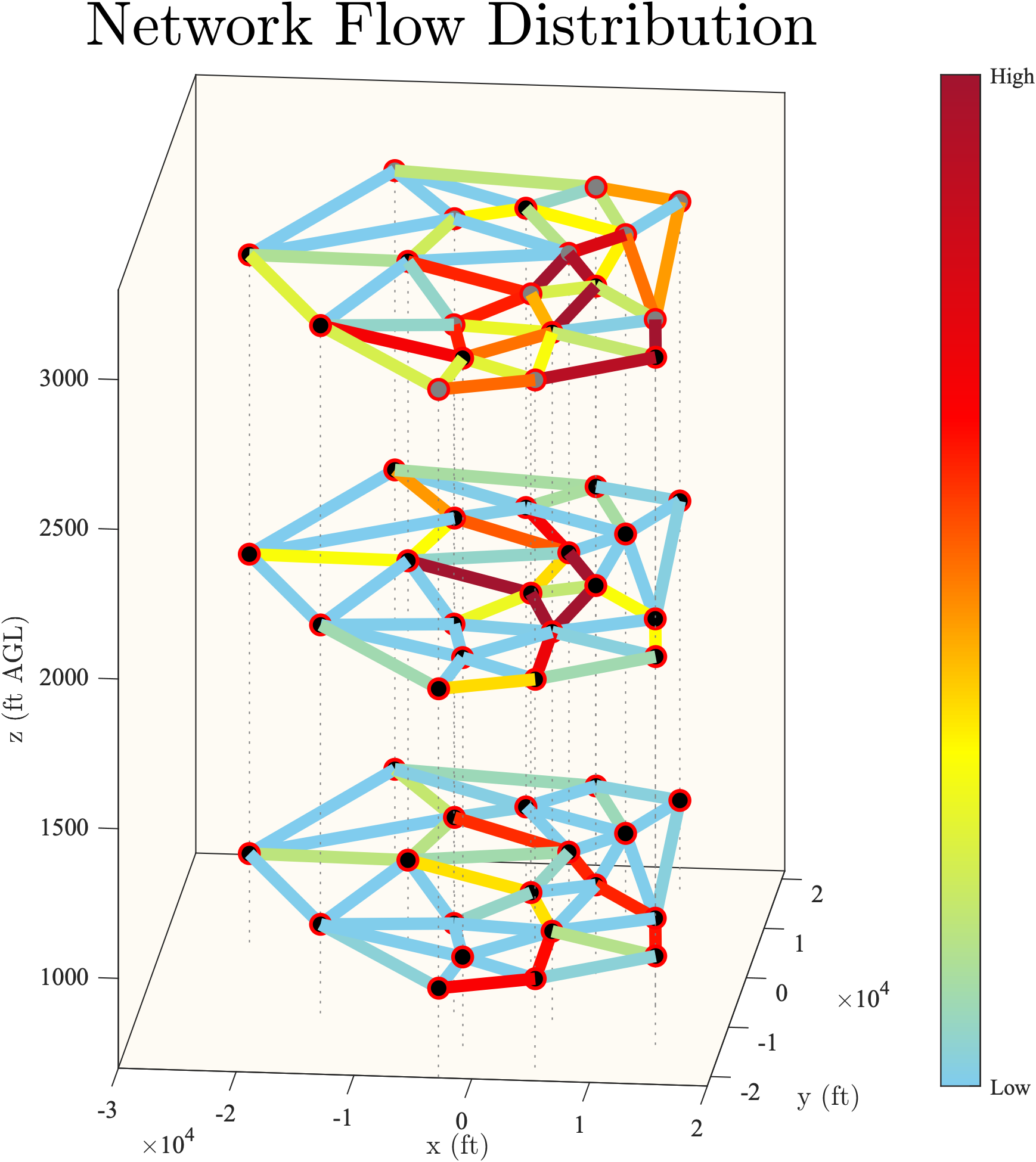}
         \hspace{0.3cm}
         \includegraphics[width=0.23\textwidth]{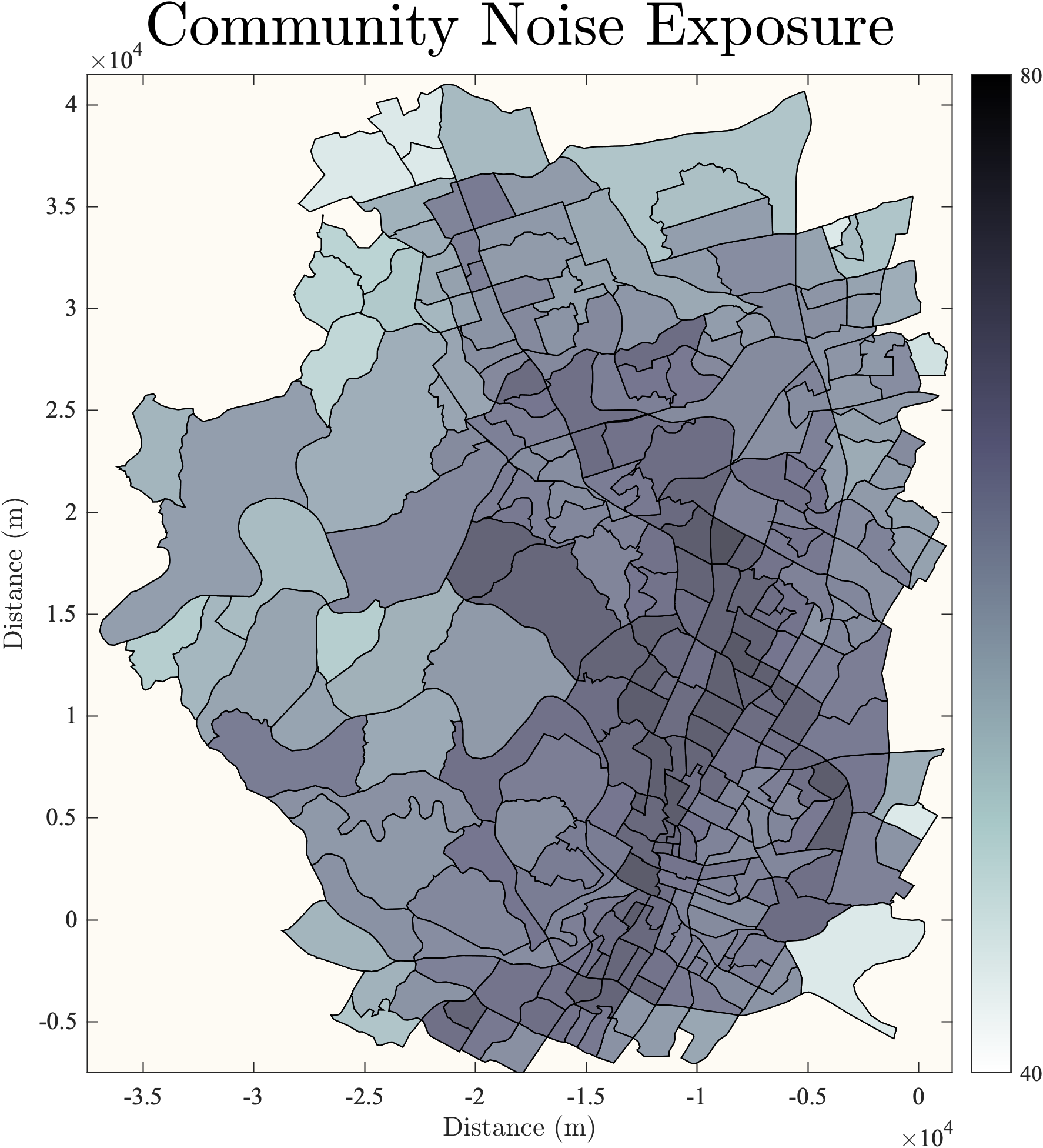}
         \hspace{0.3cm}
         \includegraphics[width=0.237\textwidth]{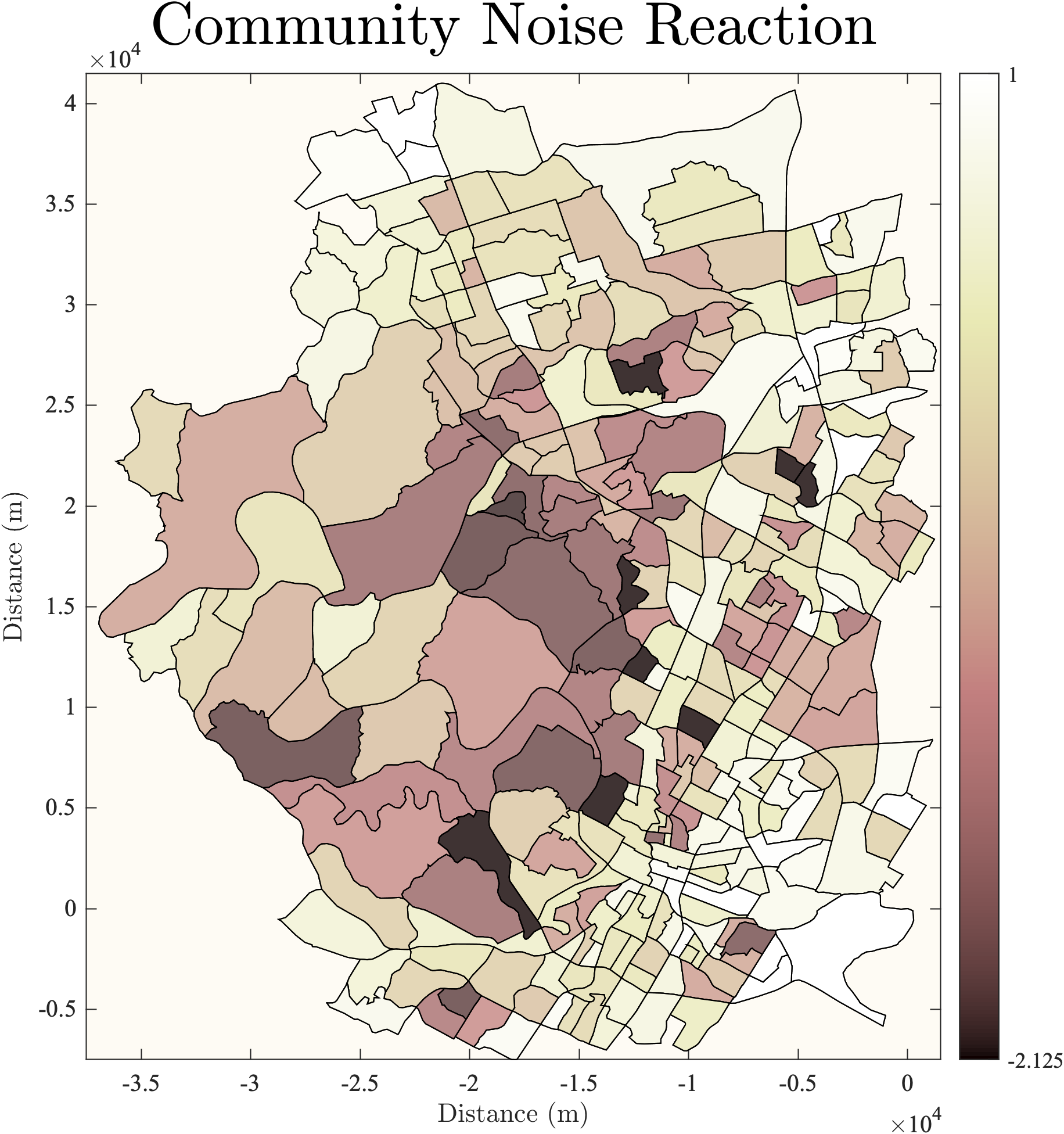}
         \hspace{0.3cm}
         \includegraphics[width=0.161\textwidth]{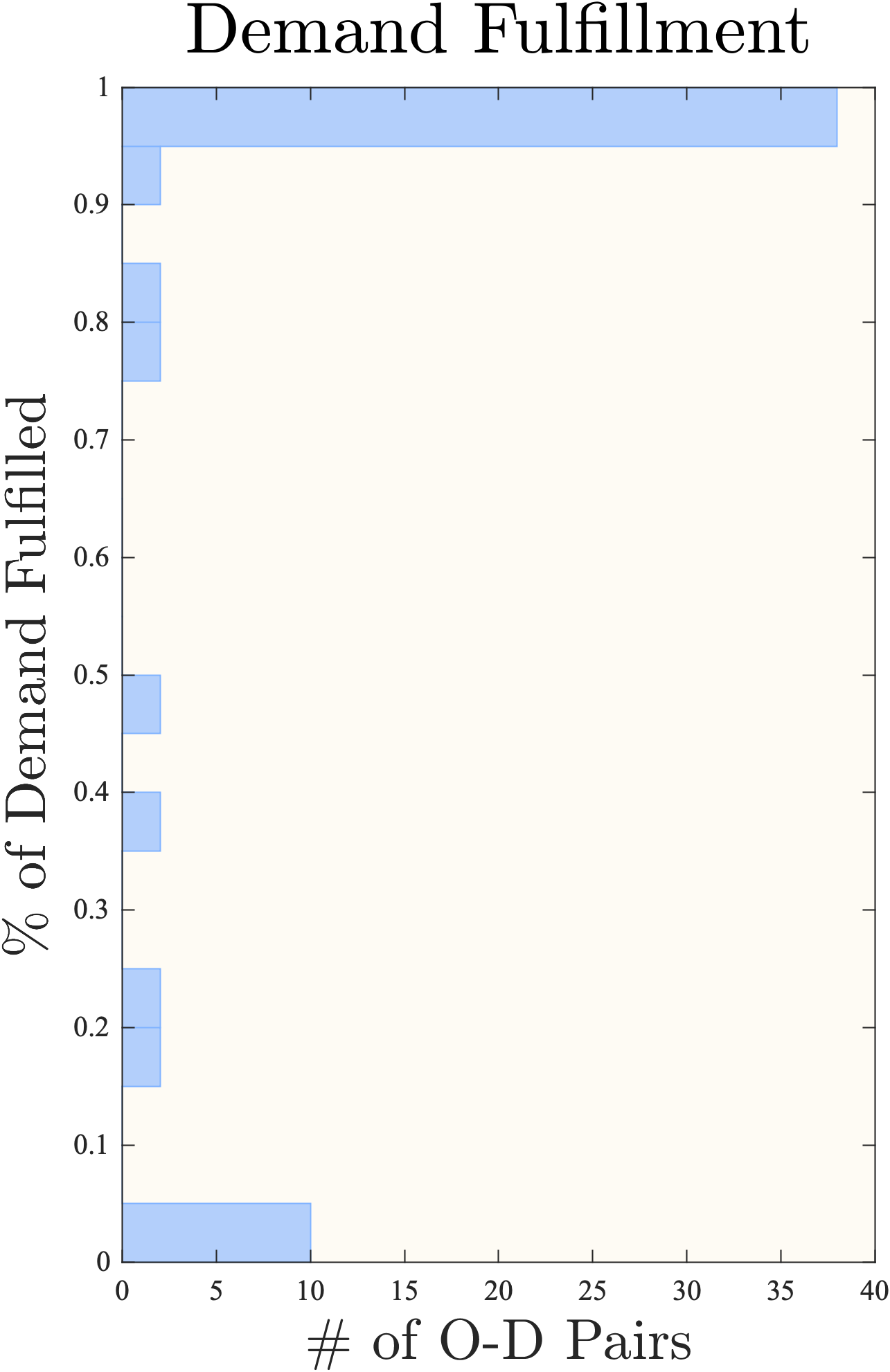}
         \caption{Maximizing demand fulfillment -- utilitarian}
         \vspace{0.15cm}
         \label{fig:viz11}
     \end{subfigure}
     \begin{subfigure}[b]{\textwidth}
         \centering
         \includegraphics[width=0.226\textwidth]{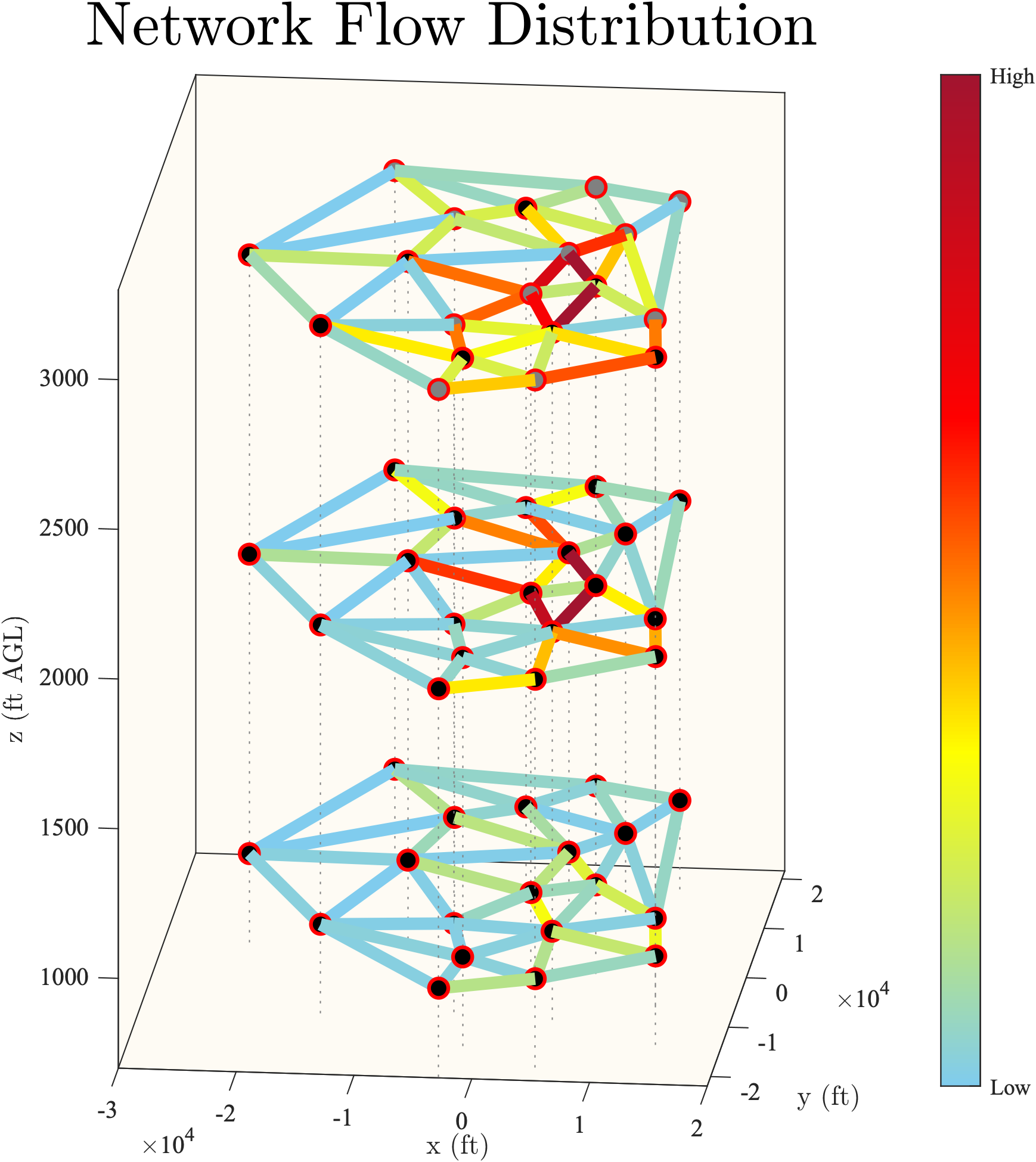}
         \hspace{0.3cm}
         \includegraphics[width=0.23\textwidth]{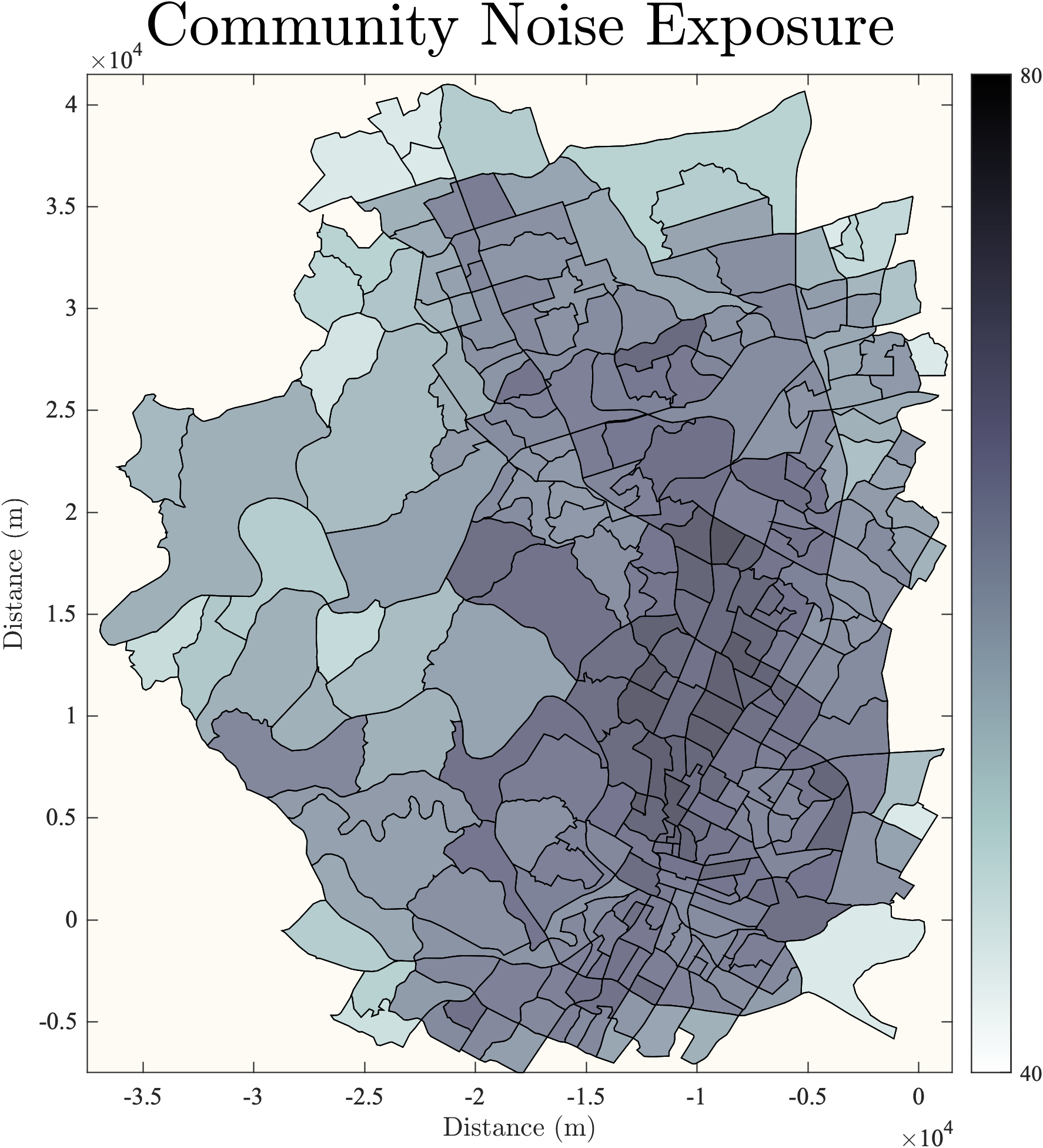}
         \hspace{0.3cm}
         \includegraphics[width=0.237\textwidth]{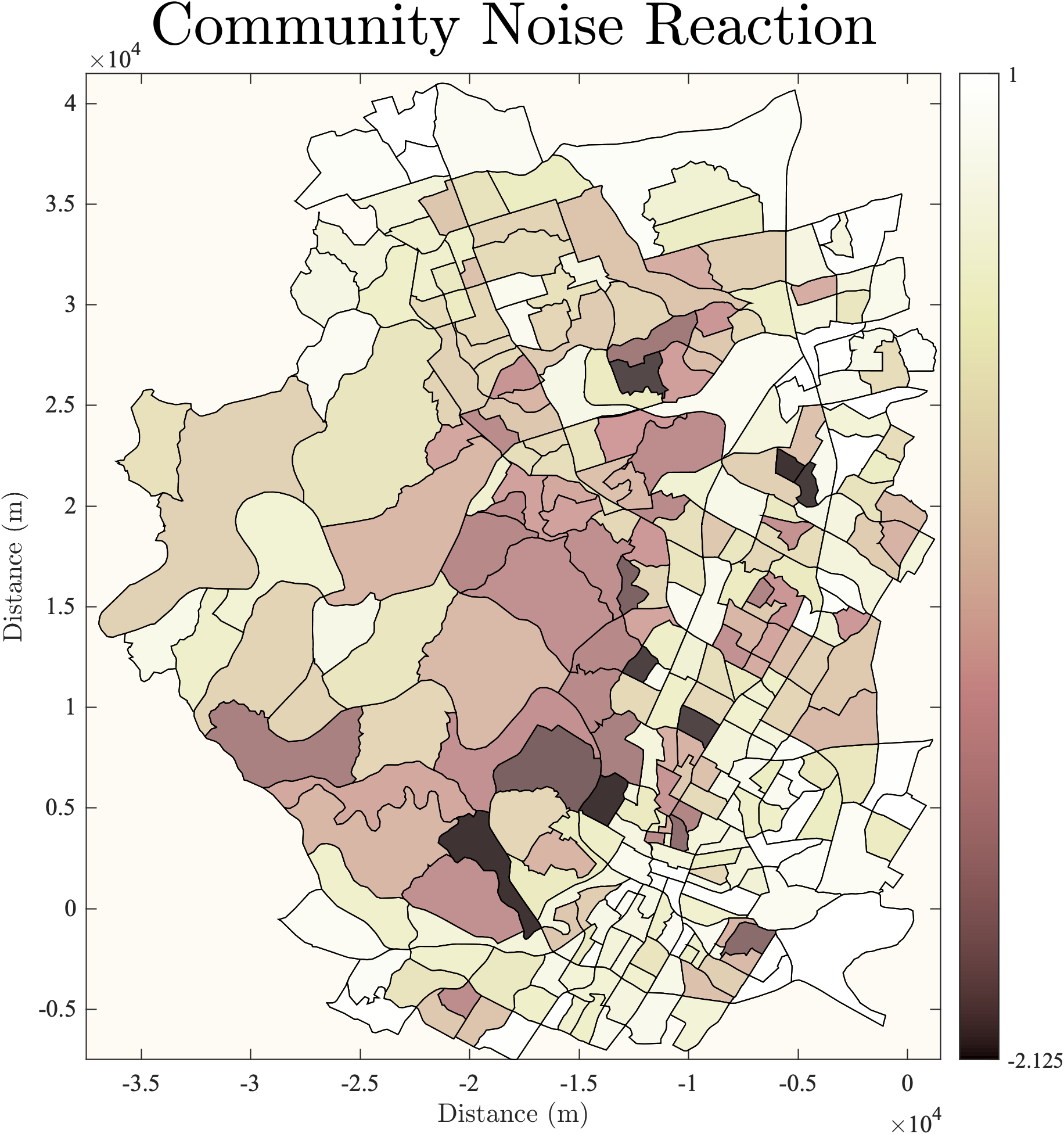}
         \hspace{0.3cm}
         \includegraphics[width=0.161\textwidth]{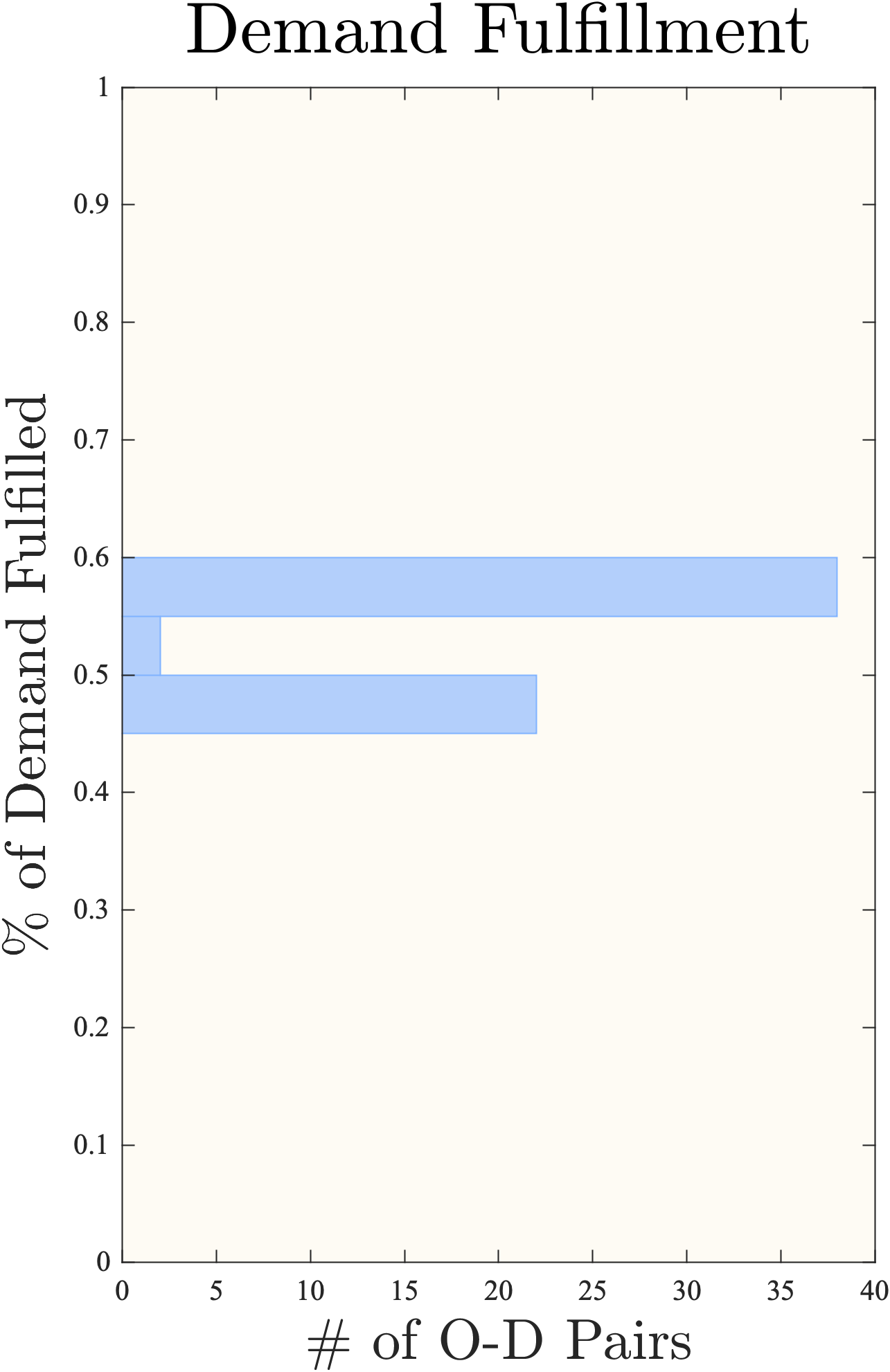}
         \caption{Maximizing demand fulfillment -- egalitarian}
         \vspace{0.15cm}
         \label{fig:viz12}
     \end{subfigure}
     \begin{subfigure}[b]{\textwidth}
         \centering
         \includegraphics[width=0.226\textwidth]{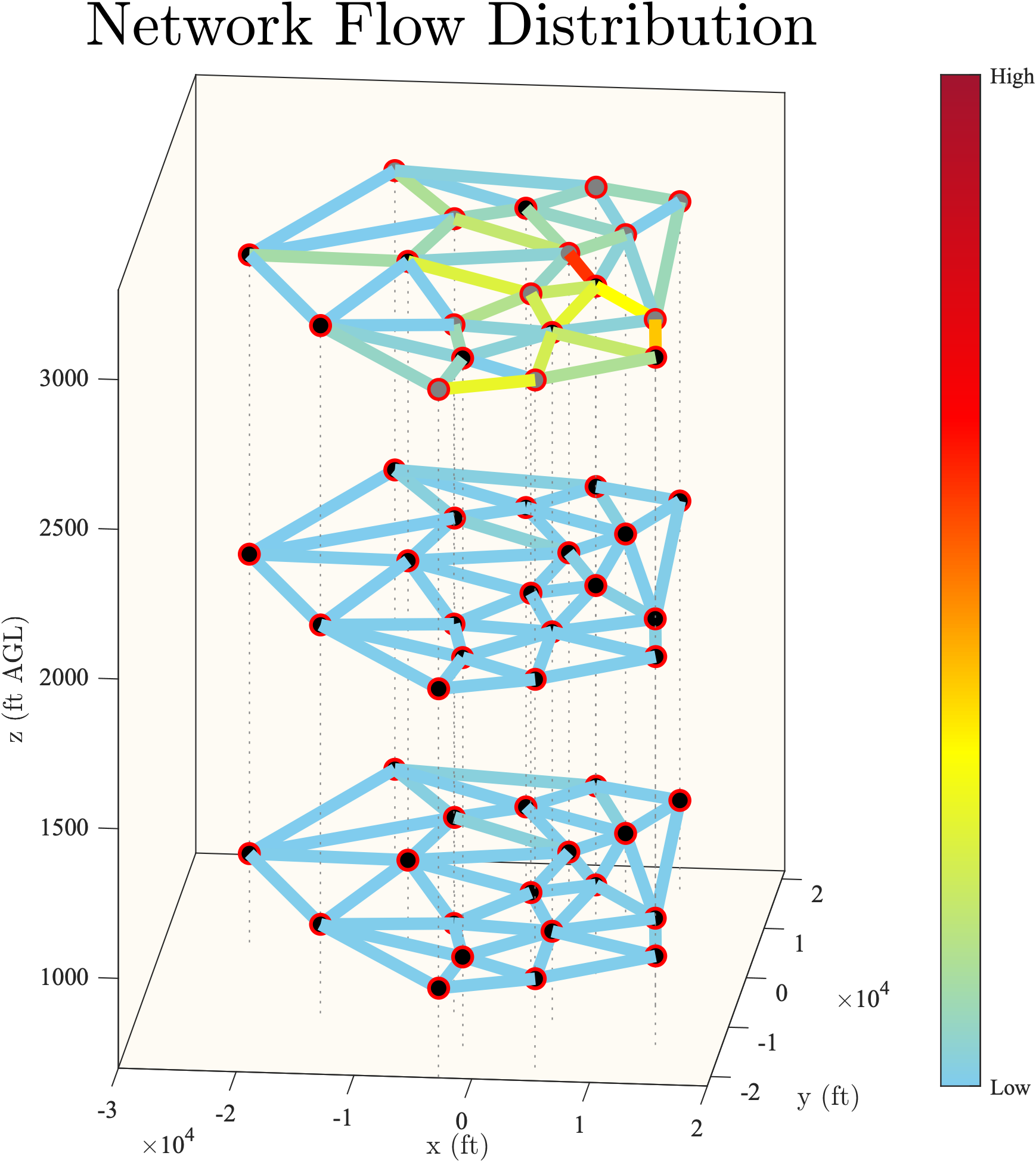}
         \hspace{0.3cm}
         \includegraphics[width=0.23\textwidth]{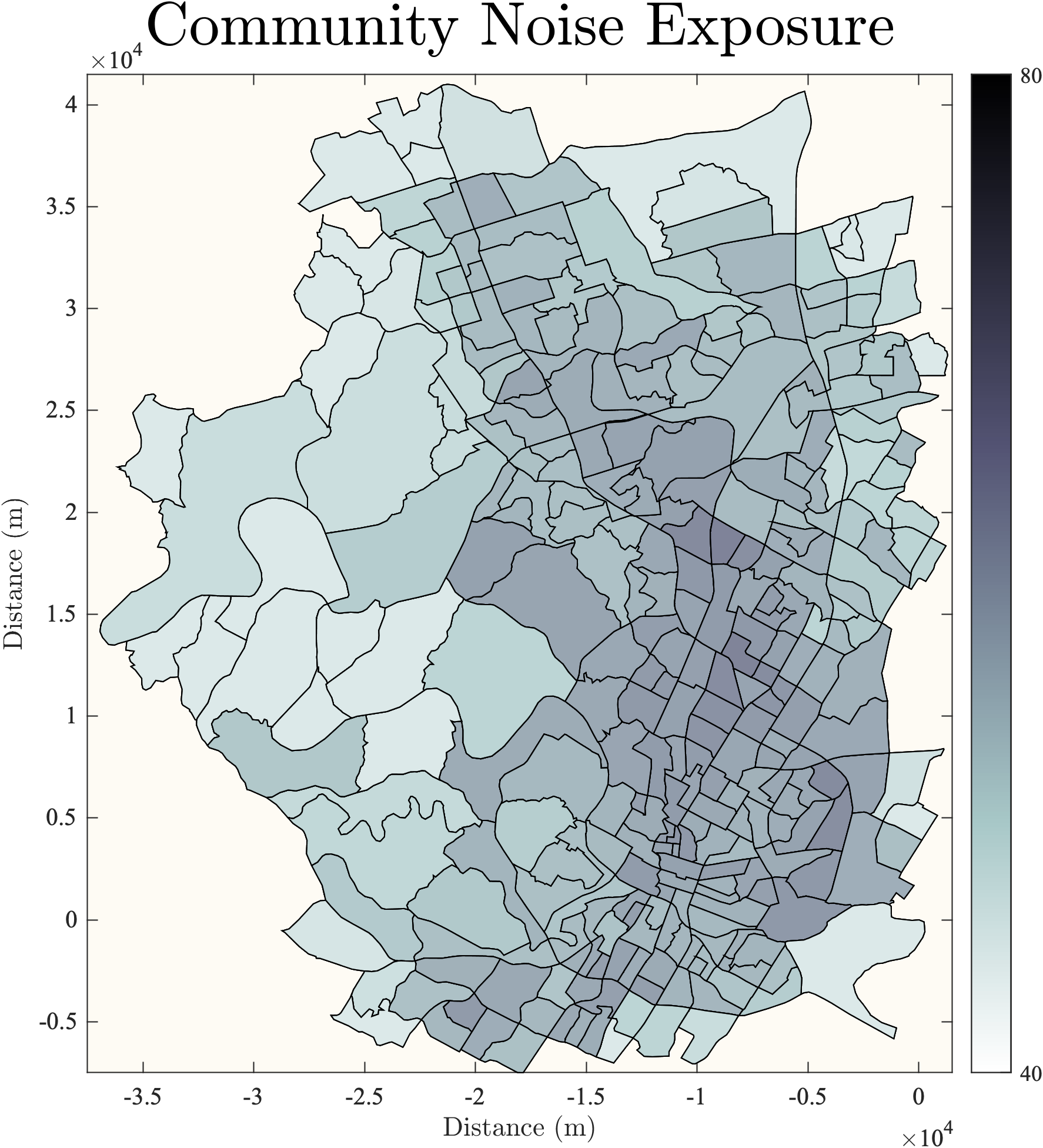}
         \hspace{0.3cm}
         \includegraphics[width=0.237\textwidth]{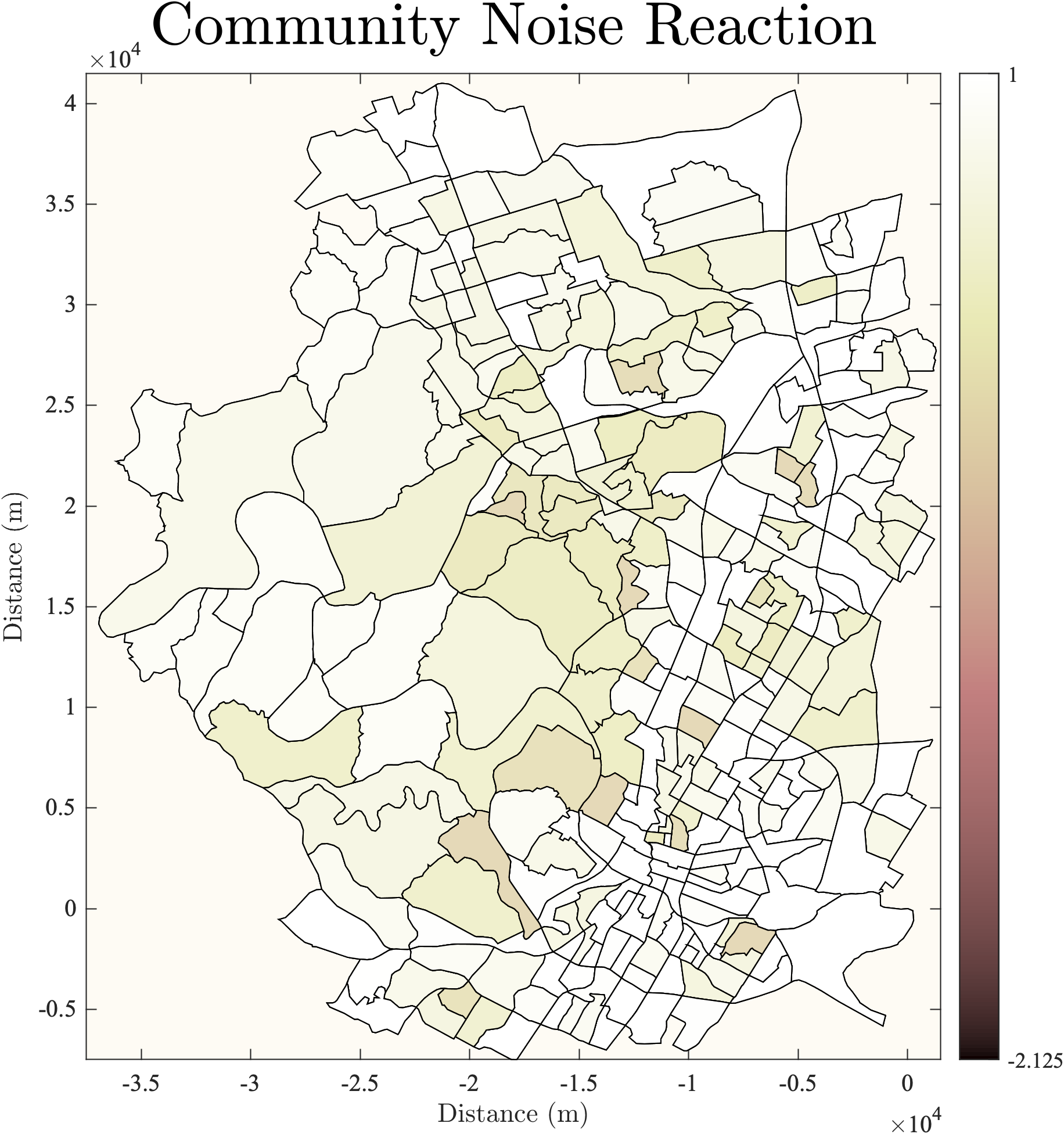}
         \hspace{0.3cm}
         \includegraphics[width=0.161\textwidth]{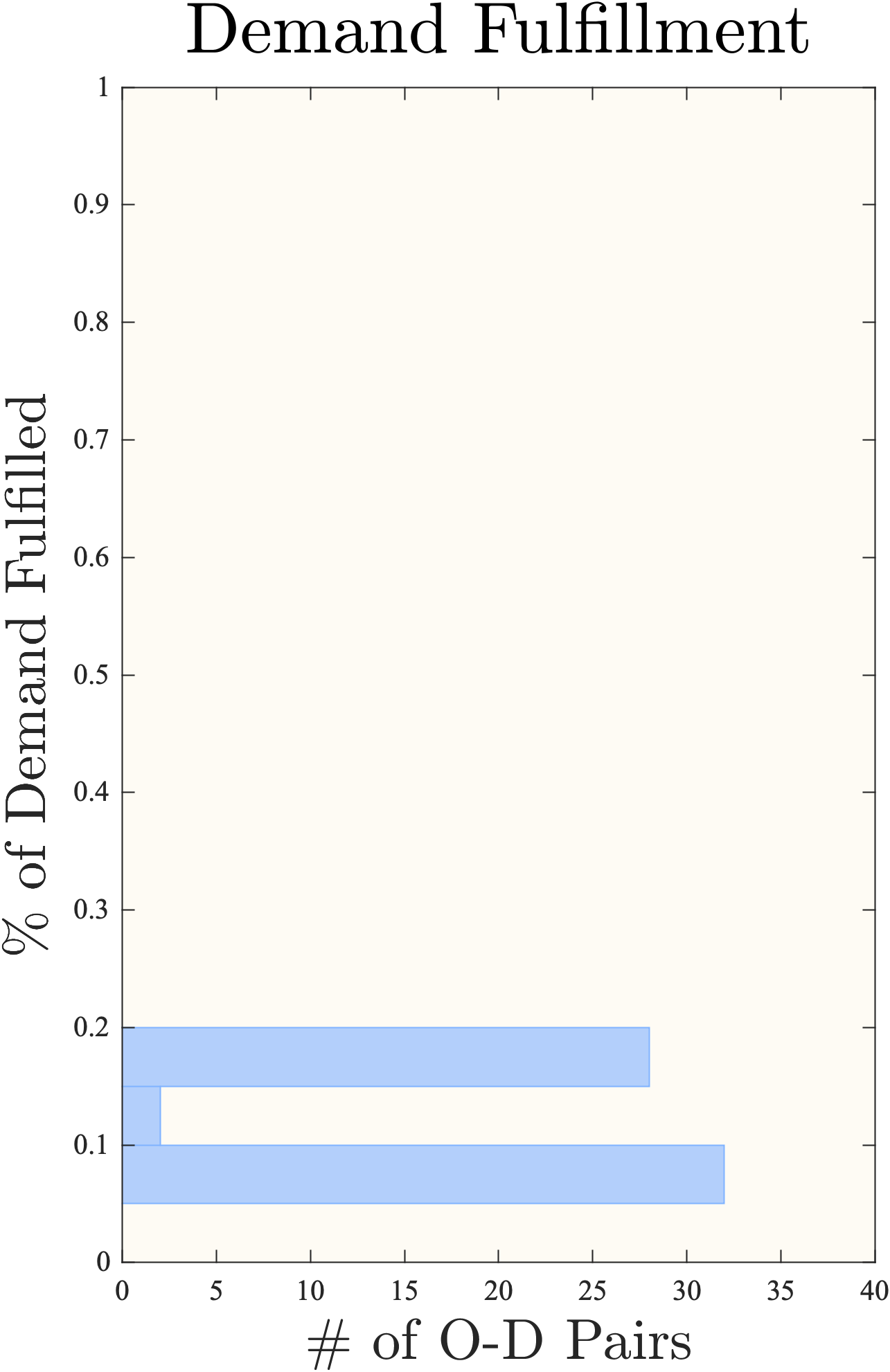}
         \caption{Low level of demand fulfillment -- example}
         \vspace{0.15cm}
         \label{fig:viz13}
     \end{subfigure}
     \caption{Visualizations of UAM traffic flow distribution and noise impact under different demand fulfillment scenarios}
     \label{fig:viz1}
\end{figure}

To intuitively demonstrate the interactions between the three decision-making aspects and the implications of fairness, we first use three groups of visualizations, Figure~\ref{fig:viz11}, \ref{fig:viz12}, and \ref{fig:viz13}. The visualizations contain 8 design results in total, while each result consists of four plots. As in Figure~\ref{fig:viz11}, the first plot depicts the air traffic flows within the three-layer UAM network in Figure~\ref{fig:3dnetwork}. We utilize color coding to indicate the flow magnitude within each link so that one can observe the traffic flow intensity and distribution in each case. The second plot depicts the overall community noise exposure in 292 Austin communities caused by UAM operations, where a progressive intensification in color shade signifies an escalation in noise impact. The third plot depicts the community noise reaction, which is a function of noise increase above a community's ambient noise level. Specifically, to quantify a community's annoyance to UAM noise, we employ a nonlinear and concave function in Figure~\ref{fig:commreact} which reflects the study results in Table~\ref{tbl:delta}. In the third plot, a darker hue indicates a higher level of community annoyance. The fourth plot is a histogram that shows the distribution of the percentage of demand fulfillment across the 62 O-D pairs.

Figure~\ref{fig:viz1} displays the first group of visualization results which focuses on different demand fulfillment scenarios. Figure~\ref{fig:viz11} shows the result when demand fulfillment is the only objective, i.e., $\omega = 1$ in \eqref{opt: multi} and \eqref{opt: app lp}. This is the result with the greatest overall demand fulfillment (utilitarian). Through the first three plots in Figure~\ref{fig:viz11}, we can observe considerable traffic flows in all three network layers, extensive noise impacts throughout the city, and intense noise reactions in numerous communities. The last plot in Figure~\ref{fig:viz11} shows that under the utilitarian criterion, the distribution of \% demand fulfillment is unfair. While over 35 O-D pairs receive 100\% demand fulfillment, 10 O-D pairs receive no demand fulfillment at all. We then apply an egalitarian criterion in the optimization of demand fulfillment and generate the result in Figure~\ref{fig:viz12}. In Figure~\ref{fig:viz12}, we observe reduced traffic flows in the network, as well as slightly reduced noise impacts and community noise reactions. More importantly, the last plot of Figure~\ref{fig:viz12} shows that every O-D pair can now receive 45\% to 60\% demand fulfillment, which is apparently fairer compared to the utilitarian case. Figure~\ref{fig:viz13} displays a scenario when the UAM system's demand fulfillment is at a relatively low level. We can see that traffic flows are light in the network, and that all O-D pairs receive only 5\% to 20\% demand fulfillment. Consequently, we notice very low noise impacts and minor noise reactions from communities.

\begin{figure}[h!]
     \centering
     \begin{subfigure}[b]{\textwidth}
         \centering
         \includegraphics[width=0.226\textwidth]{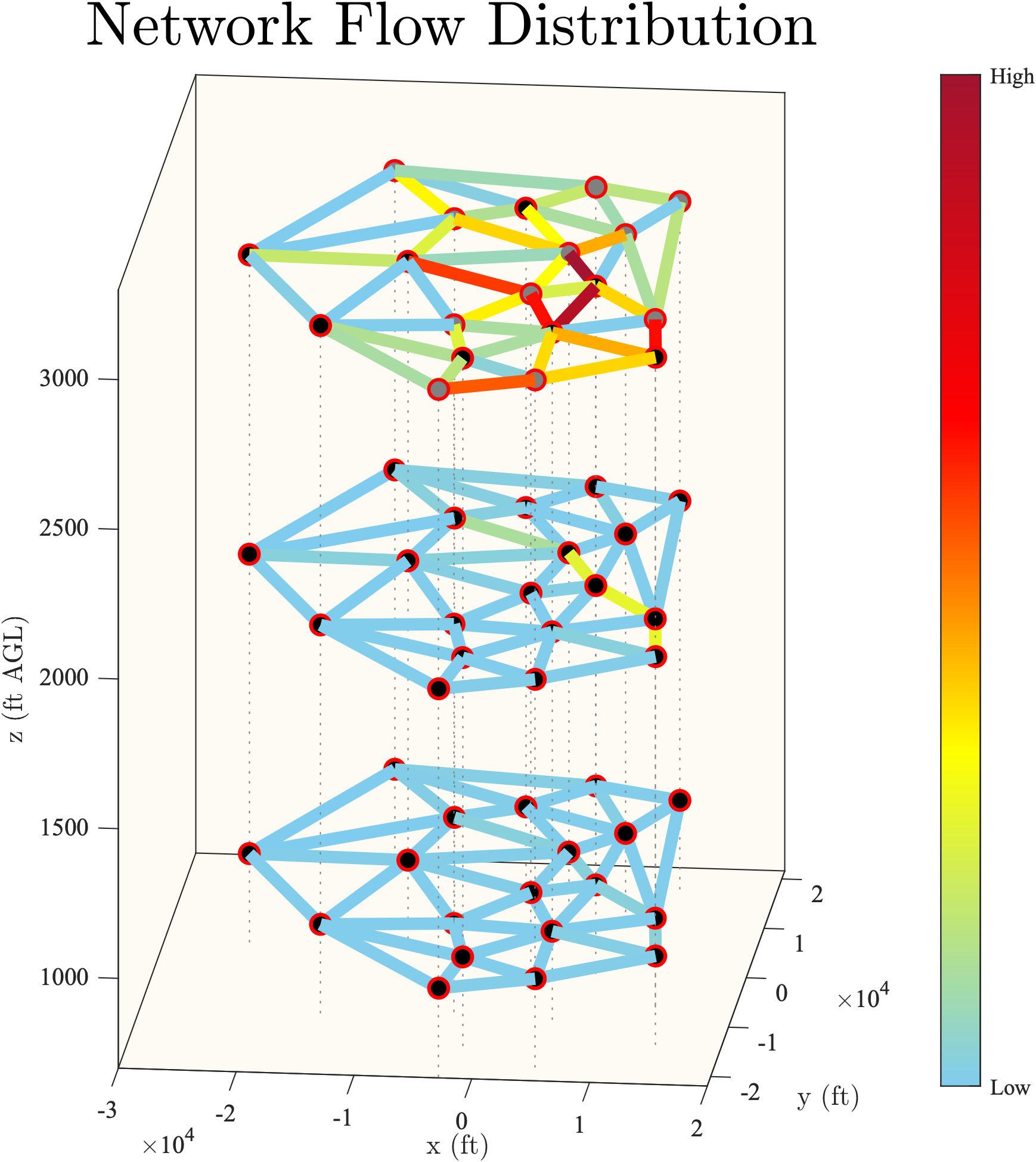}
         \hspace{0.3cm}
         \includegraphics[width=0.23\textwidth]{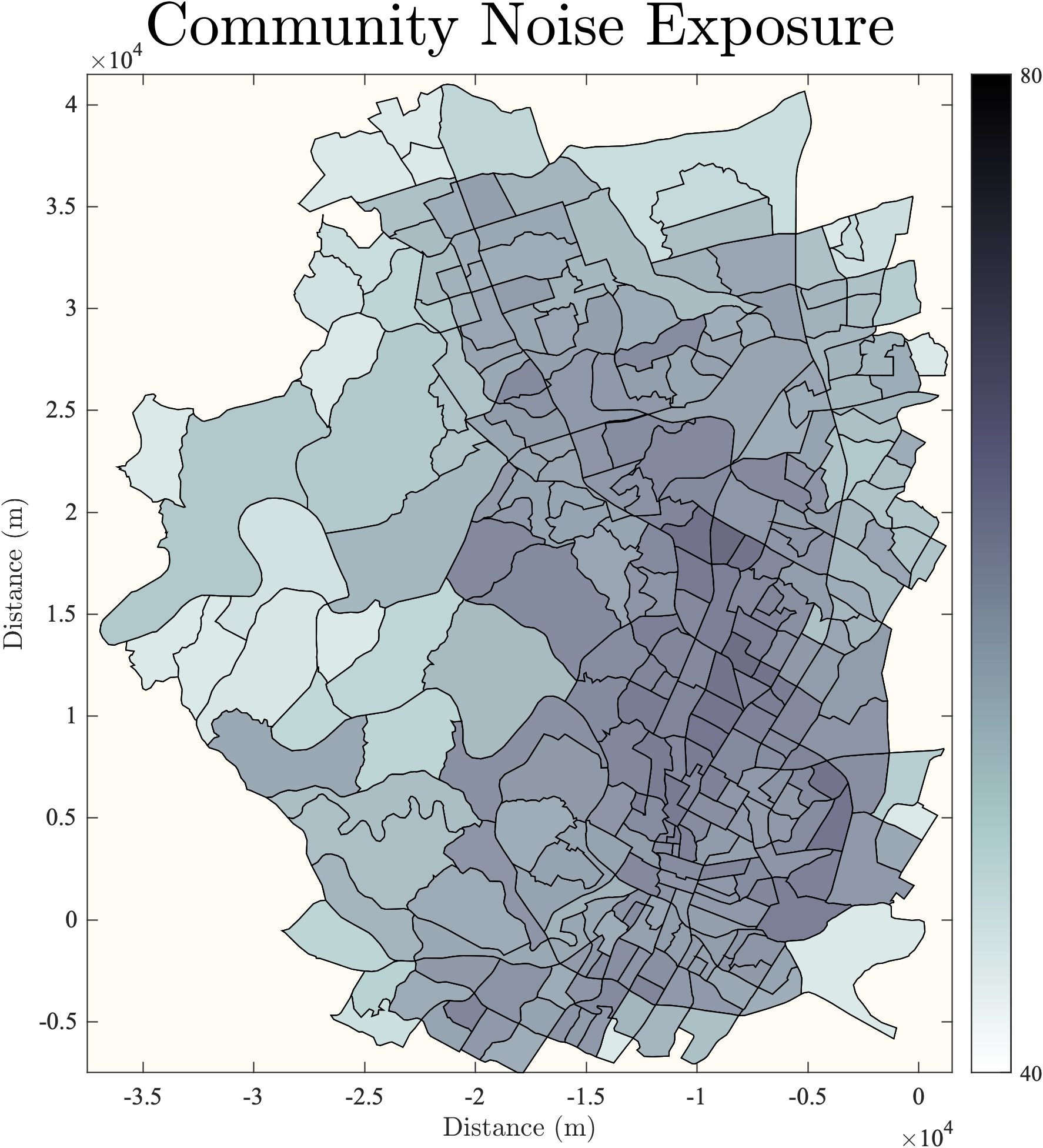}
         \hspace{0.3cm}
         \includegraphics[width=0.237\textwidth]{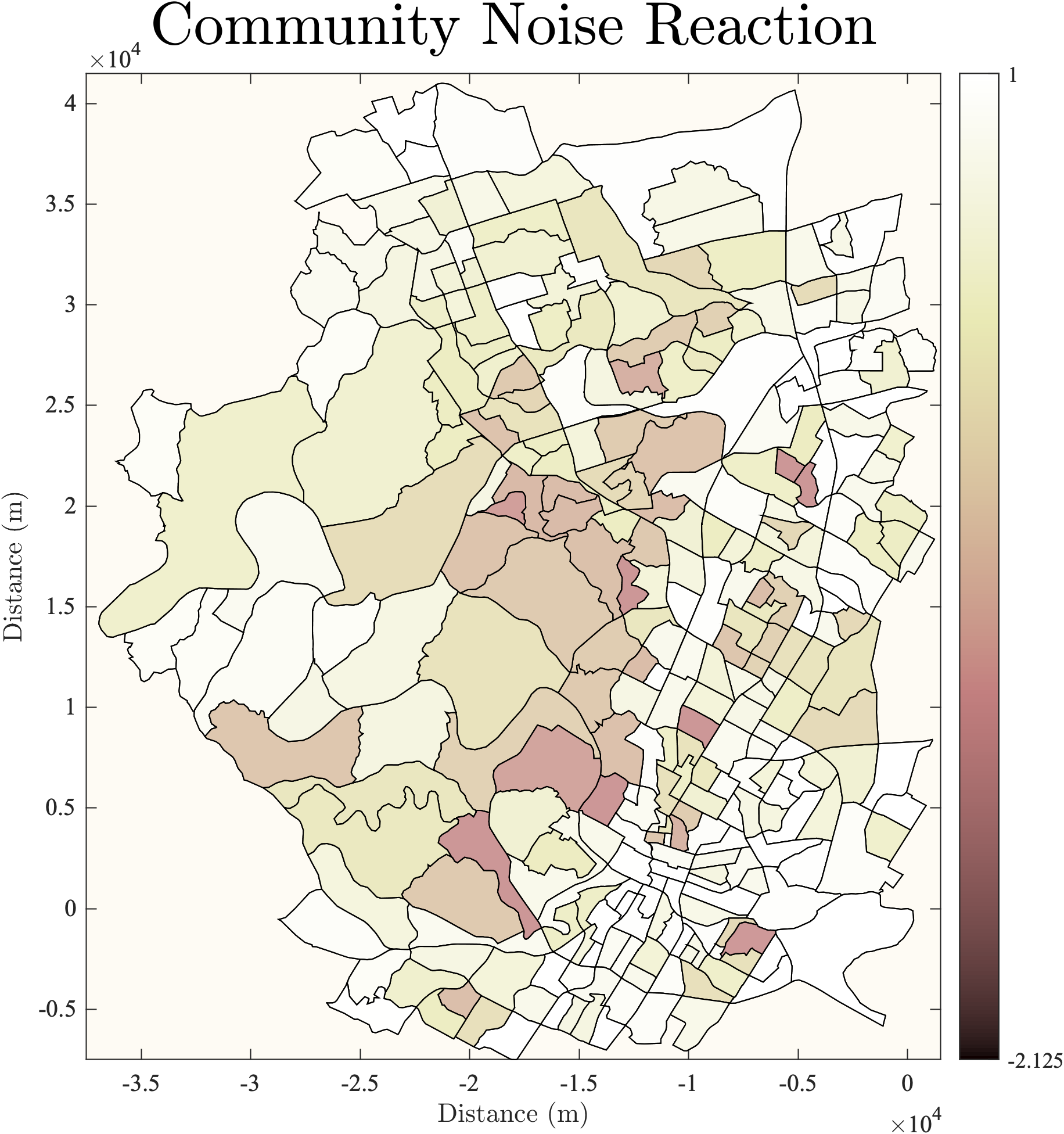}
         \hspace{0.3cm}
         \includegraphics[width=0.161\textwidth]{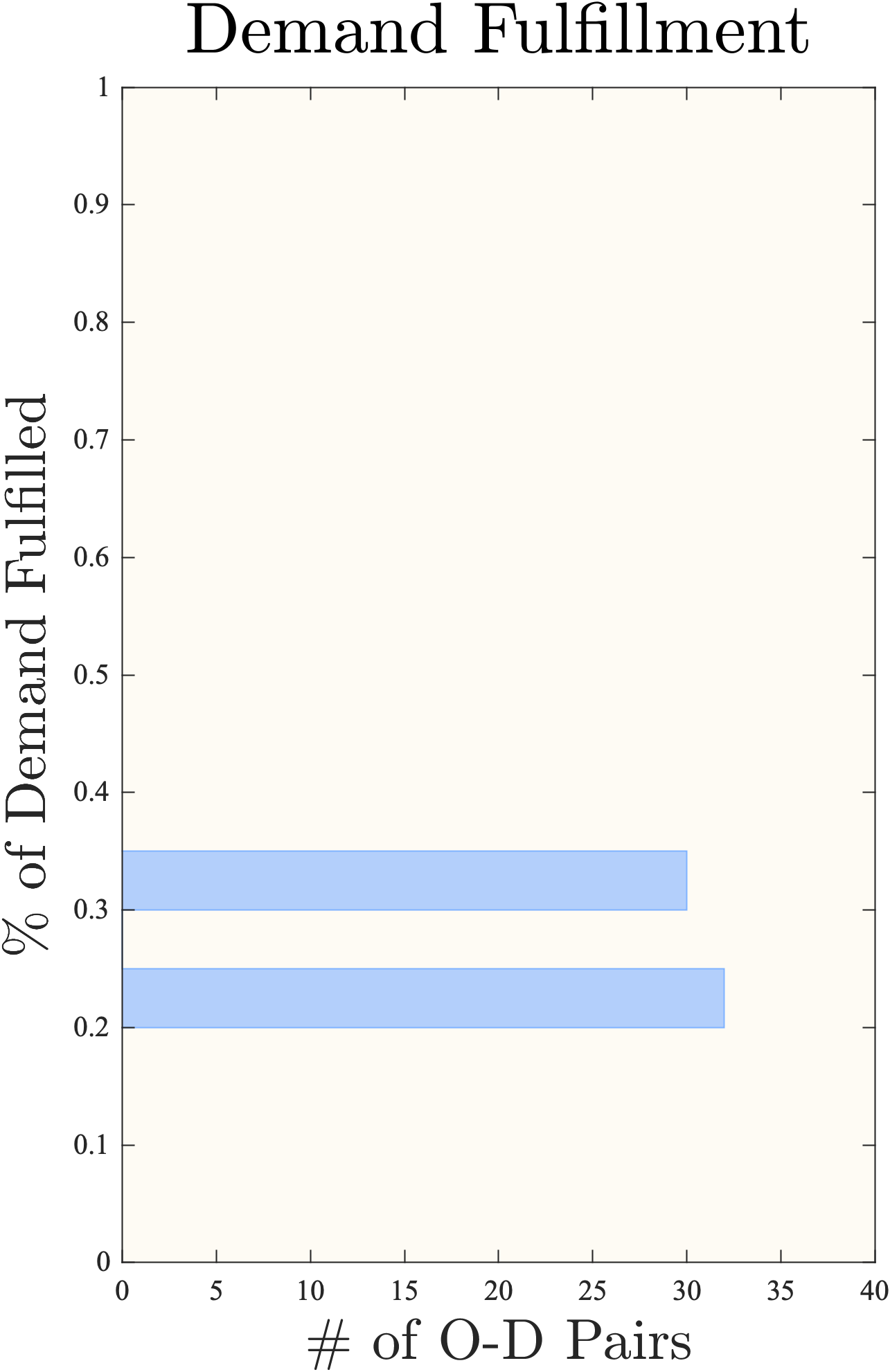}
         \caption{Minimizing noise impact under moderate demand fulfillment}
         \vspace{0.15cm}
         \label{fig:viz21}
     \end{subfigure}
     \begin{subfigure}[b]{\textwidth}
         \centering
         \includegraphics[width=0.226\textwidth]{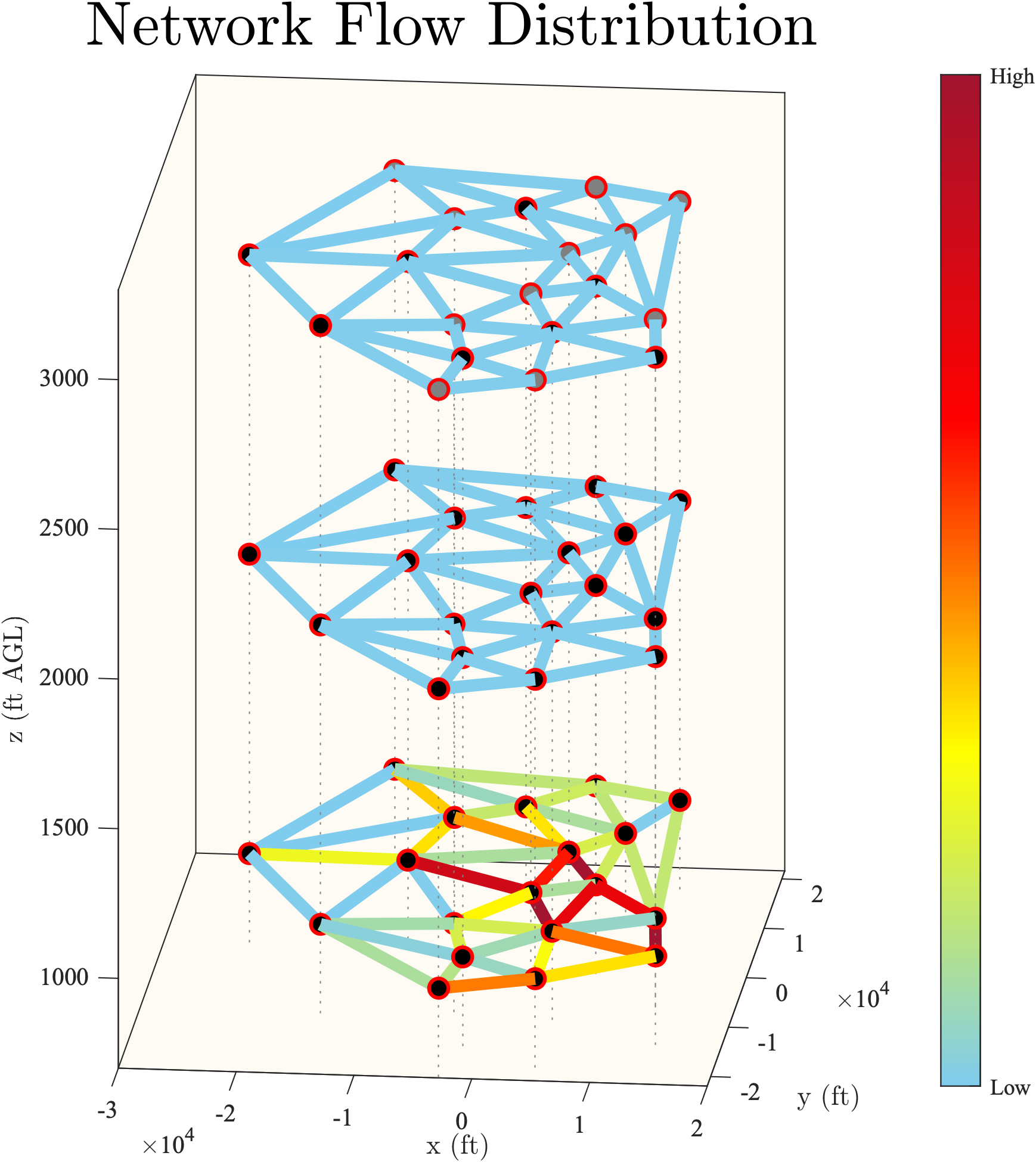}
         \hspace{0.3cm}
         \includegraphics[width=0.23\textwidth]{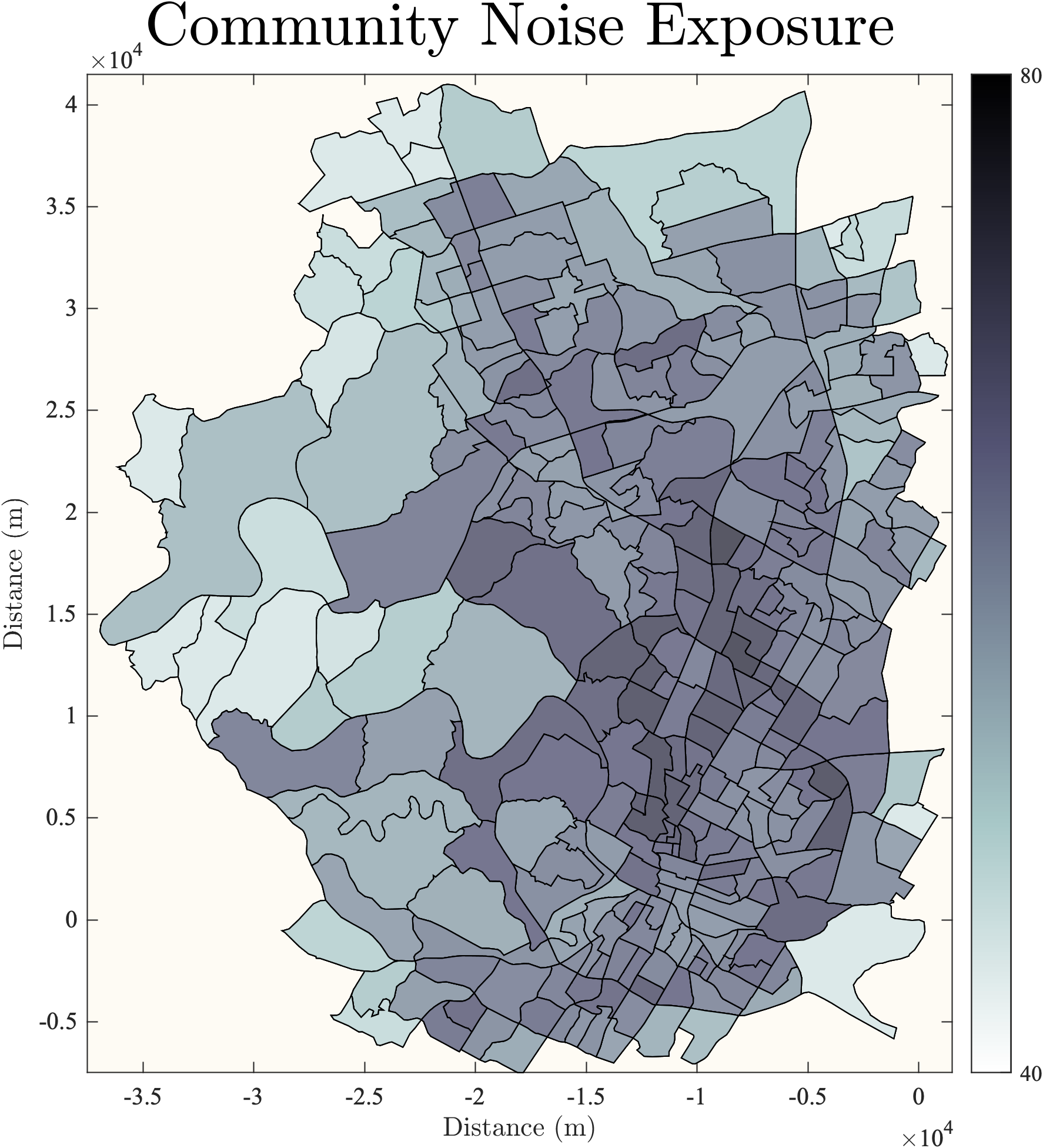}
         \hspace{0.3cm}
         \includegraphics[width=0.237\textwidth]{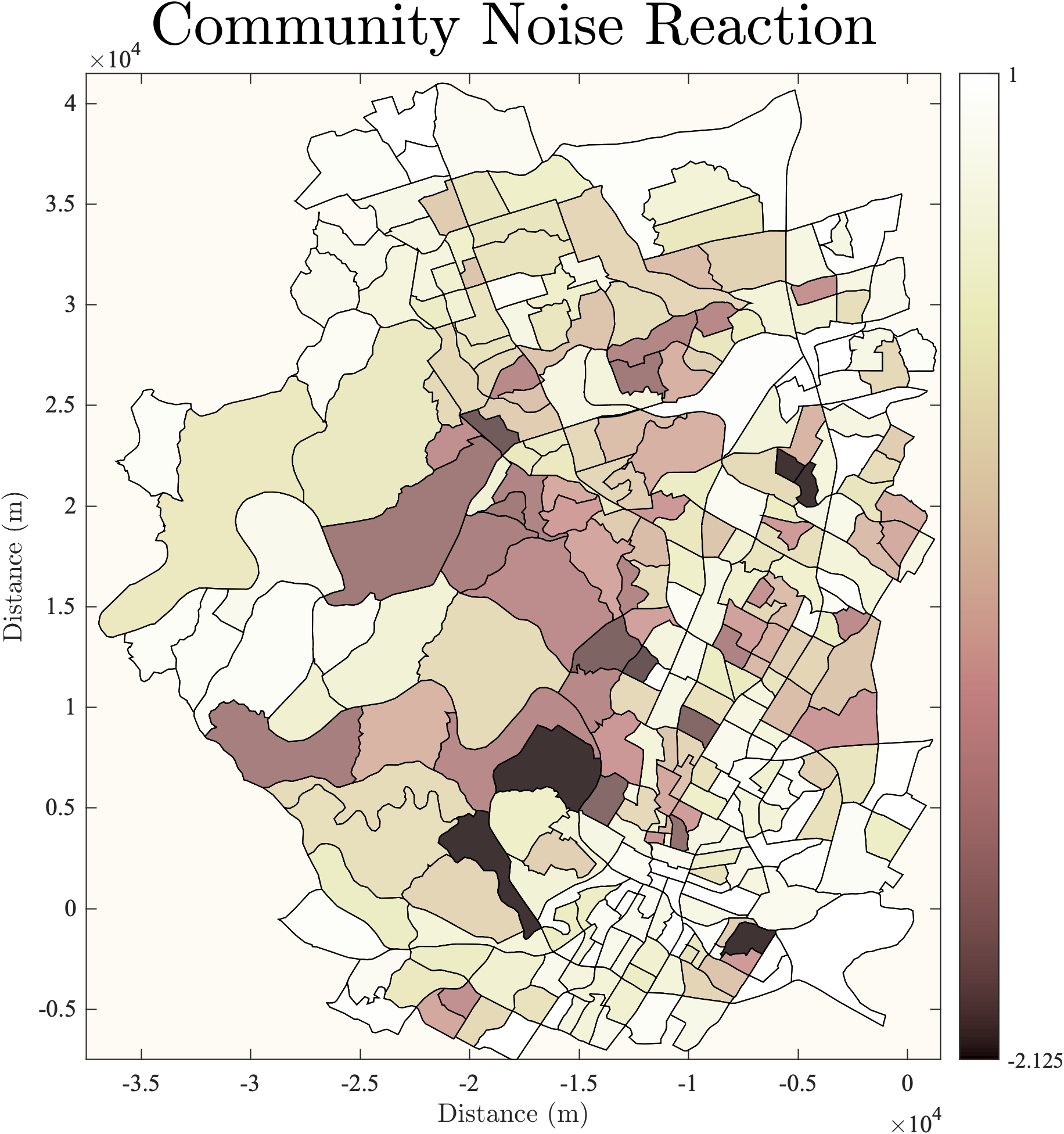}
         \hspace{0.3cm}
         \includegraphics[width=0.161\textwidth]{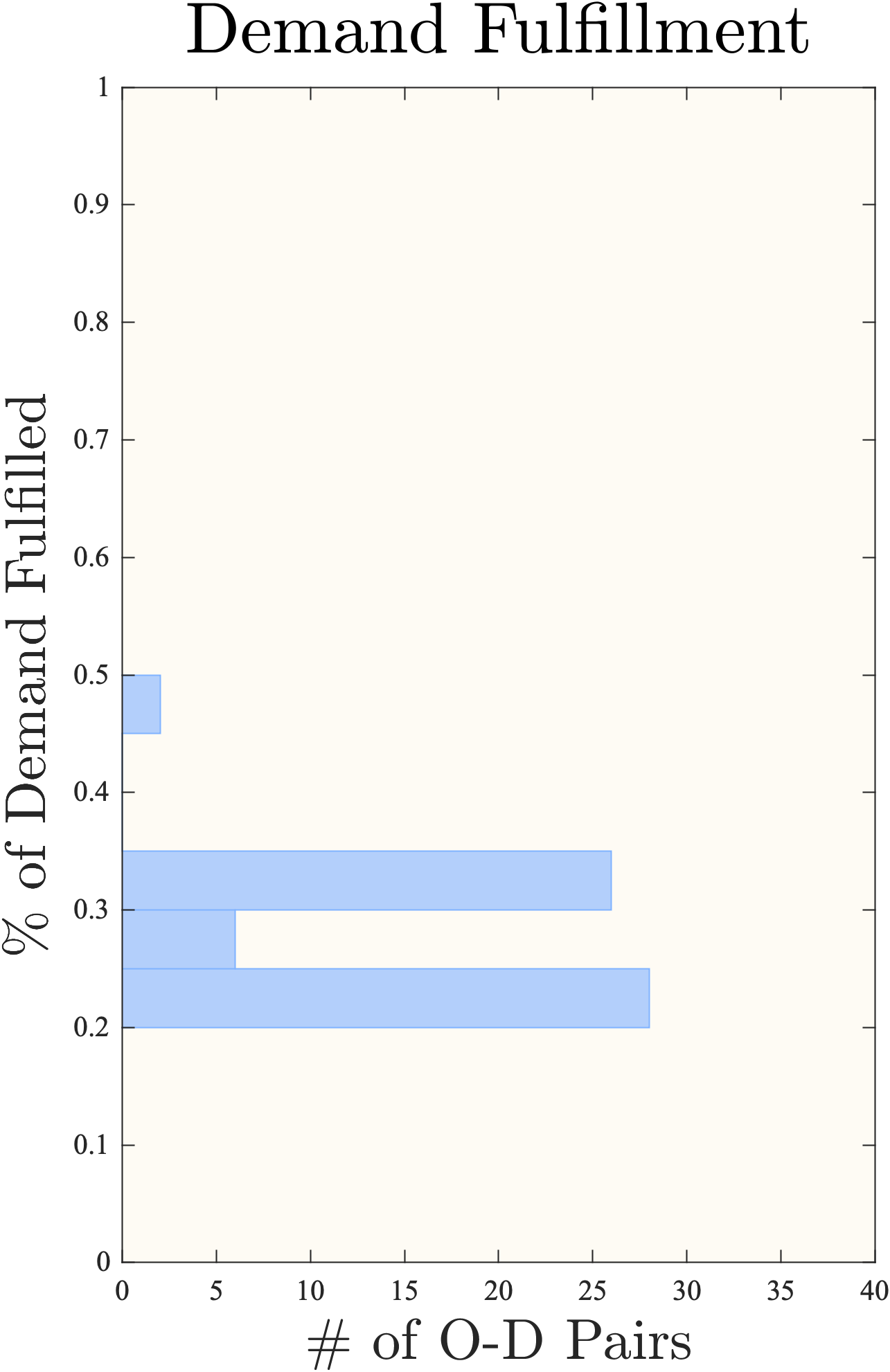}
         \caption{Minimizing energy consumption under moderate demand fulfillment}
         \vspace{0.15cm}
         \label{fig:viz22}
     \end{subfigure}
     \begin{subfigure}[b]{\textwidth}
         \centering
         \includegraphics[width=0.226\textwidth]{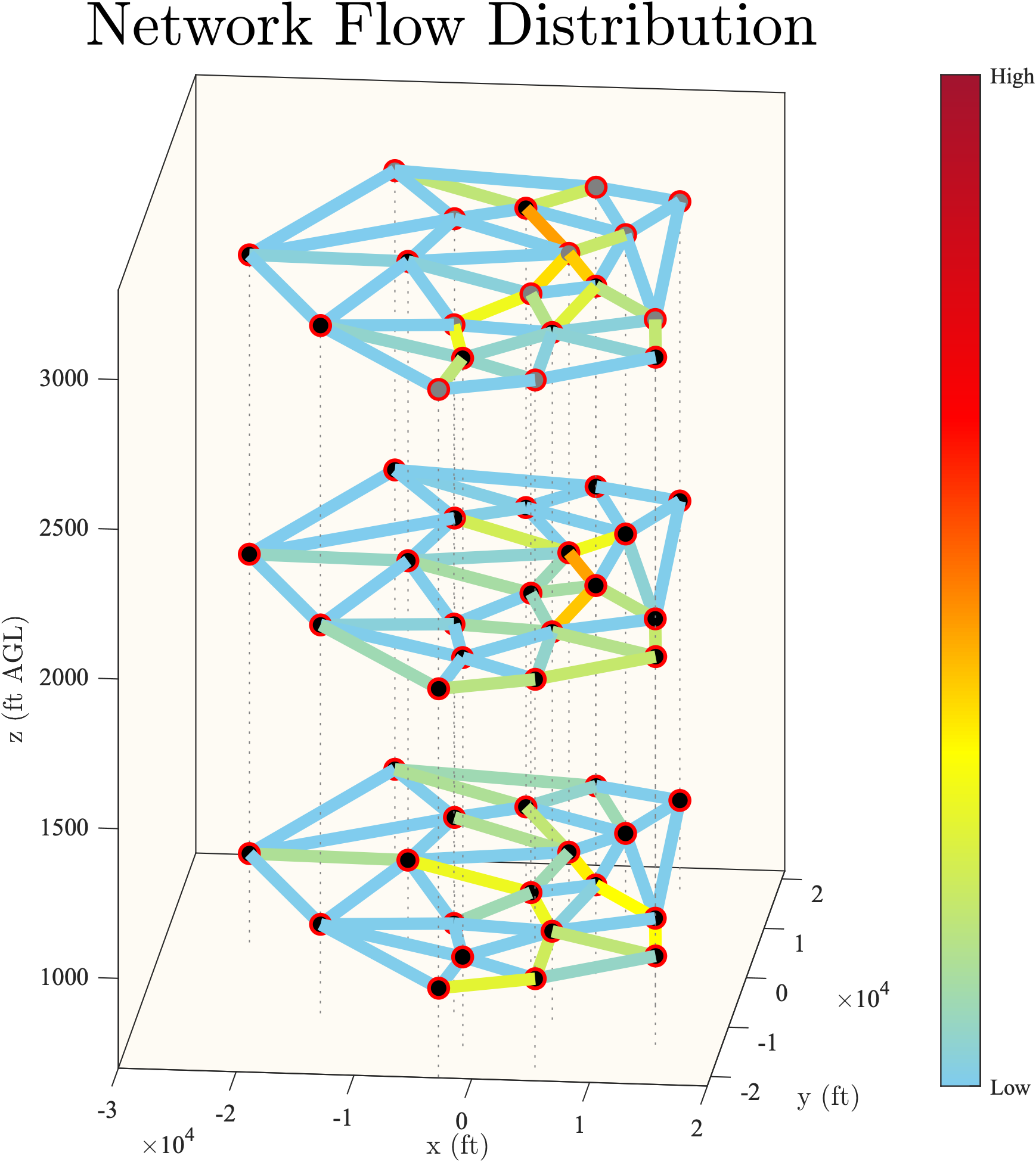}
         \hspace{0.3cm}
         \includegraphics[width=0.23\textwidth]{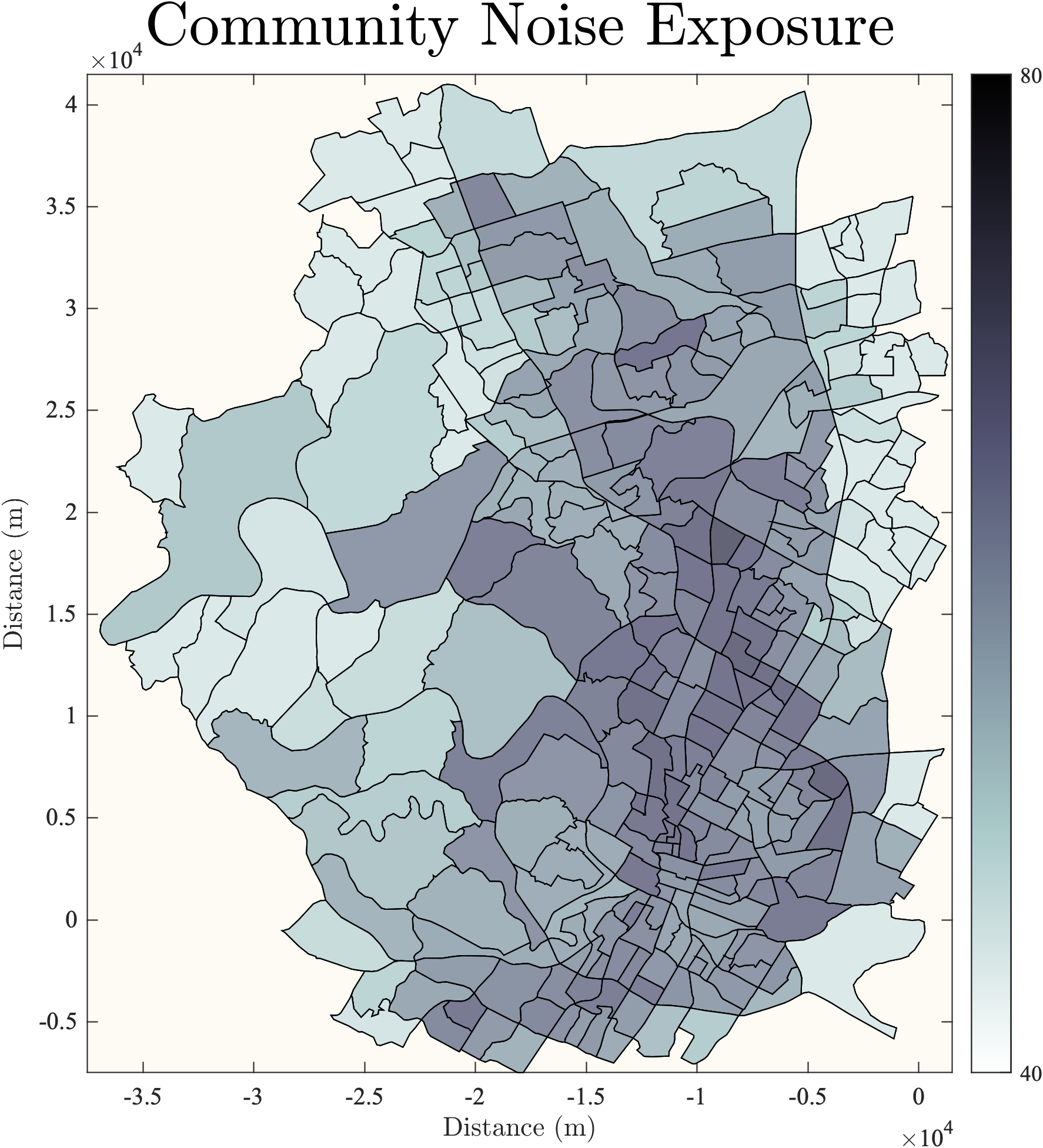}
         \hspace{0.3cm}
         \includegraphics[width=0.237\textwidth]{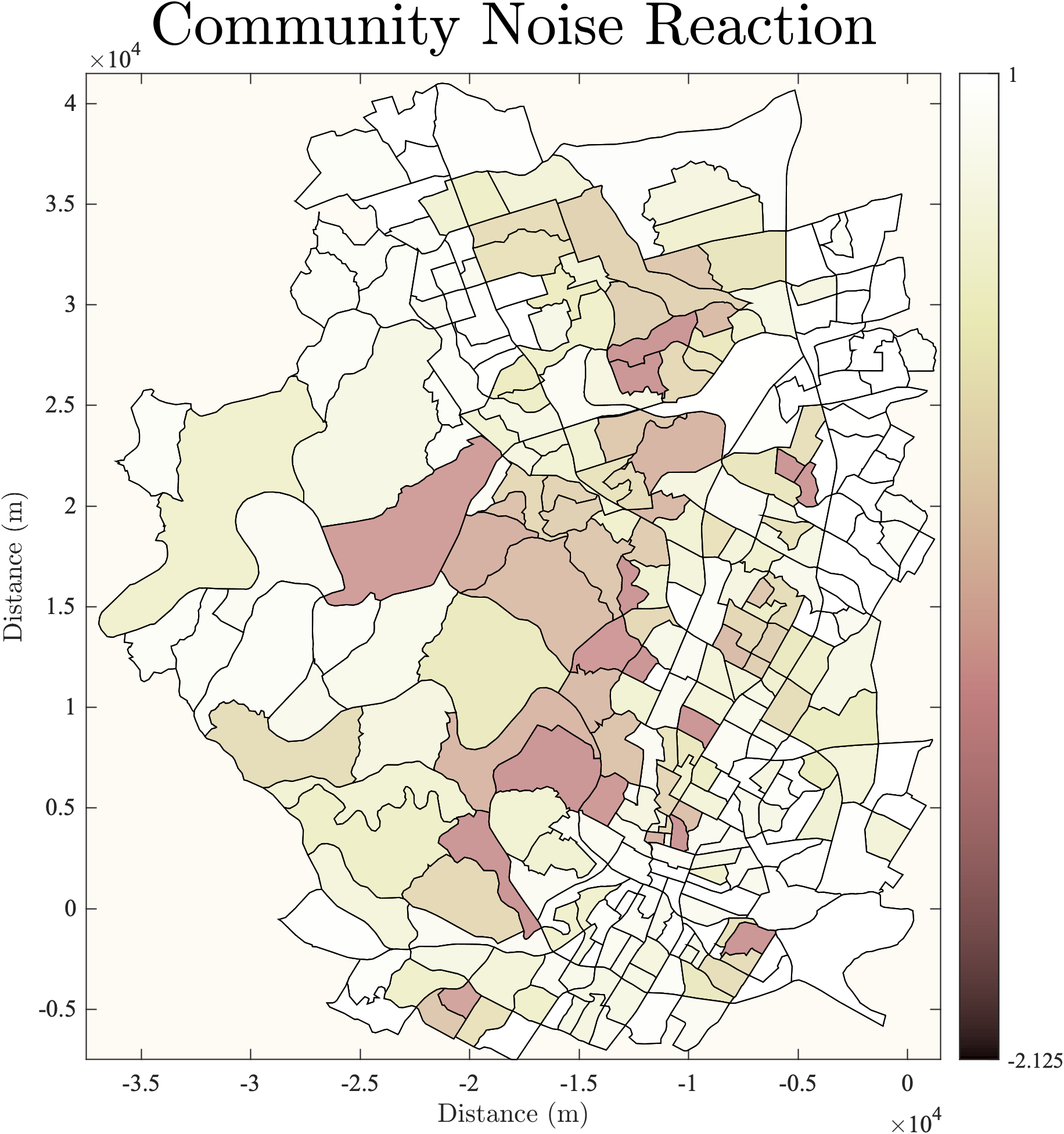}
         \hspace{0.3cm}
         \includegraphics[width=0.161\textwidth]{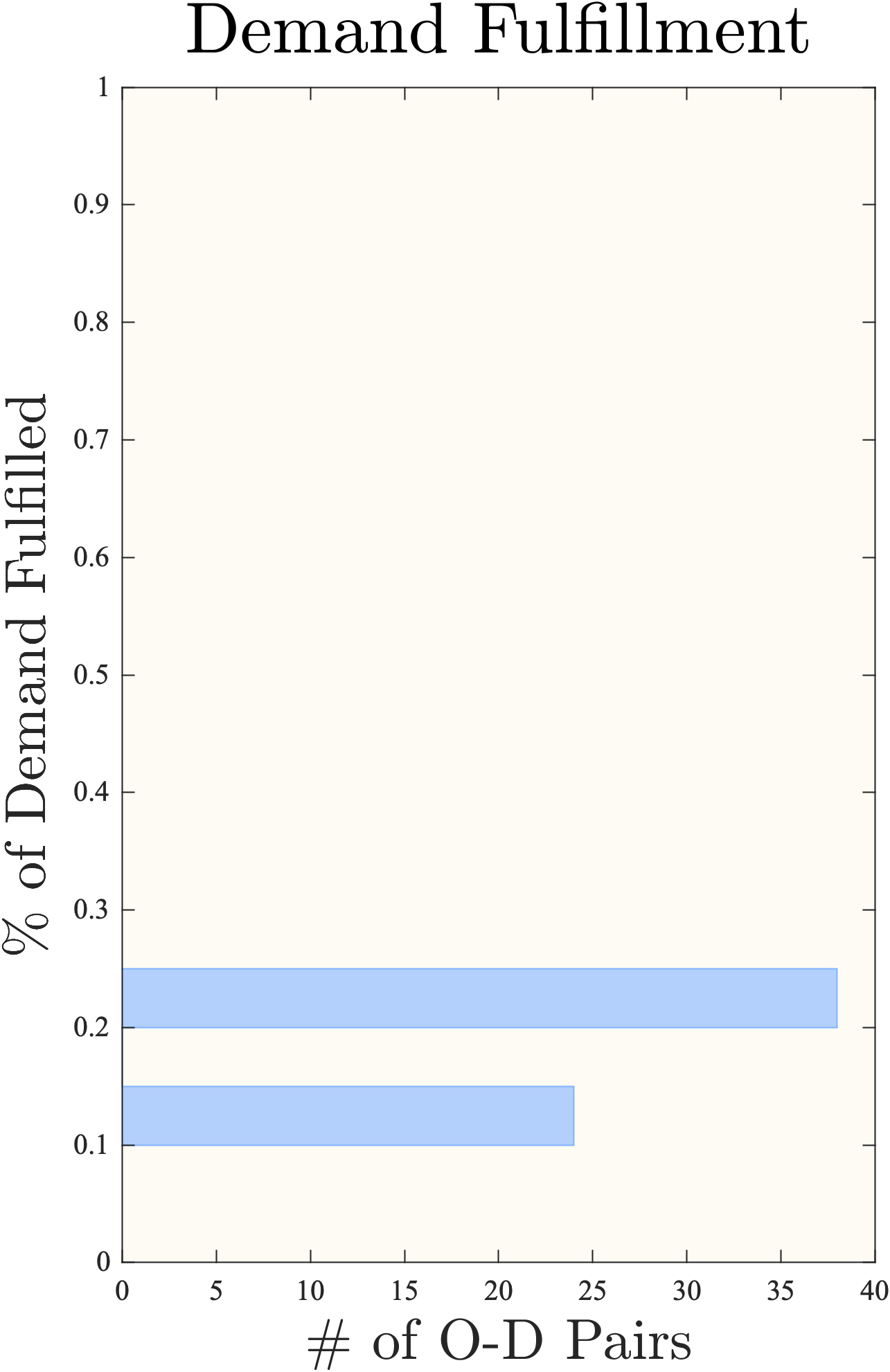}
         \caption{Considering both noise impact and energy consumption under moderate demand fulfillment}
         \vspace{0.15cm}
         \label{fig:viz23}
     \end{subfigure}
     \caption{Visualizations of UAM traffic flow distribution and noise impact under different trade-offs between noise impact and energy consumption}
     \label{fig:viz2}
\end{figure}

Figure~\ref{fig:viz2} displays a group of visualization results that highlights the trade-off between noise impact and energy consumption. Compared to results in Figure~\ref{fig:viz1}, we start to implement noise control such that the demand fulfillment outcomes in Figure~\ref{fig:viz21} to Figure~\ref{fig:viz23} are all at a moderate and constant level. Figure~\ref{fig:viz21} shows the result when noise control takes a significant role without constraint on energy consumption. Under this scenario, most air traffic flows go through the highest layer (altitude) of the network, resulting in excellent noise control effects yet high energy consumption due to climbing to the top layer. Figure~\ref{fig:viz22} shows a scenario contrary to Figure~\ref{fig:viz21}. In this scenario, we aim to provide a similar level of supply with minimal energy consumption. As a result, all air traffic flows remain in the bottom layer that is the closest to the ground surface. Unsurprisingly, this worsens both the overall noise impact and the community noise reactions. Figure~\ref{fig:viz23} presents a trade-off between Figure~\ref{fig:viz21} (noise control) and Figure~\ref{fig:viz22} (energy saving). In this scenario, we observe traffic flows in all three network layers and a noise impact outcome that is in between the two extreme cases in Figure~\ref{fig:viz21} and Figure~\ref{fig:viz22}. 

\begin{figure}[h!]
     \centering
     \begin{subfigure}[b]{\textwidth}
         \centering
         \includegraphics[width=0.226\textwidth]{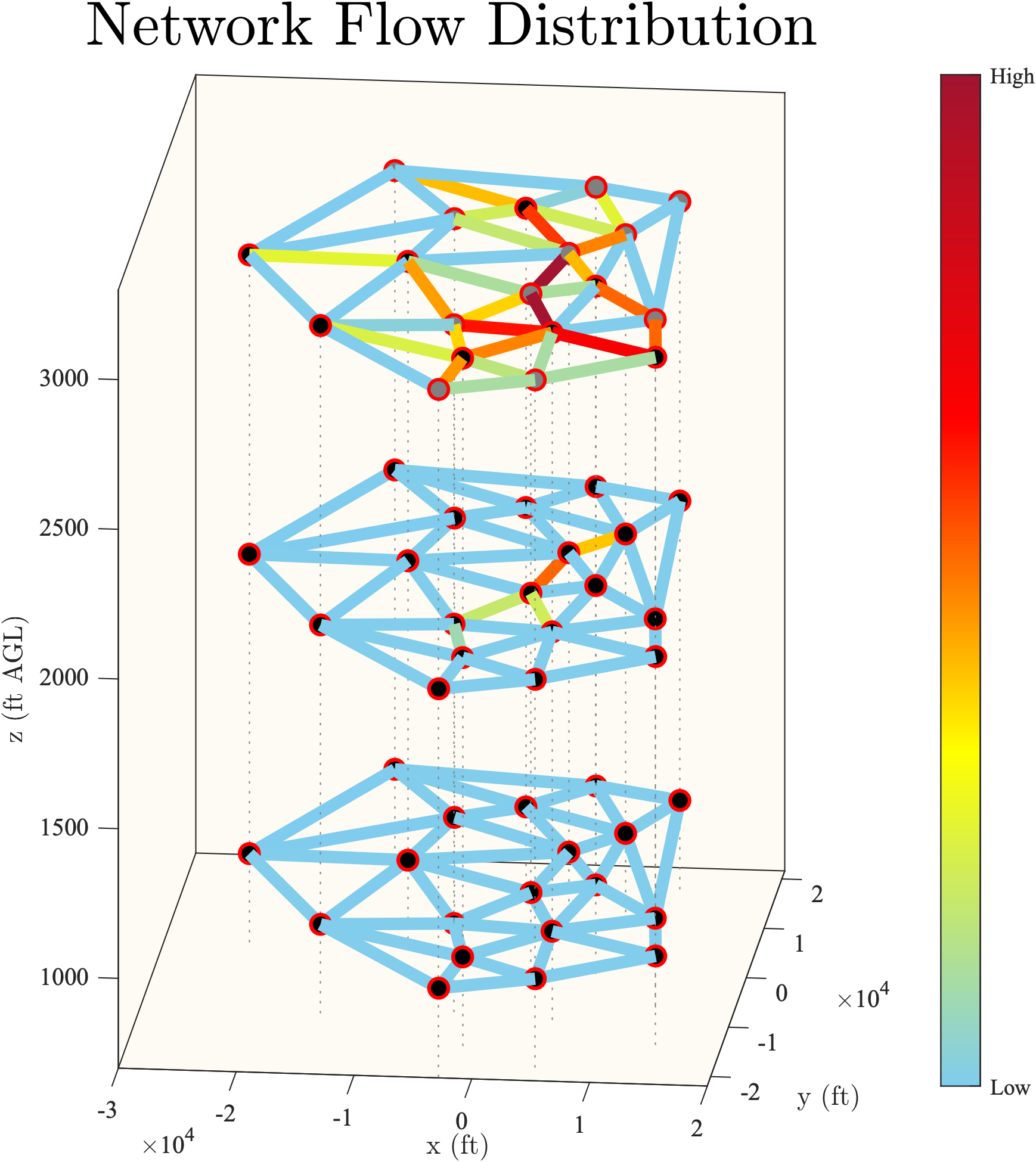}
         \hspace{0.3cm}
         \includegraphics[width=0.23\textwidth]{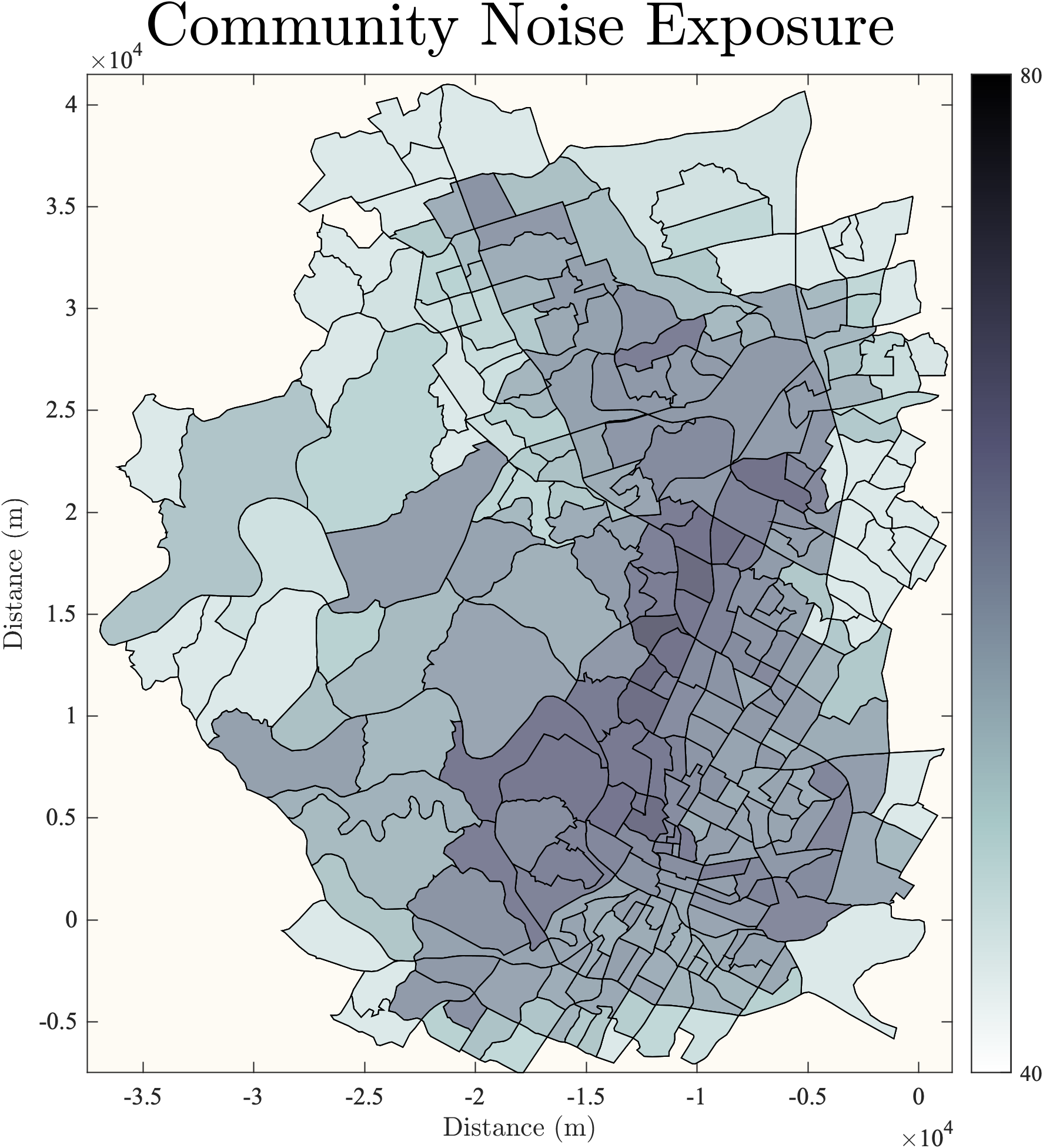}
         \hspace{0.3cm}
         \includegraphics[width=0.237\textwidth]{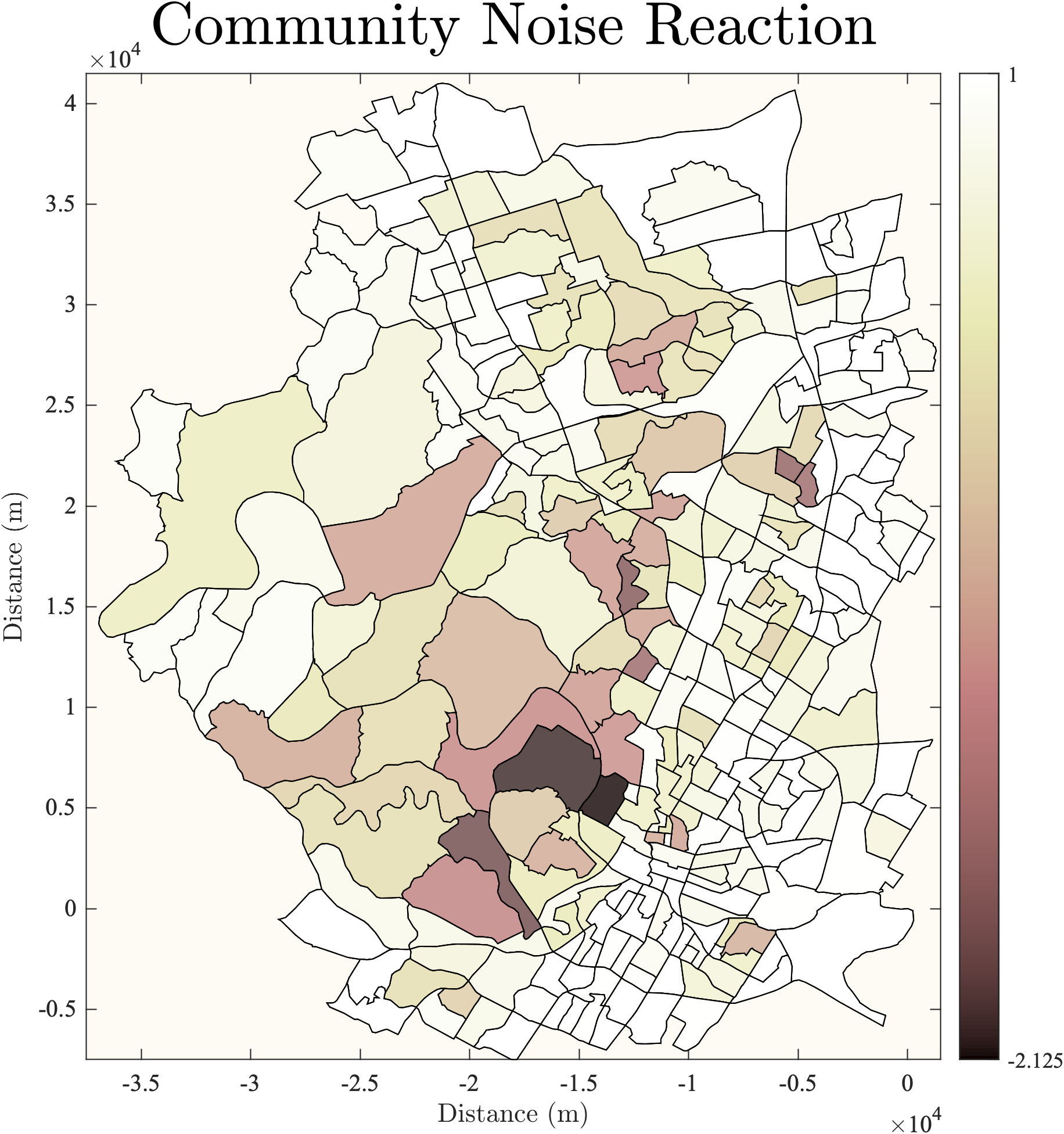}
         \hspace{0.3cm}
         \includegraphics[width=0.161\textwidth]{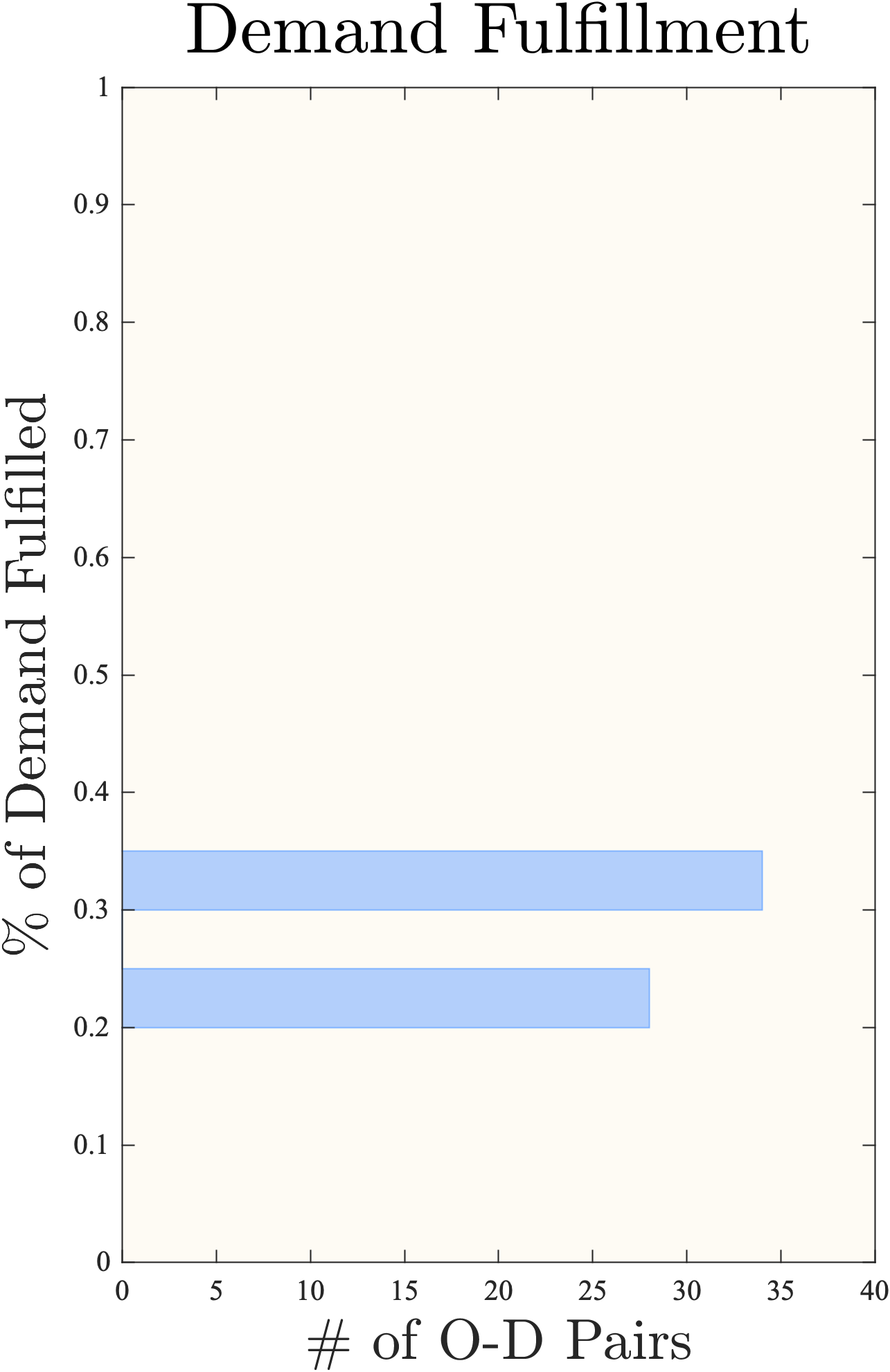}
         \caption{Controlling noise impact -- utilitarian}
         \vspace{0.15cm}
         \label{fig:viz31}
     \end{subfigure}
     \begin{subfigure}[b]{\textwidth}
         \centering
         \includegraphics[width=0.226\textwidth]{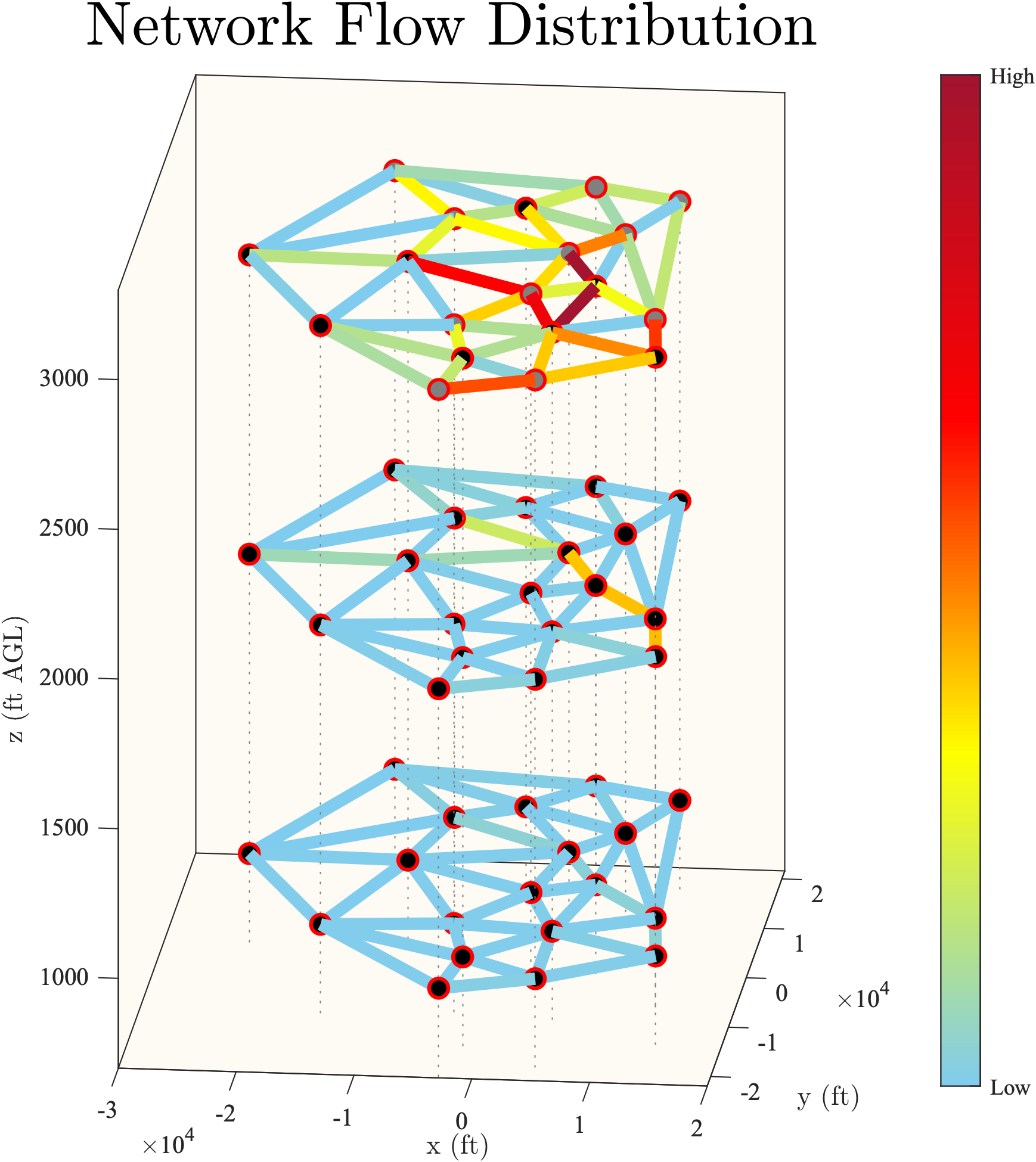}
         \hspace{0.3cm}
         \includegraphics[width=0.23\textwidth]{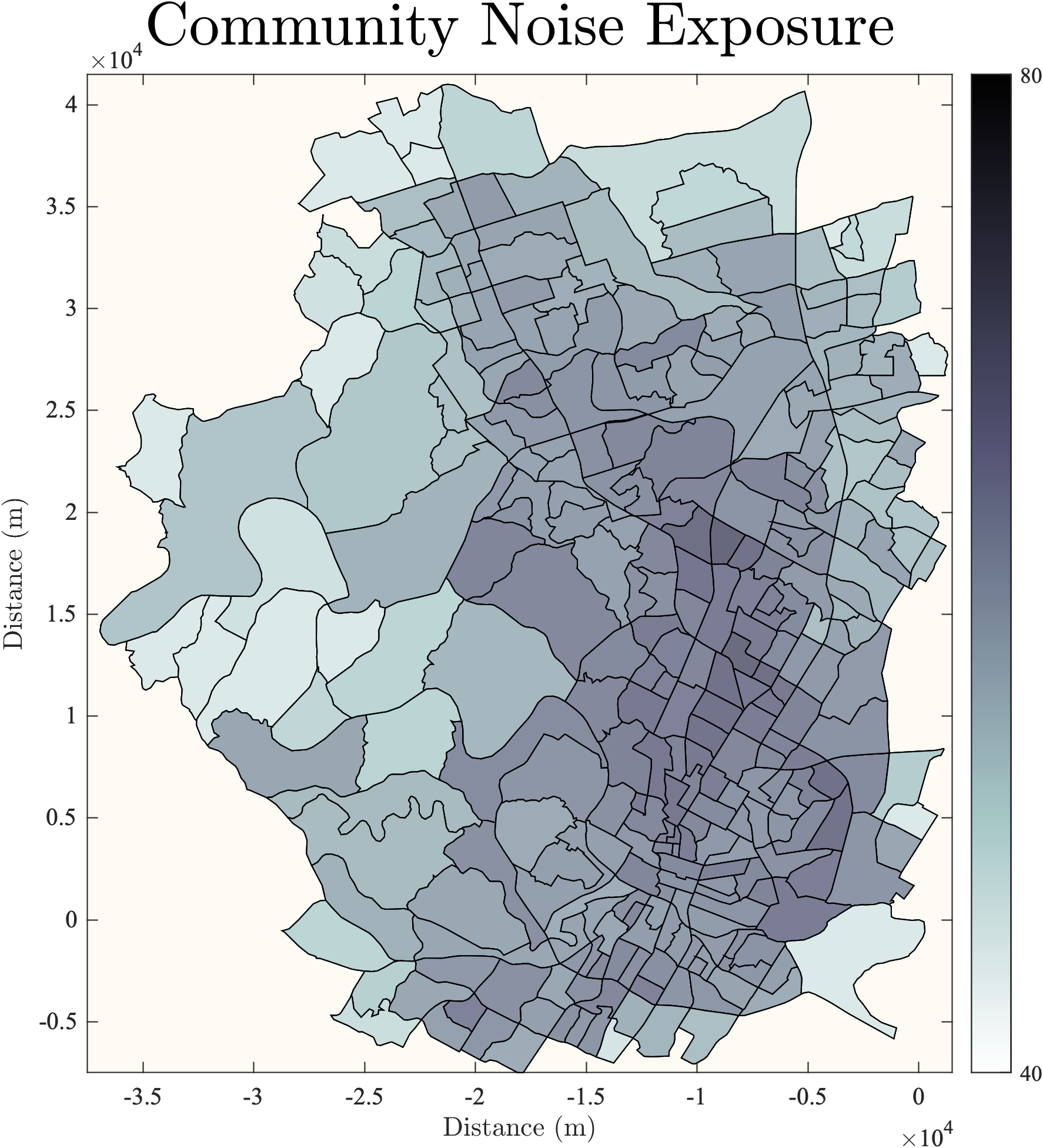}
         \hspace{0.3cm}
         \includegraphics[width=0.237\textwidth]{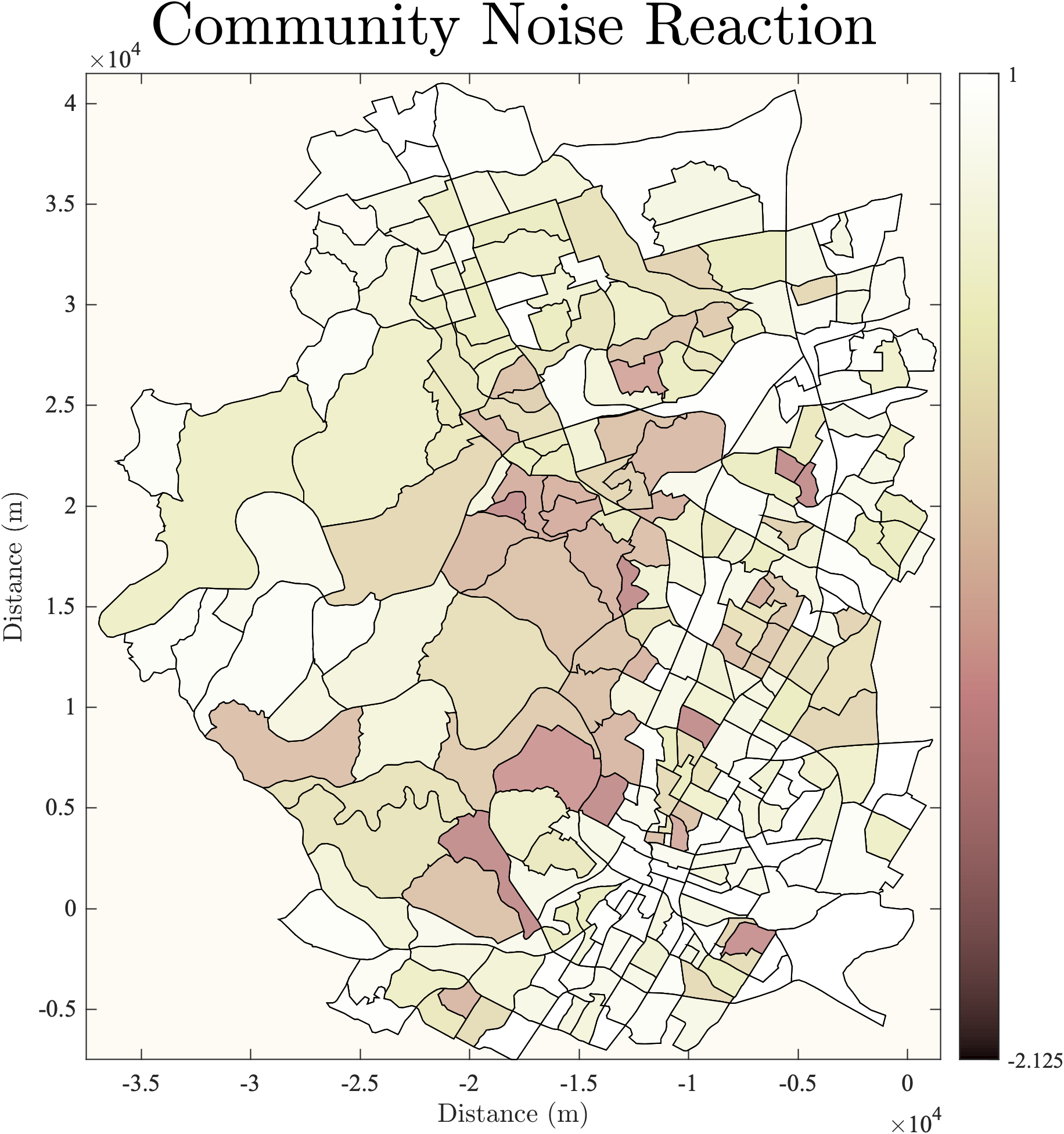}
         \hspace{0.3cm}
         \includegraphics[width=0.161\textwidth]{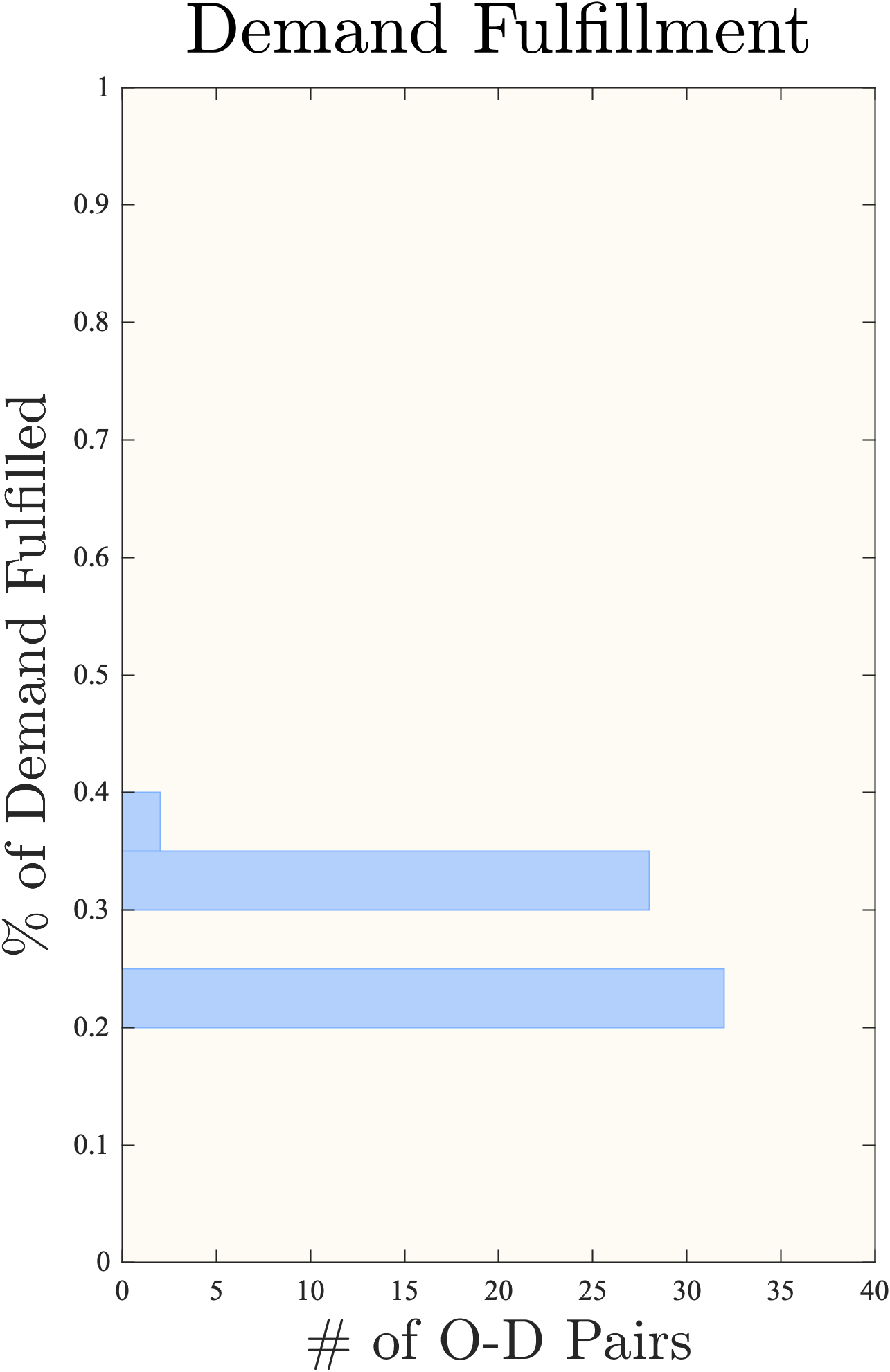}
         \caption{Controlling noise impact -- egalitarian}
         \vspace{0.15cm}
         \label{fig:viz32}
     \end{subfigure}
     \caption{Visualizations of UAM traffic flow distribution and noise impact under different noise impact fairness levels}
     \label{fig:viz3}
\end{figure}

Figure~\ref{fig:viz3} displays the final group of visualization results centered around noise fairness. Under approximately the same demand fulfillment level, Figure~\ref{fig:viz31} shows the result when a utilitarian criterion is applied to noise impact. Through the community noise reaction map, we observe that for some communities at the center of the Austin area, the noise annoyance levels are extremely high. This indicates that the current noise management scheme is unfair to those communities. On the contrary, when we apply an egalitarian criterion on community noise impact, we obtain a much more homogeneous community noise reaction result in Figure~\ref{fig:viz32}. This noise fairness concept is one of the core concepts in this study. Eventually, our objective is to achieve fairness in both demand fulfillment (across 62 O-D pairs) and community noise reaction (across 292 communities).

\subsection{Design Trade-offs}

Most sustainable system design tasks have multiple, and often conflicting, objectives. Therefore, the design trade-off is the kernel of such topics. In this study, we explore trade-offs of two distinct types. First, we conduct efficiency-fairness trade-offs for design aspects that have multiple parties/entities, including demand fulfillment and community noise reaction. Second, we conduct a trade-off between the three optimization objectives: demand fulfillment, noise control, and energy saving.

\begin{figure}[h!]
    \centering
    \includegraphics[width=0.42\textwidth]{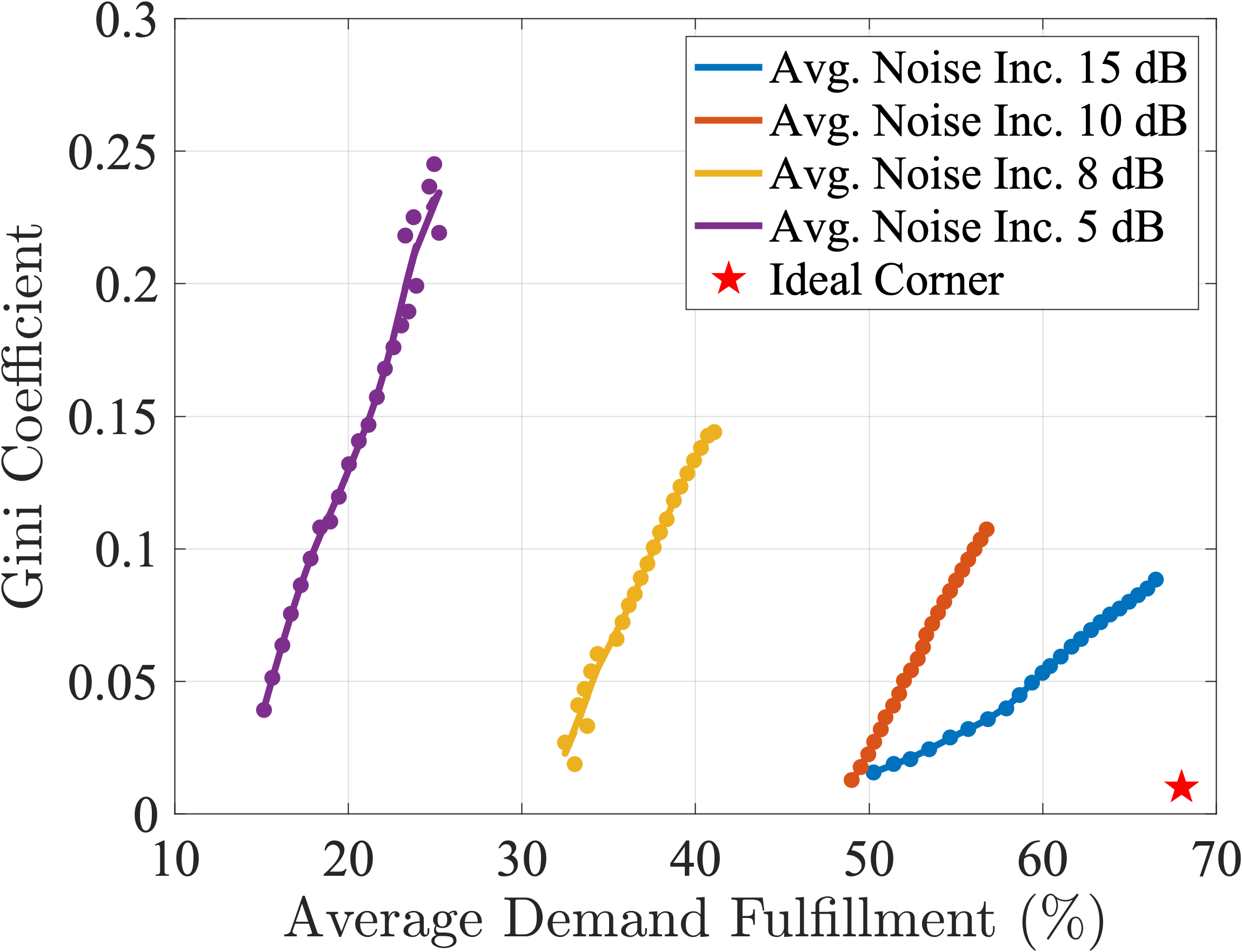}
    \hspace{1cm}
    \includegraphics[width=0.41\textwidth]{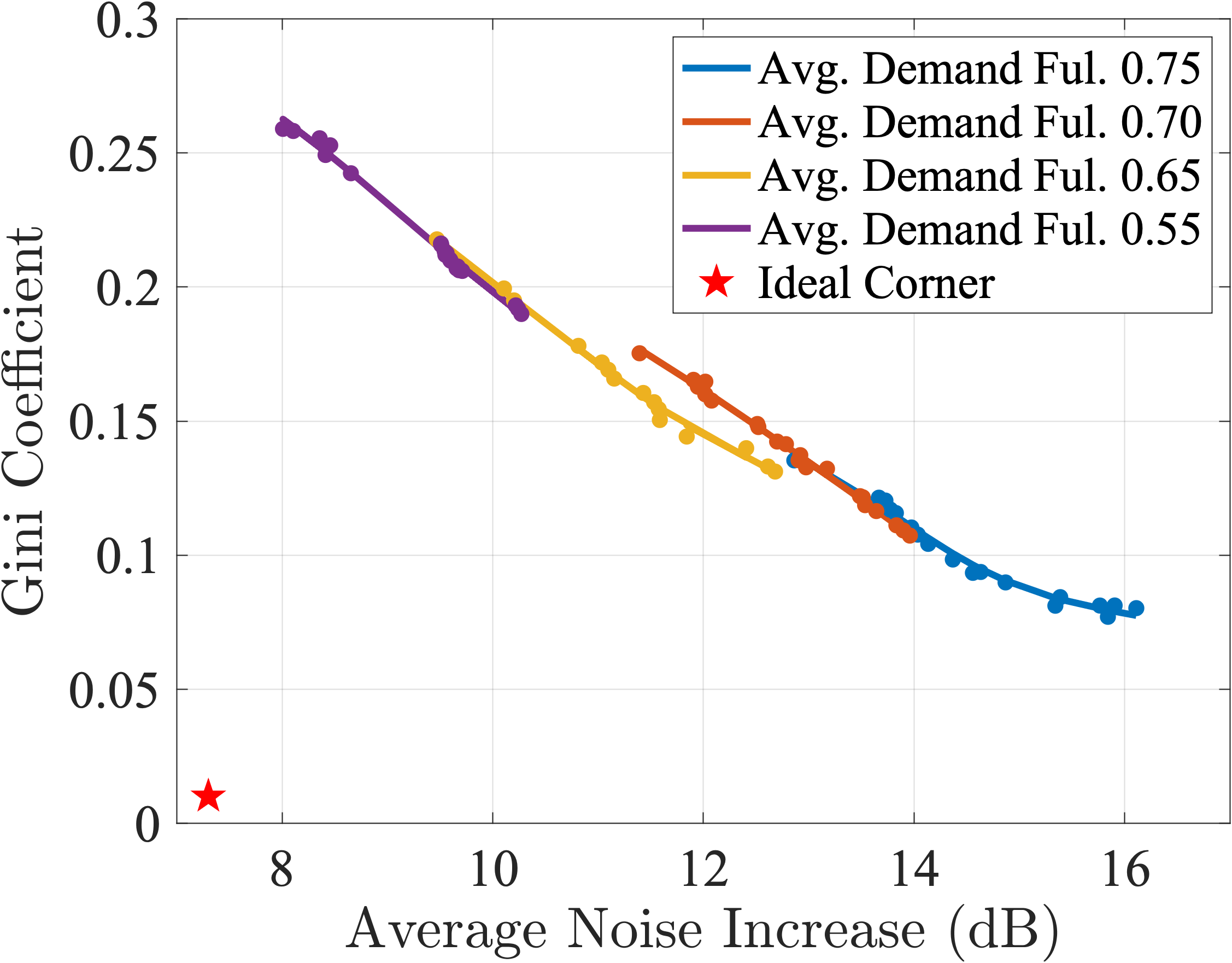}
    \caption{The efficiency-fairness trade-off for demand fulfillment (left) and noise increase (right)}
    \label{fig:eftradeoff}
\end{figure}

We first present the efficiency-fairness trade-off results for demand fulfillment and noise impact. The fundamental principle of the efficiency-fairness trade-off is the inability to concurrently achieve both system efficiency (represented by total or average benefits) and fairness (represented by equality or low variability). We employ Gini coefficient to measure inequality, where a lower Gini coefficient is indicative of enhanced equality or fairness. The left plot of Figure~\ref{fig:eftradeoff} shows the efficiency-fairness trade-off result for demand fulfillment. In the ideal case, the goal is to achieve a high average demand fulfillment along with a low Gini coefficient. This `Ideal Corner' is indicated with a red star on the plot. The left plot of Figure~\ref{fig:eftradeoff} displays the efficiency-fairness trade-off curves for demand fulfillment under four average noise increase levels. We first observe that overall, the design points get closer to the ideal point at the cost of higher average noise impacts. For every level of average noise increase, one cannot achieve a lower Gini coefficient without compromising on average demand fulfillment. Similarly, the right plot of Figure~\ref{fig:eftradeoff} shows the efficiency-fairness trade-off result for community noise increase. Aside from the unattainability of the ideal corner without concessions, an interesting observation is that the four trade-off curves, corresponding to different demand fulfillment levels, fall on a single line which captures the pronounced correlation between noise impact and demand fulfillment. Overall, the efficiency-fairness trade-off results in Figure~\ref{fig:eftradeoff} are logical and provide insights on the price of fairness in this specific problem. 

\begin{figure}[h!]
     \centering
     \begin{subfigure}[b]{\textwidth}
         \centering
         \includegraphics[width=0.4\textwidth]{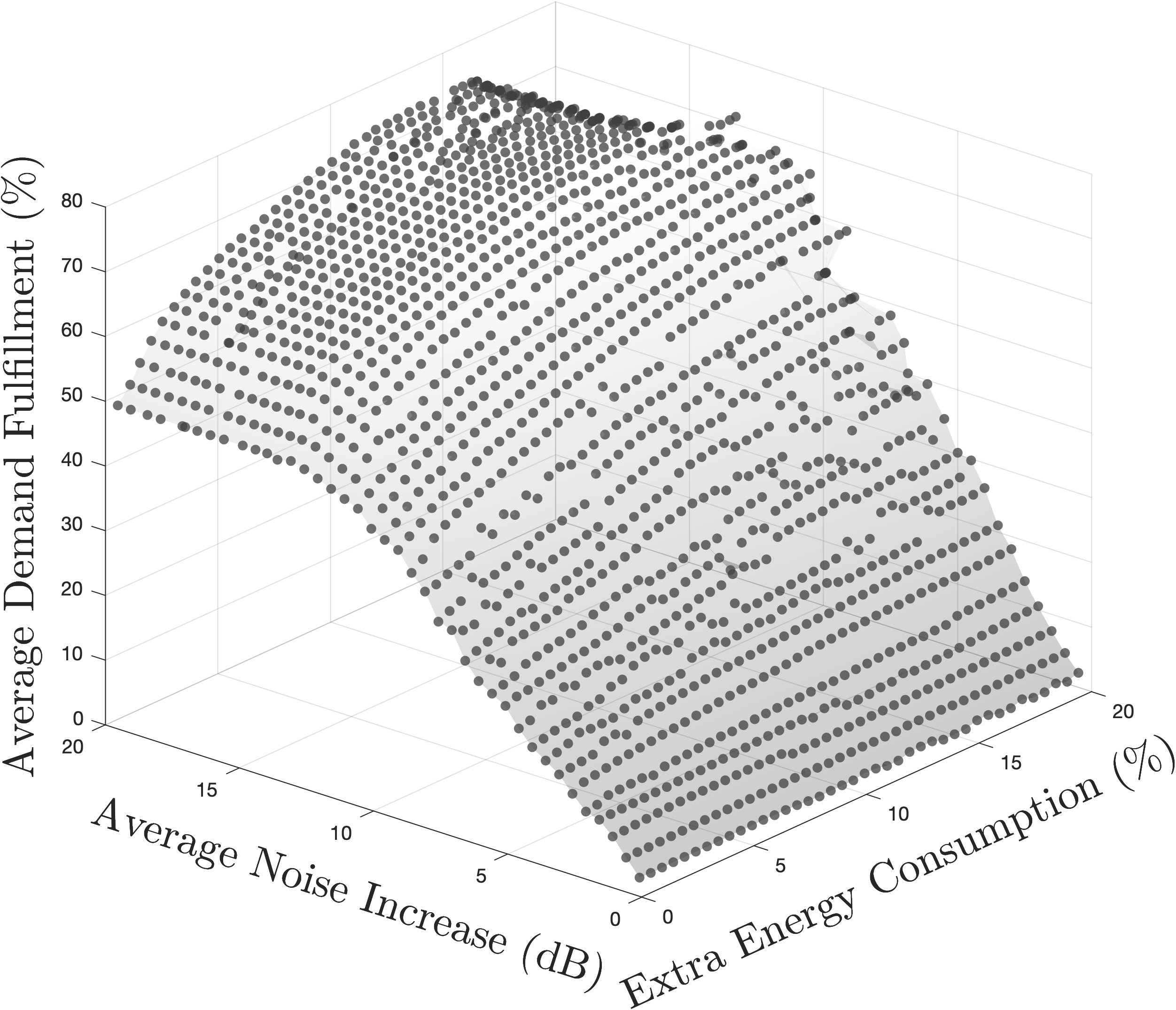}
         \hspace{1cm}
         \includegraphics[width=0.36\textwidth]{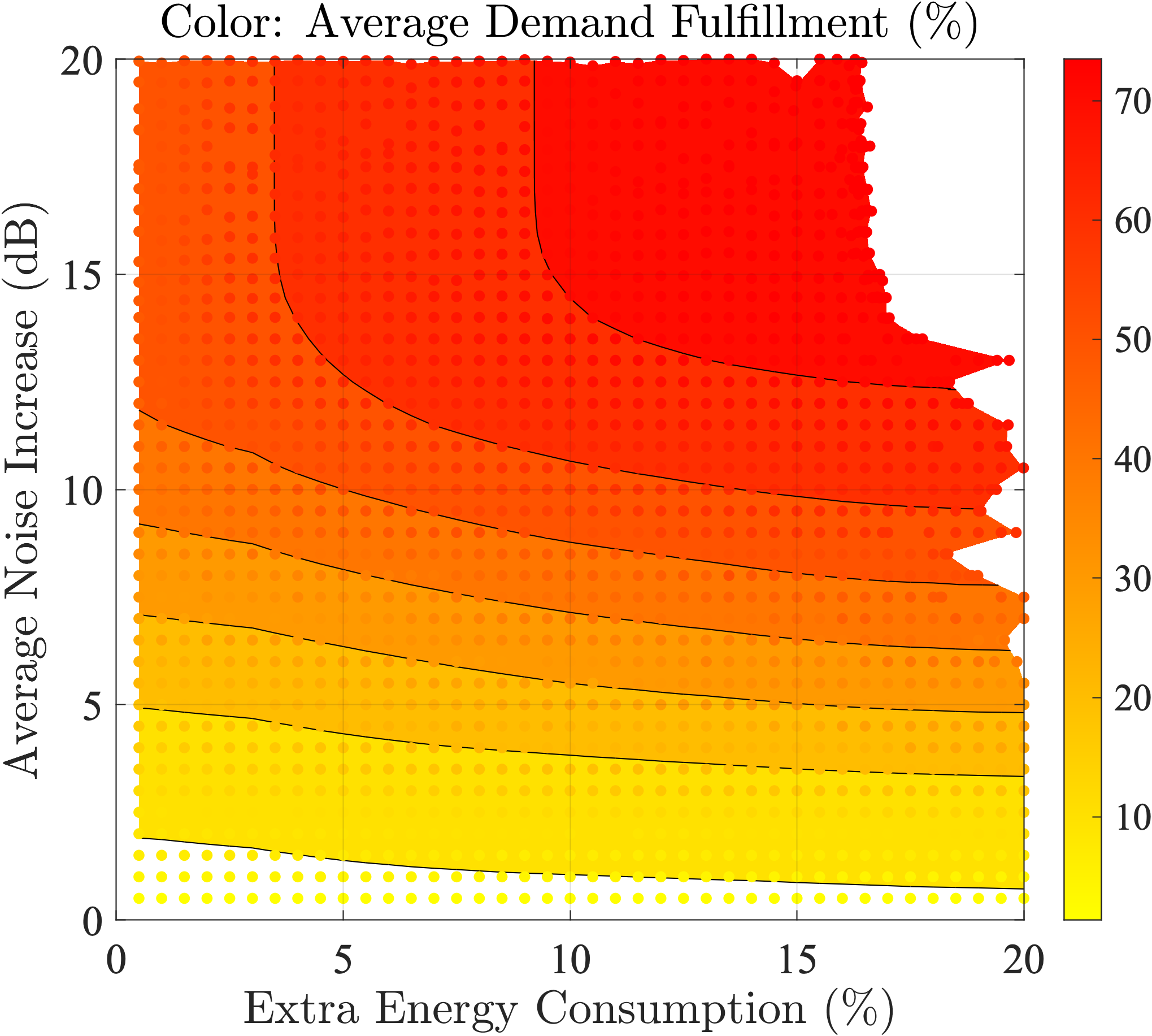}
         \caption{The three-dimensional trade-off between demand fulfillment, noise control, and energy saving}
         \vspace{0.15cm}
         \label{fig:3dtradeoffs}
     \end{subfigure}
     \begin{subfigure}[b]{\textwidth}
         \centering
         \includegraphics[width=0.4\textwidth]{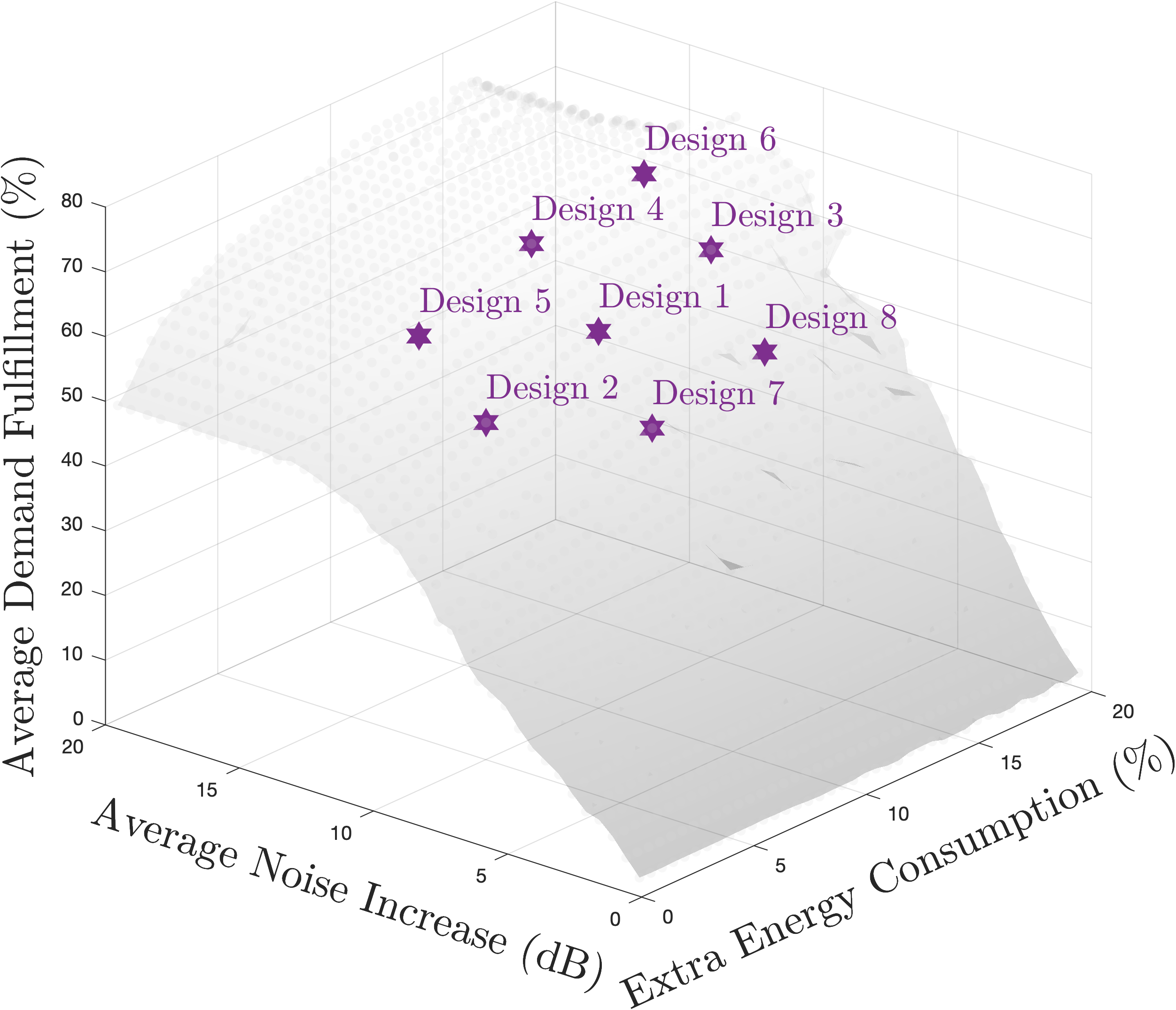}
         \hspace{1cm}
         \includegraphics[width=0.36\textwidth]{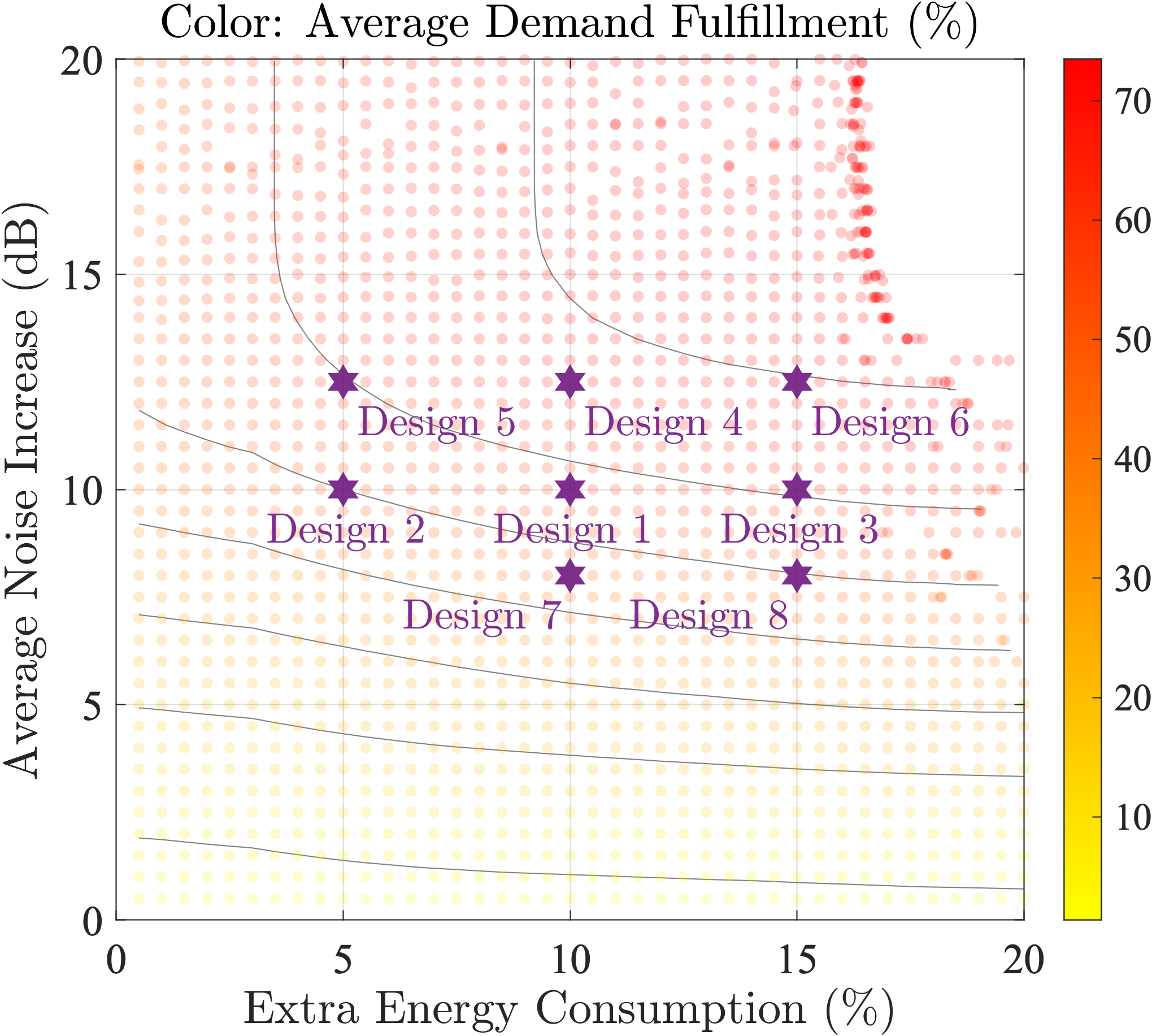}
         \caption{The eight chosen designs on the Pareto surface}
         \vspace{0.15cm}
         \label{fig:designs}
     \end{subfigure}
     \caption{The three-dimensional trade-off and design options}
     \label{fig:3dndesigns}
\end{figure}

We then investigate the trade-off between the three factors in the optimization problem: demand fulfillment, noise control, and energy saving. Here, we fully explore the design space by running 2,500 cases through optimization problem \eqref{eqn: linear app} under different parameter and constraint settings. Each design point has three attributes: average demand fulfillment (\%), average noise increase (dB), and extra energy consumption (\%). The left plot Figure~\ref{fig:3dtradeoffs} shows the distribution of generated design points in a three-dimensional space. An ideal design is expected to have high demand fulfillment, low noise increase, and low extra energy consumption. While the ideal corner (near $(0, 0, 80)$, the upper corner pointing outwards) is unattainable, the distribution of points in the left plot Figure~\ref{fig:3dtradeoffs} forms a three-dimensional Pareto surface for multi-objective decision making. The right plot of Figure~\ref{fig:3dtradeoffs} is an overhead view of the three-dimensional Pareto surface, where the magnitude of average demand fulfillment is indicated by color coding and contours. Through the contours, we observe that average noise increase is a dominating factor in determining average demand fulfillment, especially when the latter is below 50\%. When both average demand fulfillment and average noise increase are high, energy consumption begins to play a role in the three-dimensional trade-off.

\begin{figure}[htbp]
	\centering
        \includegraphics[width=0.95\textwidth]{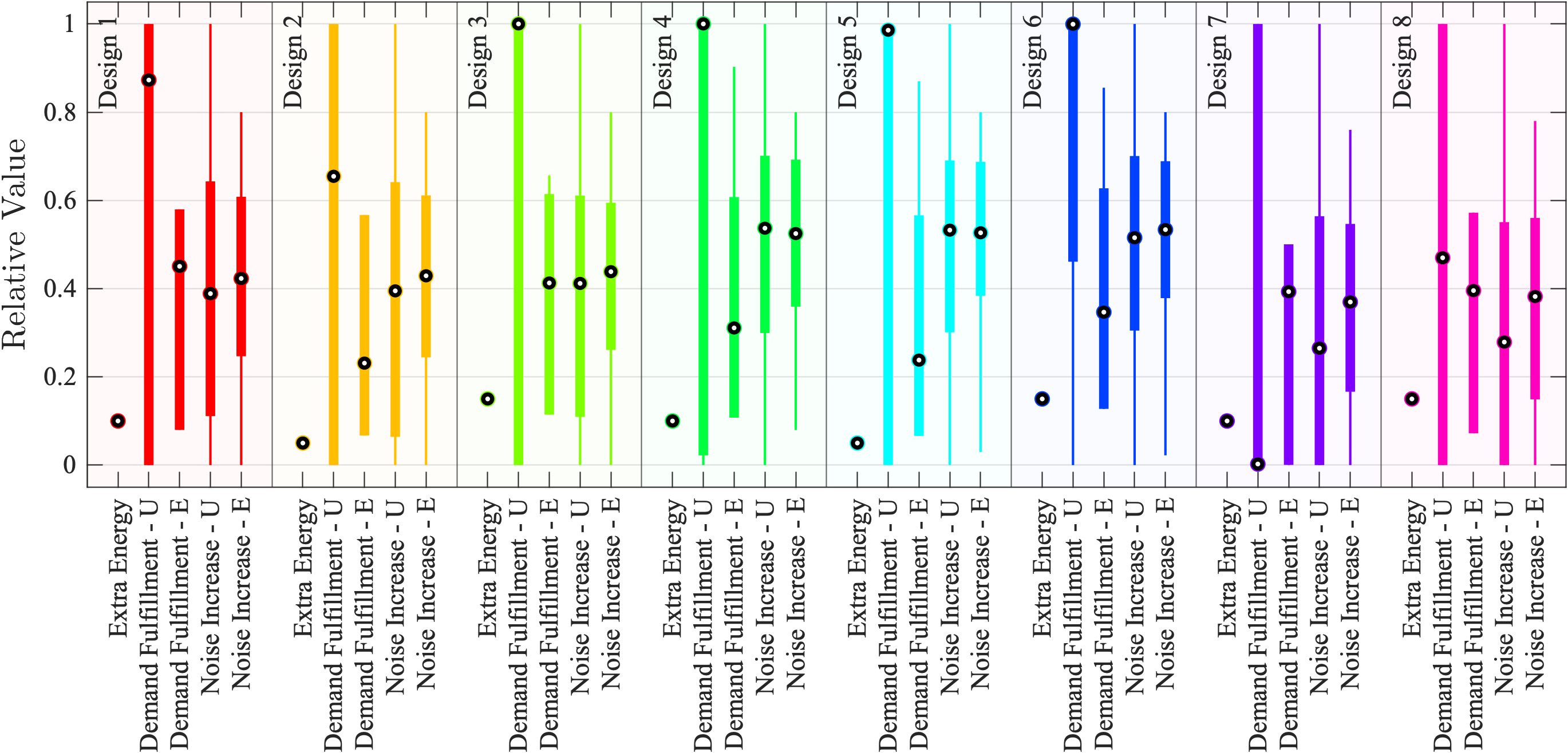}
	\caption{Boxplot comparison between the 8 chosen designs}
	\label{fig:boxplot}
\end{figure}

With the three-dimensional Pareto surface in the left plot of Figure~\ref{fig:3dtradeoffs}, we proceed to choose 8 candidate designs near the ideal corner for further analyses and comparisons. Figure~\ref{fig:designs} shows the locations of the 8 chosen designs on the Pareto surface and the contours. For each design, we further generate a utilitarian result and an egalitarian result, simultaneously for both demand fulfillment and noise increase. Figure~\ref{fig:boxplot} shows the boxplot comparison between the 8 chosen designs. Each compact boxplot displays a distribution's median (black and white dot), first quartile, third quartile, minimum, and maximum. In addition, metrics from all three objectives are normalized to be within $[0, 1]$. For each design, we display its properties on five aspects: extra energy, demand fulfillment under utilitarian criterion (U), demand fulfillment under egalitarian criterion (E), noise increase under utilitarian criterion (U), noise increase under egalitarian criterion (E). While extra energy is represented by a single value, the other four properties are represented by distributions. Within each design, one can clearly observe the efficiency-fairness trade-off through comparing between the utilitarian and egalitarian results. When we compare across the 8 utilitarian designs, Design 1 and Design 3 both have high demand fulfillment and moderate noise increase, while Design 3 consumes more energy (15\% to 10\%). When we compare across the 8 egalitarian designs, Design 1, Design 3, and Design 8 are good candidates on demand fulfillment and noise increase, while Design 1 consumes the least energy.

\begin{figure}[htbp]
	\centering
        \includegraphics[width=0.45\textwidth]{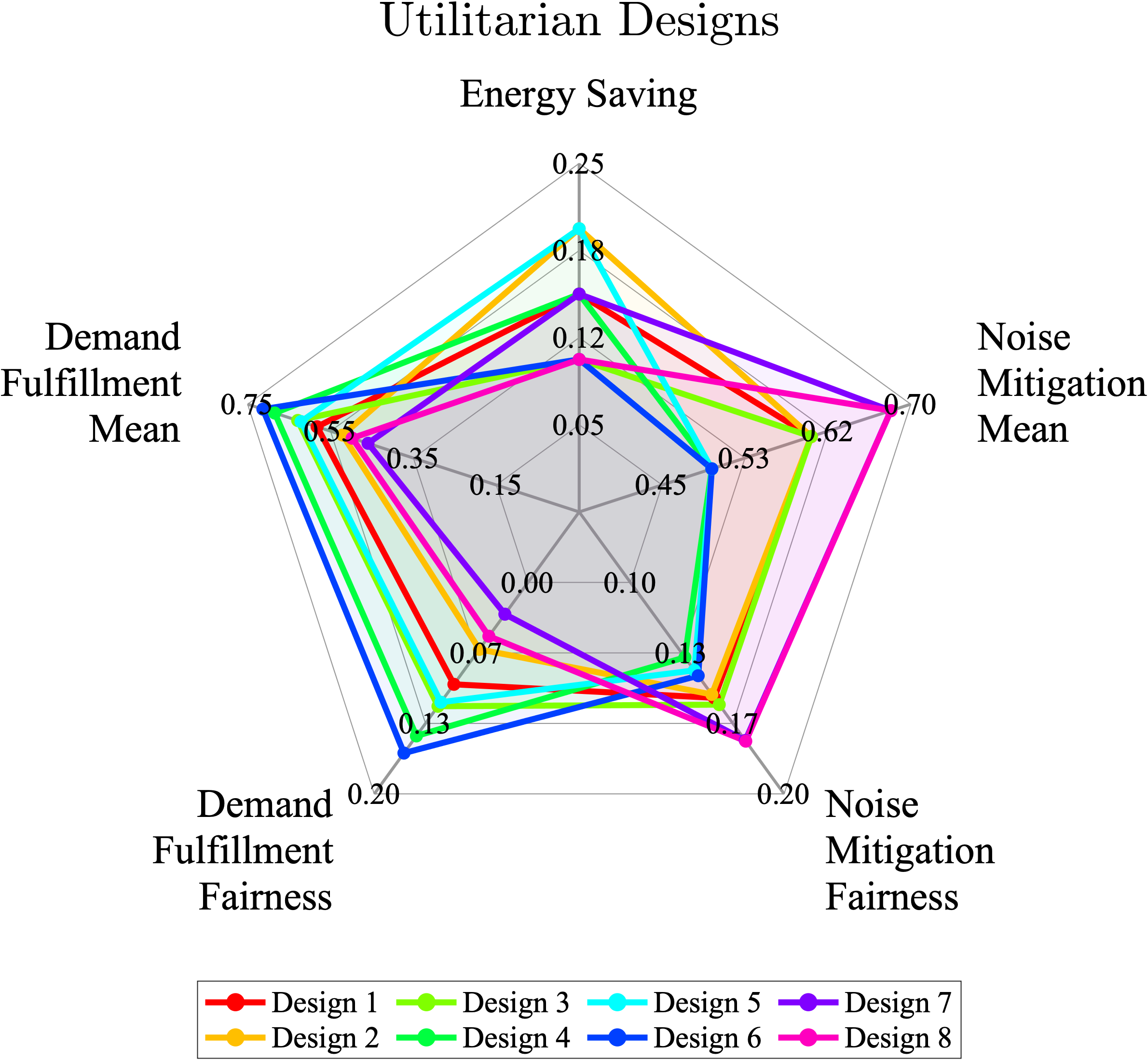}
        \hspace{1cm}
        \includegraphics[width=0.45\textwidth]{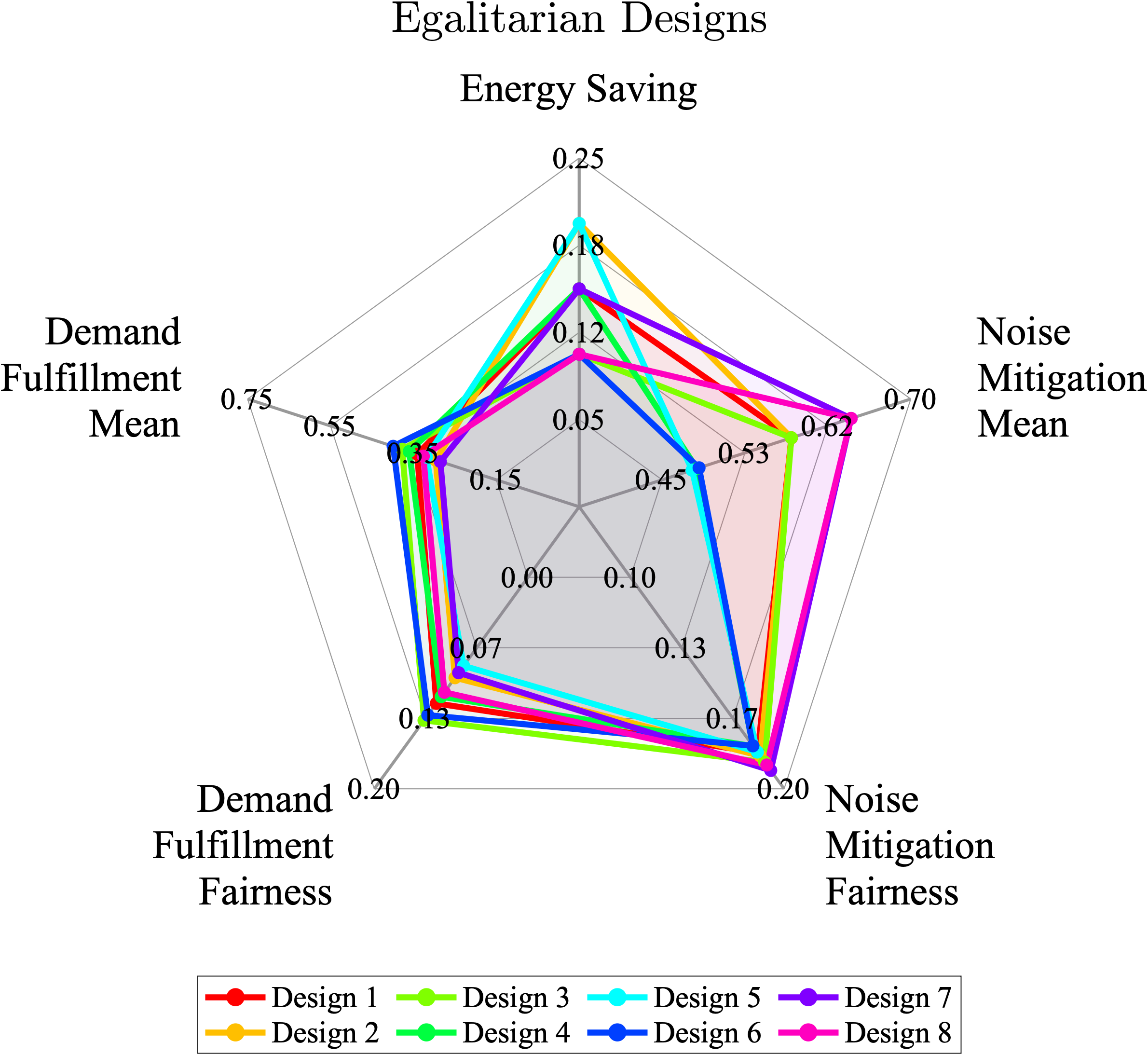}
	\caption{Radar chart comparison between the 8 chosen designs}
	\label{fig:radar}
\end{figure}

In the final set of analyses, we employ radar chart to compare the 8 designs from a slightly different angle. Figure~\ref{fig:radar} displays the two radar chart results, one for utilitarian designs, and another for egalitarian designs. Compared to Figure~\ref{fig:boxplot}, we now convert the five metrics such that a larger value is preferable for all five dimensions of Figure~\ref{fig:radar}. The five new dimensions are: demand fulfillment mean, demand fulfillment fairness, noise mitigation mean, noise mitigation fairness, and energy saving. From the left plot of Figure~\ref{fig:radar}, we can see that Design 1 (red) is the most balanced design among the utilitarian designs. Design 2 (yellow) and Design 3 (light green) are two other potential options. Among the egalitarian designs in the right plot of Figure~\ref{fig:radar}, Design 1 (red), Design 2 (yellow) and Design 8 (magenta) are potentially good options. When examining the differences between utilitarian and egalitarian designs in their entirety, the most striking observation is that in egalitarian designs, noise mitigation fairness is improved significantly at the cost of considerable reduction in demand fulfillment mean. Because of the multi-objective nature of most real-world design and planning problems, no design scheme is high performing in all dimensions. Among the 8 candidate designs, Design 1 (which is also at a `central' location on the Pareto surface) is balanced in all dimensions. Other designs can be very strong in a subset of the dimensions. For example, Design 8 has top performances in noise mitigation mean and fairness; Design 6 has top performances in demand fulfillment mean and fairness. In a scenario where each dimension is equally valued, from results in Figure~\ref{fig:radar} one could find a single quantitative metric that represents the `total' benefit of a design scheme.

\section{Remarks}\label{sec:remarks}


While the framework proposed is holistic in addressing various elements of urban air traffic management, it retains certain limitations that necessitate continued exploration and research. In this section, we briefly discuss three future avenues that will continue to advance urban air traffic control and noise management.

First, while the existing framework incorporates three key aspects--demand fulfillment, noise control, and energy consumption--there are additional perspectives in traffic flow management that could also be taken into account. For example, traffic complexity is one such perspective. The concern of traffic complexity arises as the demands in air traffic networks increase and challenge the ability of the system to maintain safe operations under various conditions~\cite{histon2002introducing}. Existing studies on air traffic complexity take airspace structures, traffic flow, and operational constraints into account, and compute the complexity metrics based on linear~\cite{wang2021air,wang2022complexity} or nonlinear dynamical systems~\cite{delahaye2003air}. Incorporating traffic complexity into the problem formulation can further improve the overall performance of air traffic management.

The second future avenue is the optimization of flight corridors. The current network structure in Figure~\ref{fig:3daustinnetwork} considers aspects such as system efficiency, safety, and navigation. Nevertheless, to maximize the potential of ambient noise making, there is room for additional optimization in the layout of flight corridors. The fundamental approach is to reconfigure the flight corridors to align more directly with the city's ambient noise patterns. Furthermore, introducing variabilities in network structures across different layers could be beneficial. The current three-layer network has the same network structure in all three layers. Adapting the network structure to be more flexible and variable could continuously enhance the system in various areas, including noise mitigation, safety, and efficiency.

The third avenue for future exploration involves incorporating additional considerations of societal impact. Aside from community noise impact, other societal impacts of UAM include third-party risk, perceived risk, privacy, etc. Through a combination of physics-based and data-driven modeling, one can also develop those societal impact maps for a city. By doing so, a more thorough spectrum of societal impact insights in complex urban environments can be achieved, enhancing UAM's societal acceptance through projects like network design, trajectory planning, and traffic flow management.

\section{Conclusion}\label{sec:conclusions}

To promote the societal acceptance of UAM and mitigate UAM's community noise impact, this paper presents a holistic optimization approach for noise-aware and equitable urban air traffic management. The overall approach concurrently considers a mix of different noise mitigation strategies (reducing operation volume, cruising at higher altitudes, ambient noise masking), conducts trade-off between three main design objectives (demand fulfillment, noise control, energy saving), and performs efficiency-fairness trade-off in aspects that involve multiple parties (demand fulfillment and noise control). Furthermore, it utilizes a city's ambient noise data to fully explore ambient noise masking, a less examined strategy in the literature. The optimization problem leverages social welfare function in objectives and seeks to optimally assign air traffic flow in a multi-layer network over a city. A numerical method involving convex-concave procedure is proposed to solve the optimization problem. A comprehensive case study using the city of Austin demonstrates the proposed approach's benefits in managing UAM's community noise impact in desired ways and performing various design trade-offs. Enhancements to the current approach can be achieved by integrating additional air traffic management aspects, optimizing flight corridors, and considering more societal impacts.

\section*{Acknowledgements}

This work was partially supported by the National Aeronautics and Space Administration (NASA) University Leadership Initiative (ULI) program under project ``Autonomous Aerial Cargo Operations at Scale'', via grant number 80NSSC21M071 to the University of Texas at Austin. Any opinions, findings, conclusions, or recommendations expressed in this material are those of the authors and do not necessarily reflect the views of the project sponsor. 

\newpage

\bibliographystyle{model1-num-names}
\bibliography{main.bib}

\newpage

\appendix

\section{Energy Consumption of eVTOL Aircraft}\label{sec:ec}

\begin{figure}[h!]
     \centering
     \begin{subfigure}[b]{0.47\textwidth}
         \centering
         \includegraphics[width=\textwidth]{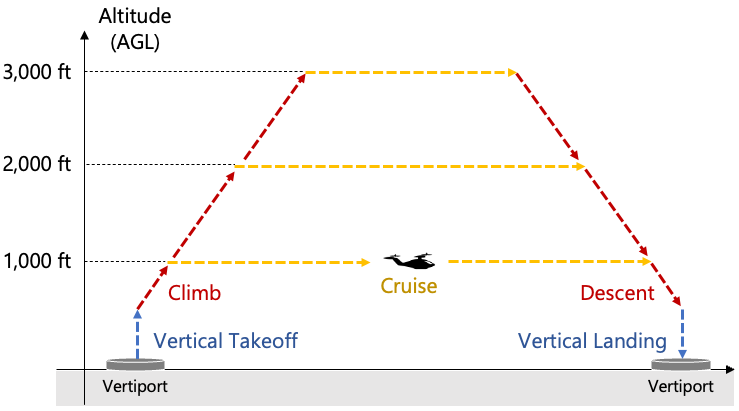}
         \vspace{0.1cm}
         \caption{The eVTOL aircraft flight profile}
         \label{fig:es1}
     \end{subfigure}
     \hspace{0.5cm}
     \begin{subfigure}[b]{0.42\textwidth}
         \centering
         \includegraphics[width=\textwidth]{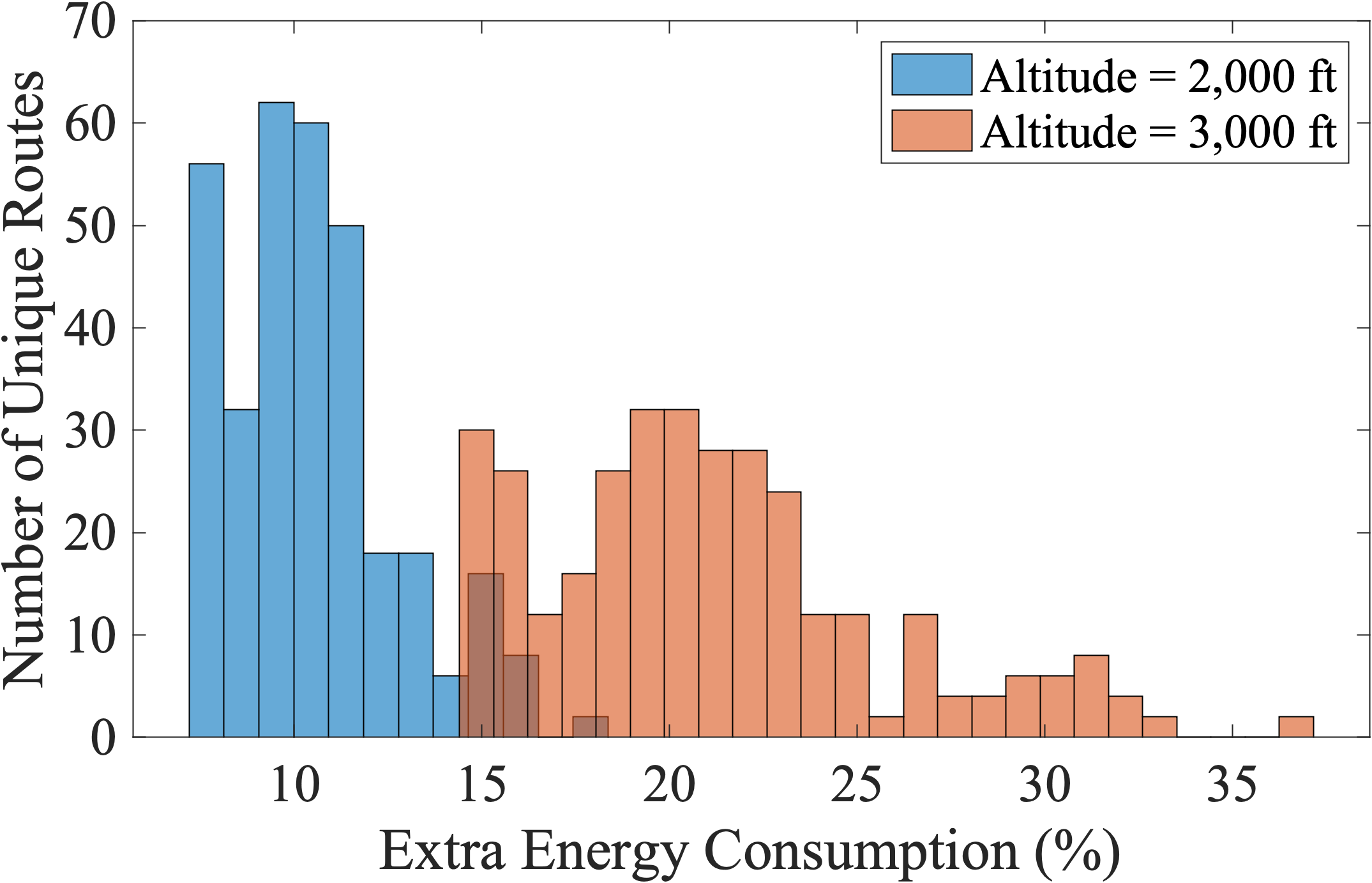}
         \caption{Distribution of Extra Energy Consumption}
         \label{fig:es2}
     \end{subfigure}
     \caption{Energy consumption information}
     \label{fig:es}
\end{figure}

The analysis of energy/power consumption during a UAM mission starts with a standard flight profile, depicted in Figure~\ref{fig:es1}. Typically, there are five segments in a UAM flight profile: vertical takeoff, climb, cruise, descent, and vertical landing. The total energy consumption of the mission is the summation of energy consumptions in all five segments. Below we analyze the energy consumption model for each segment.

Both the vertical takeoff and landing segments can be approximated as hover, which is the most energetically intensive segment of the flight profile~\cite{kasliwal2019role}. Using the momentum theory for helicopter~\cite{johnson2012helicopter}, we compute the hover power of an eVTOL aircraft with:
\begin{equation}
    P_{\text{hover}} = \frac{mg}{\eta_h} \sqrt{\frac{\delta}{2 \rho}} \,.
\end{equation}
where $m$ (kg) is the aircraft takeoff mass, $g = 9.81 ~\text{m}/\text{s}^2$ is the acceleration due to gravity, $\eta_h$ is the hover system efficiency, $\delta$ is the disk loading, and $\rho$ is the air density. Based on estimated data for a  tilt-propeller configuration eVTOL aircraft~\cite{stoll2014conceptual,stoll2022development}, we assume $\delta = 580~\text{N}/\text{m}^2$, $m = 1800~\text{kg}$, $\eta_h = 0.7$, and have:
\begin{equation}
    P_{\text{hover}} = \frac{1800~\text{kg} \cdot 9.81 ~\text{m}/\text{s}^2}{0.75} \sqrt{\frac{580~\text{N}/\text{m}^2}{2 \cdot 1.225~ \text{kg}/\text{m}^3}} = 362.3~\text{kW} \,.
\end{equation}

The AGL altitude where vertical takeoff ends (and vertical landing starts) depends on the environment and topography surrounding the vertiport. Here we assume that for both hover segments, the aircraft takes 30 seconds to climb/descent 250 ft (76 m). The energy consumption for vertical takeoff and landing is
\begin{equation}
    E_{\text{hover}} = P_{\text{hover}} \cdot t_{\text{hover}} = 362.3~\text{kW} \cdot 60~\text{s} = 21.7~\text{MJ} 
\end{equation}
which is constant for missions at all three altitudes in this study.

Next, we consider the climb segment. We assume a small flight path angle ($\gamma = 10~\text{deg}$) and that weight is approximately equal to lift during this process. We have:
\begin{equation}
    P_{\text{climb}} = \frac{TV_{\text{climb}}}{\eta_c} = \frac{V_{\text{climb}}}{\eta_c} (mg \sin{\gamma} + D) 
\end{equation}
where $\eta_c$ is the climb system efficiency, and the drag $D$ is
\begin{equation}
\begin{aligned}
    D = \frac{1}{2} \rho V_{\text{climb}}^2 S \left(C_{D_0} + K C_L^2\right) &= \frac{1}{2} \rho V_{\text{climb}}^2 S \left(C_{D_0} + \frac{1}{4 C_{D_0} (L/D)_{\text{max}}^2} \frac{(m g)^2}{(1/2 \rho V_{\text{climb}}^2 S)^2}\right)\\
    &= \frac{1}{2} \rho V_{\text{climb}}^2 S C_{D_0} + \frac{1}{4 C_{D_0} (L/D)_{\text{max}}^2} \frac{(m g)^2}{1/2 \rho V_{\text{climb}}^2 S}
\end{aligned}
\end{equation}
where $C_{D_0}$ is the coefficient for skin friction and form drags of the aircraft (zero-lift drag coefficient), $S$ is the reference area, and $(L/D)_{\text{max}}$ is the maximum lift-to-drag ratio. Then, we have the climb power of an eVTOL aircraft:
\begin{equation}
    P_{\text{climb}} = \frac{V_{\text{climb}}}{\eta_c} \left(mg \sin{\gamma} + \frac{1}{2} \rho V_{\text{climb}}^2 S C_{D_0} + \frac{1}{4 C_{D_0} (L/D)_{\text{max}}^2} \frac{(m g)^2}{1/2 \rho V_{\text{climb}}^2 S}\right) \,.
\end{equation}

Note that when climbing from a low altitude near the ground surface to the cruising altitude, the air density $\rho$ is not a constant and decreases with altitude. However, when the altitude is below 10,000 ft MSL, a linear model can accurately represent the change in air density. Consequently, we use the mid-point air density $\rho_{\text{mid-h}}$ to represent the air density during climb. For example, when the cruising altitude is $h =$ 2,000 ft MSL, $\rho_{\text{mid-h}}$ is the air density at $(2000-250)/2 + 250 = 1125$ ft. In this work, we assume that the climb rate is ROC = 1,000 ft/min.

For the cruise segment, the power is modeled in a similar way as:
\begin{equation}
    P_{\text{cruise}} = \frac{TV_{\text{cruise}}}{\eta_r} = \frac{V_{\text{cruise}}}{\eta_r} D = \frac{V_{\text{cruise}}}{\eta_r} \left(\frac{1}{2} \rho V_{\text{cruise}}^2 S C_{D_0} + \frac{1}{4 C_{D_0} (L/D)_{\text{max}}^2} \frac{(m g)^2}{1/2 \rho V_{\text{cruise}}^2 S}\right) 
\end{equation}
where $\eta_r$ is the cruise system efficiency. Lastly, for the descent segment, we assume that the engine power is reduced to 40\% of the power required for the cruise segment. In this work, we assume that $\eta_c = 0.75$, $\eta_r = 0.8$, $(L/D)_{\text{max}} = 20$, $C_{D_0} = 0.03$, $S = 30~\text{m}^2$. With $V_{\text{climb}} = 1000/ (\sin(10~\text{deg}) \cdot 60) = 96~\text{ft/s}$, $V_{\text{cruise}} = 135~\text{ft/s}$, we have the following power estimations:
\begin{itemize}
    \item At cruising altitude 1,000 ft:
    \begin{equation}
        P_{\text{climb}} = 154.1~\text{kW}, \quad P_{\text{cruise}} = 40.8~\text{kW}, \quad P_{\text{descent}} = 16.3~\text{kW}
    \end{equation}
    \item At cruising altitude 2,000 ft:
    \begin{equation}
        P_{\text{climb}} = 154.0~\text{kW}, \quad P_{\text{cruise}} = 39.9~\text{kW}, \quad P_{\text{descent}} = 16.0~\text{kW}
    \end{equation}
    \item At cruising altitude 3,000 ft:
    \begin{equation}
        P_{\text{climb}} = 154.0~\text{kW}, \quad P_{\text{cruise}} = 39.0~\text{kW}, \quad P_{\text{descent}} = 15.6~\text{kW}
    \end{equation}
\end{itemize}

We then calculate the energy consumption for the 4 segments. The energy consumption for climb is
\begin{equation}
\begin{aligned}
    E_{\text{climb}} &= P_{\text{climb}} \cdot t_{\text{climb}}\\
    &= \frac{V_{\text{climb}}}{\eta_c} \left(mg \sin{\gamma} + \frac{1}{2} \rho_{\text{mid-h}} V_{\text{climb}}^2 S C_{D_0} + \frac{1}{4 C_{D_0} (L/D)_{\text{max}}^2} \frac{(m g)^2}{1/2 \rho_{\text{mid-h}} V_{\text{climb}}^2 S}\right) \frac{(h-250)}{\text{ROC}}
\end{aligned}
\end{equation}

For the cruise segment, we need to first calculate the actual cruise distance. During the climb and descent segments, since the rate of climb/descent is 1,000 ft/min, the horizontal speed is $1000/\tan(10~\deg) = 5671$ ft/min = 94.5 ft/s. Therefore, the distance traveled for climb and descent combined is 
\begin{equation}
    d_{\text{climb+descent}} = 94.5~\text{ft/s} \cdot \frac{(h-250)\cdot 60}{1000} s \cdot 2 = 11.34(h-250)~\text{ft}
\end{equation}

The remaining distance for cruise is
\begin{equation}
    d_{\text{cruise}} = d_{\text{total}} - d_{\text{climb+descent}}
\end{equation}

Therefore, the energy consumption for cruise is
\begin{equation}
    E_{\text{cruise}} = P_{\text{cruise}} \cdot t_{\text{cruise}} = \frac{V_{\text{cruise}}}{\eta_r} \left(\frac{1}{2} \rho V_{\text{cruise}}^2 S C_{D_0} + \frac{1}{4 C_{D_0} (L/D)_{\text{max}}^2} \frac{(m g)^2}{1/2 \rho V_{\text{cruise}}^2 S}\right) \frac{d_{\text{cruise}}}{V_{\text{cruise}}}
\end{equation}

Lastly, the energy consumption for descent is
\begin{equation}
    E_{\text{descent}} = 0.4 P_{\text{cruise}} \cdot t_{\text{descent}} = \frac{0.4 V_{\text{cruise}}}{\eta_r} \left(\frac{1}{2} \rho V_{\text{cruise}}^2 S C_{D_0} + \frac{1}{4 C_{D_0} (L/D)_{\text{max}}^2} \frac{(m g)^2}{1/2 \rho V_{\text{cruise}}^2 S}\right) \frac{(h-250)}{\text{ROC}}
\end{equation}

Finally, the total energy consumption for the entire mission is
\begin{equation}
    E_{\text{total}} = E_{\text{hover}} + E_{\text{climb}} + E_{\text{cruise}} + E_{\text{descent}}
\end{equation}

Through the analysis, we observe that when operating at a higher altitude, the total energy consumption $E_{\text{total}}$ could either increase or decrease, depending on the total travel distance. On one hand, cruising at a higher altitude requires greater energy consumptions during climb ($E_{\text{climb}}$) and descent ($E_{\text{descent}}$). On the other hand, the thinner air density at a higher altitude results in reduced drag and therefore less energy consumption during cruise $E_{\text{cruise}}$ (this applies to aircraft configurations that rely mainly on fixed wings to generate lift during cruise). Hence, when the cruise distance is long enough, flying at a higher altitude may become beneficial in energy consumption. Overall, within the range of UAM, flying at a higher altitude still in general requires more energy. Figure~\ref{fig:es2} displays the distributions of extra energy consumption (in \%) when cruising at 2,000 ft and 3,000 ft AGL. Depending on the route distance, cruising at 2,000 ft rather than 1,000 ft AGL results in between 7.28\% to 18.34\% extra energy; cruising at 3,000 ft rather than 1,000 ft AGL results in between 14.70\% to 36.97\% extra energy. In the longer Regional Air Mobility (RAM) operations, this extra energy consumption could become energy-saving due to the larger portion of the cruise segment.

\end{document}